\documentclass[a4paper,11pt]{article}
\pdfoutput=1
\usepackage{style_files/jheppub}
\usepackage{breakurl}
\usepackage{mathrsfs}
\usepackage{amsmath}
\usepackage{amsfonts}
\usepackage{mathrsfs}
\usepackage{slashed}
\usepackage{epigraph}
\usepackage{fancyhdr}
\usepackage{amssymb}
\usepackage{stackengine}
\usepackage{graphicx}
\usepackage{longtable}
\usepackage{makecell}
\usepackage{amscd}
\usepackage{tabu}
\usepackage{bm}
\usepackage{color}
\usepackage{hyperref} 
\hypersetup{linktocpage=true}
\usepackage{fancyhdr}
\usepackage{cleveref}
\def\beq{\begin{equation}} 
\def\eeq{\end{equation}} 
\def\bea{\begin{eqnarray}} 
\def\eea{\end{eqnarray}} 
\def\nn{\nonumber}

\newcommand{\spar}[1]{\left\lbrack #1 \right\rbrack}
\newcommand{\rpar}[1]{\left( #1 \right)}
\title{Extending the Analysis of Electroweak Precision Constraints in Composite Higgs Models}
\author[a,b,1]{Diptimoy Ghosh}
\author[b,2]{Matteo Salvarezza}
\author[c,3]{Fabrizio Senia}
\affiliation[a]{\normalfont{Department of Particle Physics and Astrophysics, 
Weizmann Institute of Science, Rehovot 76100, Israel}}
\affiliation[b]{\normalfont{Dipartimento di Fisica, Universit\`a di Roma ``La Sapienza" and INFN, Roma, Italy}}
\affiliation[c]{\normalfont{Scuola Normale Superiore and INFN, Piazza dei 
Cavalieri 7, 56126 Pisa, Italy\\[0.5mm]}}
\emailAdd{diptimoy.ghosh@weizmann.ac.il}
\emailAdd{matteo.salvarezza@roma1.infn.it}
\emailAdd{fabrizio.senia@sns.it}
\abstract{In this paper we present a detailed analysis of the Electroweak Precision Observables (EWPO) in composite Higgs models based on the coset 
$SO(5)/SO(4)$. In our study we include both the fermionic top partners and the spin-1 resonances and consider their possible interplay as well.  In order to 
achieve calculability we use the assumptions of  i) partial Ultra Violet Completion (PUVC) following \cite{ContinoRho} and, ii) absence of sizeable effects 
from physics above the cut-off.
Apart from the EWPO, we also take into account the constraints from the top quark, $Z$ and the Higgs boson masses whenever they can be predicted 
in terms of the model parameters. After presenting our analytic results (often, in certain limits) and discussing their symmetry properties, we also perform 
detailed fits of 
the model parameters following the Bayesian approach. We show the posterior probability distributions of the parameters in various scenarios and provide 
with analytic understanding whenever possible. We find that in certain cases the EWPO  allow the compositeness scale to lie well below $1$ TeV.  Moreover, 
fermionic top partners of mass around $1$ TeV and spin-1 resonances of mass around $2-3$ TeV are  consistent with the precision data. }
\begin{document} 
\maketitle
\flushbottom
%
\section{Introduction}
\label{sec:intro}

The precise measurements of the Electroweak Precision Observables (EWPO) generically pose a serious challenge to any theory of Electroweak Symmetry 
Breaking (EWSB), in particular to theories with strong dynamics at the TeV scale. Composite Higgs models \cite{Kaplan:1983fs,Dimopoulos:1981xc,Banks:1984gj,
Kaplan:1983sm, Georgi:1984ef, Georgi:1984af, Dugan:1984hq,Contino:2003ve,Agashe:2004rs,Contino:2006nn,Contino:2010rs,ArkaniHamed:2002qx,ArkaniHamed:2002qy,Mrazek:2011iu,Gripaios:2009pe} are interesting representative of this class of theories (for a recent review, see \cite{Panico:2015jxa}).  An interesting feature of 
these theories is the requirement of fermionic top partners with mass below $\sim ~ {\rm TeV}$ in order to generate the correct Higgs mass without large 
tuning~\cite{Matsedonskyi:2012ym,Panico:2012uw,Pomarol:2012qf,Redi:2012ha,Croon:2015wba}.  The existence of light top partners also guarantees  interesting signatures at the Large Hadron 
Collider (LHC)~\cite{DeSimone:2012fs,Matsedonskyi:2014mna,Azatov:2015xqa}. However, no such signal of New Physics (NP) has been seen in the 8 TeV run 
of the LHC which, in turn, puts stringent limits on the masses of the fermionic top partners. Currently, the lower bound stands around $800~ {\rm GeV}$~
\cite{Chatrchyan:2013uxa,TheATLAScollaboration:2013sha,TheATLAScollaboration:2013oha,ATLAS:2013ima,TheATLAScollaboration:2013jha,CMS:vwa}, 
although the exact number is slightly model and species (i.e., electric charge) dependent~\cite{Matsedonskyi:2014mna}. 

The composite nature of the Higgs boson in these models also predicts that the coupling of the Higgs boson to the electroweak gauge bosons should be rescaled 
by some factor $a$ with respect to the Standard Model (SM) prediction. 
Such deviations result in shifts to the EWPO: for example, the Peskin and Takeuchi parameters $\hat{S}$ and $\hat{T}$ \cite{Peskin:1991sw, Barbieri:2004qk} 
receive corrections of the order $(\Delta\hat{S} \, , \Delta\hat{T}) \sim (+ \, , -) g^2/16\pi^2(1-a^2)\log(\Lambda/M_H)$~\cite{Barbieri:2007bh}. This  already poses a 
rather strong bound on the rescaling $a$. In addition, spin-1 resonances of mass $M_\rho$ yield sizeable tree-level contributions through mass mixings with the 
EW gauge bosons,
\beq
\Delta \hat{S} \sim \frac{M_W^2}{M_\rho^2}.
\eeq
They sum up to the positive shift coming from the Higgs\footnote{The shift arising from the composite Higgs contribution is positive if $a$ is 
less then unity. 
For models where the rescaling parameter $a$ can be greater than unity, see \cite{Falkowski:2012vh,Bellazzini:2014waa}.}, 
making the agreement with Electroweak (EW) precision 
data even worse. Given this picture, it is interesting to investigate if additional contributions can improve the agreement with the data. 
Such contributions arise 
at the one-loop level due to the spin-1 and/or fermionic resonances.

The aim of this paper is to perform an extensive one-loop analysis of electroweak precision constraints in the minimal SO(5)/SO(4) model including both the 
lowest-lying fermionic and spin-1 resonances as well as to study their interplay. These states are assumed to be lighter and more weakly interacting than 
the other composite states at the cutoff (such a spectrum can be naturally obtained in some ``5D realisation" of composite Higgs models, see for 
example ref.~\cite{Carena:2002dz} ). Note that these working assumptions of 1) a mass gap between the lightest resonances and the cut-off and 
2) calculability (i.e., weak coupling of the lightest resonances)  might not be always realised by the underlying strong dynamics and therefore, our results are not applicable in 
general. However, we expect our analysis to provide an approximate description of the contributions from composite resonances arising in the full theory.

The paper is organised as follows. In section \ref{sec:theoretical framework} we will present the relevant part of the Lagrangian with the definition of the various 
terms. We will also make some remarks about the symmetry properties that are relevant for our purpose. In the following section we will outline the details 
of our calculation for the various contributions to the EWPO.\footnote{Part of these contributions were previously calculated in \cite{PanicoGrojean} and 
\cite{ContinoSalvarezza}.}  The details of our numerical fit will be discussed in section \ref{sec:fit}. We will present our final results in section \ref{sec:results} 
and conclude thereafter.

\section{Theoretical Framework}
\label{sec:theoretical framework}
%
\subsection{The model}
\label{ssec:the model}
%

In this section we will present the basic structure of the composite Higgs models of our interest and show the various relevant parts of the Lagrangian. 
These models consist of two distinct sectors interacting with each other: a composite, strongly-interacting sector is formed at the TeV scale by a strong 
Ultraviolet (UV) dynamics and couples to an elementary, weakly-interacting sector represented by the SM fields. 

For the composite sector we will adopt the choice of the minimal composite Higgs model, based on the global symmetry breaking pattern 
$SO(5) \otimes U(1)_X \rightarrow  SO(4) \otimes U(1)_X$, where the $U(1)_X$ factor is needed to reproduce the weak hypercharge of fermions. 
The breaking of the global symmetry $SO(5) \rightarrow  SO(4)$ at some scale $f$ gives rise to 4 Nambu-Goldstone-Bosons (NGBs)
denoted by $\pi^a \,, a=1,2,3,4$. These NGBs form a 4-dimensional vectorial representation of the unbroken group $SO(4)$. As $SO(4)$ is isomorphic 
(at the Lie algebra level) to $SU(2) \otimes SU(2)$ (usually called as $SU(2)_L$ and $SU(2)_R$ respectively), the NGBs can also be 
thought as a bi-doublet (i.e., $(\bf{2},\bf{2})$) of $SU(2)_L \otimes SU(2)_R$. 
One of the basic ingredients of the composite Higgs scenario is the identification of these NGBs with the Higgs doublet. Hence, it is clear that 
$SO(5)/SO(4)$ is the minimal coset that has only one Higgs doublet and also contains the custodial symmetry protecting the $\hat{T}$ 
parameter\footnote{Additional pseudo-NGBs in the low energy spectrum can be 
present if one goes beyond the minimal coset. In such cases, depending on the model, the 
EW precision constraints can be very different. For example, in composite Higgs models with 
two Higgs doublets, the custodial symmetry protecting the $T$ parameter can be broken 
and contribution to the T parameter can be present already at tree level - this is the 
case of the cosets $SO(6)/SO(4) \times SO(2)$, $SU(5)/SU(4)$ and $SU(5)/SU(4) \times U(1)$. In this scenarios, the presence 
of discrete symmetries must be invoked in order to protect the T parameter \cite{Mrazek:2011iu,Bertuzzo:2012ya}. Other cosets
which give rise to two Higgs doublets, $SO(9)/SO(8)$ and $Sp(6)/Sp(4) \times SU(2)$ for instance, 
instead, ensure an effective custodial symmetry at tree level \cite{Bertuzzo:2012ya}. On the other hand, for models with no more 
than one Higgs doublet, like the next-to-minimal coset $SO(6)/SO(5)$ \cite{Gripaios:2009pe}, no dramatic changes 
are expected to occur to the $S$ and $T$ parameters compared to the minimal coset studied 
in this paper.}
In this construction, the rescaling factor $a$ is linked to the separation of scales $\xi$ by,
\beq
a = \sqrt{1 - \xi}, 
\hspace{1 cm}
\xi = \rpar{\frac{v}{f}}^2 \, ,
\eeq
where, $v \approx 246$ GeV is the Vacuum Expectation Value (VEV) of the Higgs doublet. 
We will adopt the formalism of Callan, Coleman, Wess and Zumino (CCWZ) \cite{Coleman:1969sm, Callan:1969sn} to build a simplified low-energy effective 
Lagrangian which describes the phenomenology of the lowest-lying fermionic and spin-1 composite resonances. Such a model has a natural cutoff $\Lambda$, 
which should be identified as the mass scale of the heavier states. The masses of the SM fermions are assumed to be generated by their linear coupling to the 
composite sector, i.e., the so-called partial compositeness mechanism \cite{Kaplan:1991dc, Contino:2006nn}. In this paper we will only consider the top quark 
to be massive and all the other quarks will be assumed to be massless. 
Partial compositeness will be implemented choosing heavy fermions into $\bf{4}$ and $\bf{1}$ dimensional representations of the unbroken $SO(4)$, and 
both the $t_L$ and the $t_R$ fields will be assumed to be partially composite. On the other hand, spin-1 resonances are introduced as triplets of $SU(2)_L$ 
and $SU(2)_R$, namely $(\bf{3},\bf{1})$ and $(\bf{1},\bf{3})$ of $SU(2)_L \times SU(2)_R$. 

We will use the criterion of Partial Ultra Violet Completion (PUVC) \citep{ContinoRho} as our rule of thumb in  order to estimate the size of the various operators in 
the effective Lagrangian. According to PUVC, all couplings in the composite sector must not exceed, and preferably saturate, the sigma model coupling 
$g^* \equiv \Lambda/f$ at the cut-off scale $\Lambda$. The only exception to this prescription is applied to the coupling strength $g_X$ associated 
(by naive dimensional analysis) to the masses of the various resonances $M_X \sim g_X f$. For these couplings we demand $g_{X} < g^*$. These 
exceptions will also fix the order of magnitude of the masses of the resonances to the  scale $M_X < \Lambda$. On the other hand, the elementary sector 
is not constrained by PUVC, thus for the operators involving elementary fields we assume a power-counting rule in which fields and derivative expansions 
are controlled by powers of $1/\Lambda$. Couplings involving the elementary sector are generally expected to be smaller than the ones purely related 
the composite strongly-interacting sector.

We will now discuss the construction of the model in detail. We will closely follow Refs.~\cite{ContinoRho,PanicoGrojean} to which we refer for more details. 
In the standard CCWZ language, one constructs the Goldstone matrix $U$ out of the 4 NGBs as
\beq
U=\exp \spar{i\sqrt{2}\frac{\pi_a(x)}{f}T^{\hat{a}}}, 
\eeq
where $T^{\hat{a}}$ are the $SO(5)/SO(4)$ broken generators and $f$ denotes the scale of the global symmetry breaking $SO(5) \rightarrow  SO(4)$ 
(note that the NGBs are neutral under $U(1)_X$). Starting from $U$ one can define the CCWZ symbols $d_\mu \in SO(5)/SO(4)$ and $E_\mu \in SO(4)$, where 
the $E_\mu$ structure can be further decomposed into $E_\mu^{L} \in SU(2)_L$ and $E_\mu^{R} \in SU(2)_R$. 
The SM gauge group is contained in a different $SO(4)^\prime \sim SU(2)_L^\prime\otimes SU(2)_R^\prime$ which is rotated by an angle $\theta$ (degree of 
misalignment) with respect to the unbroken $SO(4)$. It turns out that with this construction we have the relation\cite{ContinoRho},
\beq
\xi = \sin^2(\theta). 
\eeq

Using the $d_\mu$ and the $E_\mu$ structures one can only form one invariant operator at the level of two derivatives,
\beq
\mathcal{L}_\pi^{(2)} = \frac{f^2}{4} {\rm Tr} \spar{d_\mu d^\mu},
\eeq
which contains the mass terms for the electroweak gauge bosons:
\beq
\mathcal{L}_\pi^{(2)} \supset \frac{1}{8}f^2 \xi
\rpar{ g_{\text{el}} W_\mu^a - g^\prime_{\text{el}} \delta^{a3} B_\mu }^2 .
\eeq

Notice that, since the elementary EW gauge bosons can mix with heavy spin-1 resonances, their gauge couplings $(g_{\text{el}}, \, g^\prime_{\text{el}})$ 
must be in general different from the couplings $(g, \, g^\prime)$ of the physical states.

Fermionic resonances in the $\bf{1}$ and $\bf{4}$ dimensional representations of $SO(4)$ will be denoted by $\Psi_1$ and $\Psi_{4}$ respectively.
As mentioned before, the elementary fermions couple linearly to the composite sector following the paradigm of partial compositeness. In order to give mass 
to the top quark we must assign a $U(1)_X$ charge equal to $2/3$ to the composite fermions.\footnote{On the other hand, in order to generate the 
bottom mass one needs a $X=-1/3$ representation.} 
Hence, $\Psi_4$ and $\Psi_1$ transform as ${\bf4_{2/3}}$ and ${\bf1_{2/3}}$ of the unbroken $SO(4) \otimes U(1)_X$ respectively. 
Given these quantum numbers, the $\Psi_4$ and $\Psi_1$ fields can be written in terms of the electric charge eigenstates as
\beq
\Psi_4 = \frac{1}{\sqrt{2}} \rpar{ \begin{array}{c}
iB-iX^{5/3}\\
B+X^{5/3}\\
iT+iX^{2/3}\\
-T+X^{2/3}\\
0
\end{array} }
,\hspace{1cm}
\Psi_1 = \rpar{ \begin{array}{c}
0\\
0\\
0\\
0\\
\tilde{T}
\end{array}}.
\eeq

In the above notation, both the fields $\Psi_4$ and $\Psi_1$ symbolically transform as $\Psi_{1,4} \rightarrow h \Psi_{1,4}$ 
(note that $h \Psi_1 = \Psi_1$) where $h$ is an element of $SO(4)$.

When both  $\Psi_4$ and $\Psi_1$ are present, the Lagrangian of the composite fermions at leading order in the derivative expansion can be written as
\beq
\mathcal{L}_\Psi = \overline{\Psi}_4 \spar{i\rpar{\slashed{\nabla}+i\frac{2}{3}\slashed{B}} - M_4}\Psi_4 + 
\overline{\Psi}_1 \spar{i\rpar{\slashed{\partial}+i\frac{2}{3}\slashed{B}} - M_1}\Psi_1 + \rpar{ c_d \overline{\Psi}_1 \slashed{d} \Psi_4 + \rm h.c },
\eeq
where $\nabla_\mu = \partial_\mu + i E_\mu$ is the standard CCWZ covariant derivative. For simplicity and to avoid possible $CP$-violating 
effects from the composite sector, we assume $c_d$ to be real in our analysis. 
For future convenience, it is useful to define the couplings
\beq
g_i \equiv \frac{M_i}{f}, \hspace{0.6cm} i=1,4.
\eeq
Our working assumption requires that $g_i < g*$ while PUVC sets $|c_d| \lesssim 1$. 

The elementary  quark doublet $q_L=(t_L,b_L)^T$ is a ${\bf2_{2/3}}$ of the rotated $SU(2)_L^\prime \otimes U(1)_X$ having $T_\theta^{(R)\,3} = -1/2$. It is thus embedded in an incomplete $\bf{4_{2/3}}$ of $SO(4)^\prime \otimes U(1)_X$ as
\beq
q_L^{\bf5} = \Delta_L q_L = \frac{1}{\sqrt{2}} \rpar{\begin{array}{c}ib_L\\b_L\\it_L\\-\cos\rpar{\theta}t_L\\-\sin\rpar{\theta}t_L\end{array}} \, \, \,\,
\text{where}, \hspace{0.8cm}
\Delta_L = \frac{1}{\sqrt{2}} \rpar{ \begin{matrix}
0& i\\
0& 1\\
i& 0\\
-\cos\rpar{\theta}& 0\\
-\sin\rpar{\theta}& 0
\end{matrix}}\, .
\eeq
Similarly, $t_R$ transforms as $\bf{1_{2/3}}$ and is written as
\beq
t_R^{\bf5} = \Delta_R t_R =  \rpar{\begin{array}{c}0\\0\\0\\-\sin\rpar{\theta}t_R\\\cos\rpar{\theta}t_R\end{array}} \,\,\,\, 
\text{where}, \hspace{0.8cm}
\Delta_R = \rpar{ \begin{matrix}
0\\
0\\
0\\
-\sin\rpar{\theta}\\
\cos\rpar{\theta}
\end{matrix}}.
\eeq

As mentioned earlier, we only generate a mass for the top quark and the bottom quark remains exactly massless. The mixing Lagrangian for the top 
quark for the general case in which both $\Psi_4$ and $\Psi_1$ are present can be written as
\beq
\mathcal{L}_{mix} = y_{L4}f \overline{q}_L^{\bf 5}U\Psi_4 + y_{L1}f \overline{q}_L^{\bf 5}U\Psi_1 + 
y_{R4}f \overline{t}_R^{\bf 5}U\Psi_4 + y_{R1}f \overline{t}_R^{\bf 5}U\Psi_1 + \rm h.c. \, \, .
\label{eq:Lmix}
\eeq

In order to show the various representations involved, the above operators are often written as
\beq
\begin{split}
\mathcal{L}_{mix} &= y_{L4}f \rpar{\overline{q}_L}^\alpha \rpar{\Delta_L^\dagger}^\alpha_I U_{Ij}\rpar{\Psi_4}_j + y_{L1}f 
\rpar{\overline{q}_L}^\alpha \rpar{\Delta_L^\dagger}^\alpha_I U_{I5}\tilde{T}  \\ 
&+ y_{R4}f \; \overline{t}_R \rpar{\Delta_R^\dagger}_I U_{Ij} \rpar{\Psi_4}_j + y_{R1}f \; \overline{t}_R \rpar{\Delta_R^\dagger}_I 
U_{I5}\tilde{T} + \rm h.c. \, ,
\end{split}
\eeq
where $I=1\dots5$, $j=1\dots4$ and $\alpha=1,2$ are respectively $SO(5)$, $SO(4)$ and $SU(2)_L^\prime$ indices. Notice that PUVC does not provide 
any information about the $y_i$ couplings.

The spin-1 resonances are introduced as two triplets i.e.,  $({\bf 1,3})$ and $({\bf 3,1})$ of the unbroken $SU(2)_L \otimes SU(2)_R$. They are neutral 
under $U(1)_X$ and are introduced as vector fields transforming like a gauge field. When both of them are assumed to be present, their leading order 
Lagrangian in the derivative expansion can be written as
\beq
\mathcal{L}_\rho = \sum\limits_{r=L,R} -\frac{1}{4 g_{\rho\,r}^2}\text{Tr}\spar{\rho^{r\;\mu\nu}\rho_{\mu\nu}^{r}}+
\frac{1}{2}\frac{M_{\rho\,r}^2}{g_{\rho\,r}^2} \text{Tr}\spar{\rpar{\rho_\mu^r - E_\mu^r}^2}+\alpha_{2\,r}\text{Tr}\spar{\rho^{r\;\mu\nu}f_{\mu\nu}^{r}},
\label{eq:rho Lagrangian}
\eeq
where $f_{\mu\nu}^r$ is another CCWZ structure obtained from the field strength of the EW gauge bosons (see eq.(10) of Ref.~\cite{ContinoRho}). 
The above Lagrangian generates both kinetic and mass mixings between composite vectors and elementary EW gauge bosons. We define the 
following parameters for future convenience,
\beq
a_{\rho\,r} \equiv \frac{M_{\rho\,r}}{g_{\rho\,r} f} \, , 
\hspace{0.8cm}
\beta_{2\,r} \equiv \alpha_{2\,r} g_{\rho\,r}^2 \, ,
\hspace{1cm} (r=L,R).
\label{arho-beta2}
\eeq
While PUVC requires $a_{\rho\,r} \lesssim 1$, the coefficients $\alpha_{2\,r}$ are not constrained by such criterion as they couple the elementary and 
the composite sectors. However, as discussed in \citep{ContinoRho}, large values of $\alpha_{2\,r}$ can result in wrong-sign kinetic terms in the 
physical basis of spin-1 fields. 
For example, if only one vector resonance is present, this gives a consistency condition $|\alpha_{2\,r}| \lesssim \ 1/g_{\text{el}}g_{\rho\,r}$.
Moreover, as pointed out in Ref. \cite{Contino:2015gdp}, when $\alpha_{2\,r}$ are included the estimate of the contributions to the 
$\hat{S}$ parameter coming from the states above the 
cutoff gets an additional correction which is increased by a factor $\alpha_{2\,r}^2g_{\rho\,r}^4(\Lambda^2/M_\rho^2)$. Such enhancement is 
avoided by the choice $|\alpha_{2\,r}| \lesssim a_{\rho\,r}/g_{\rho\,r}g_*$ which is stronger than the consistency condition mentioned above.

One may also consider the four-derivative operators
\beq
Q_{1\,r} = \text{Tr}\rpar{\rho^{r\;\mu\nu}i\spar{d_\mu,d_\nu}}, 
\eeq
whose coefficients, $\alpha_{1\,r}$, are constrained by PUVC with a similar bound, $|\alpha_{1\,r}| \lesssim 1 / g^* g_{\rho\,r} $. However, their impact on 
EWPO is purely at the one-loop level, whereas the operators in \eqref{eq:rho Lagrangian} contribute at the tree level to the $\hat{S}$ parameter. 
Thus,  these operators are not considered in our calculations.\footnote{The authors of ref.~\cite{Azatov:2013ura} used a different basis for the 
four-derivative operators, namely 
$Q_{1\,r}^\prime = \text{Tr}\rpar{\rho^{r\;\mu\nu}i\spar{d_\mu,d_\nu}}$ and $Q_{2\,r}^\prime = \text{Tr}\rpar{\rho^{r\;\mu\nu} E^r_{\mu\nu}}$.  
Their basis is related to our basis by the relation $E_{\mu\nu} = f_{\mu\nu}^+ - i\spar{d_\mu,d_\nu}$. For our purpose we find it inconvenient 
to use their basis because in that case the PUVC bounds on $\alpha_{1\,r}^\prime$ would be correlated with the bound on $\alpha_{2\,r}^\prime$.  
Hence, even at the price of enhancing the contribution from the cut-off, it will not  be possible to explore a large $\alpha_{2\,r}^\prime$ with 
much smaller $\alpha_{1\,r}^\prime$. Moreover, the operators $Q_{1\,r}$ give rise to tree-level contributions to the decay 
$h\rightarrow Z\gamma$, hence it is expected to be one loop suppressed under the requirement of minimal coupling \cite{Azatov:2013ura}.}

If the fermion and the vector resonances are considered together, one can write new operators which couple directly the two sectors. The leading ones in the 
derivative expansion read,
\beq
\mathcal{L}_{\Psi\rho} = \sum\limits_{r=L,R} c_r \overline{\Psi}_4 \rpar{\slashed{\rho}^r-\slashed{E}^r}\Psi_4.
\label{eq:rhopsi Lagrangian}
\eeq

Again using PUVC we have $|c_r| \lesssim 1$. Just for completeness, below we also give the kinetic terms for the 
elementary fields,
\beq
\mathcal{L}_{\text{el}} = -\frac{1}{4 g_{\text{el}}^2}\text{Tr}\spar{W^{\mu\nu}W_{\mu\nu}} -\frac{1}{4 g_{\text{el}}^{\prime\,2}}\text{Tr}\spar{B^{\mu\nu}
B_{\mu\nu}} + \overline{q}_L i\slashed{D}q_L + \overline{t}_R i\slashed{D}t_R \, .
\eeq

Before closing this subsection, we would like to make the readers aware that the coset $SO(5)/SO(4)$ also implies the 
existence of axial-like resonances corresponding to the broken generators and transforming as a (2,2) of the unbroken 
$SO(4)$. Typically, the mass of such states are also expected to be close to  that of the two triplets and hence, should 
ideally be considered in a full analysis. However, the (2,2) resonances do not contribute to the $S$, $T$ parameters and $Zbb$ 
vertex at the tree level. This can be understood by noting that the (2,2) resonances do not mix with the $W$ and $Z$ bosons 
because of the difference in their quantum numbers (at one-loop level, they can in principle contribute). We have neglected 
these resonances in the present work.

\subsection{Symmetries}
\label{ssec:symmetries}

In this section we will discuss some approximate symmetries and symmetry-related issues which are relevant for our purpose.

The so-called $P_{LR}$ parity plays an important role in the EWPO as it is related to corrections to the $Z\overline{b}_Lb_L$ coupling \cite{ContinoZbb}. 
The action of this discrete symmetry is to exchange the generators of the unbroken $SU(2)_{L}$ and $SU(2)_{R}$ subgroups, and change the sign of the 
first three broken generators. It can be defined by the following action on the fields\footnote{The transformation properties of fermionic resonances under 
$P_{LR}$ can be obtained by expressing the ${\bf(2,2)}$ multiplet in a matrix form (in analogy with NG bosons),
%
$
\tilde{\Psi}_4 = T^{\hat{a}} \rpar{\Psi_4}_a
$ where, $a=1\dots4.$}
\begin{alignat}{3}
\pi^a &\rightarrow \eta^a\pi^a, &\qquad
B &\rightarrow -B, &\quad
T &\leftrightarrow -X^{2/3} , \nonumber \\
\rho_\mu^{L\;a} &\leftrightarrow \rho_\mu^{R\;a}, &\qquad
X^{5/3} &\rightarrow -X^{5/3}, &\quad
\widetilde{T} &\rightarrow \widetilde{T},
\end{alignat}
where, $\eta^a = \rpar{-1,-1,-1,+1}$. Using the above rules one can derive
\beq
d_\mu^a \rightarrow \eta^a d_\mu^a,
\hspace{1 cm}
E_\mu^{L\;a} \leftrightarrow E_\mu^{R\;a}.
\hspace{1cm}
\eeq

Following  the above definition of $P_{LR}$ one can check that it is an exact symmetry of the composite sector if masses and couplings 
of $\rho_\mu^L$ are equal to those for $\rho_\mu^R$.\footnote{The symmetry is lost when operators beyond the leading chiral order in the 
NG bosons Lagrangian are introduced. 
It is possible to enforce the symmetry at all orders in the chiral expansion by promoting the global symmetry group to $O(5)$ \cite{ContinoRho}.} 
When the composite sector couples to the elementary sector, this symmetry gets explicitly broken by the gauge couplings $\rpar{g, \, g^\prime}$ and the 
four $y_{L/Ri}$ ($i=1, \, \dots4$) mass mixings whose insertions are needed to generate a nonzero $\delta g_L$. This property can lower the degree of 
divergence of the various contributions in our model as we will discuss in subsection \ref{ssec:Zbb}.

The Custodial symmetry is represented by the subgroup $SU(2)^{\prime}_{L+R}$ spanned by the generators $T^{L\;a}_\theta + T^{R\;a}_\theta$.\footnote{As 
the custodial group also belongs to the unbroken subgroup, such combination is equal to $T^{L\,a} + T^{R\,a}$.} 
It is explicitly broken by the gauge coupling $g^\prime_{\text{el}}$ and the left mixings $y_{Li}$. The insertion of powers of those mixings makes the pure 
fermionic contribution to the $\hat{T}$ parameter finite (see subsection \ref{ssec:T} below).

One interesting case, the two-site model \cite{Panico2site1,Panico2site2}, is obtained by including the full 5-plet of fermionic resonances as well as 
both the vector triplets and enforcing the relations,
\begin{alignat}{3}
y_{L1} &= y_{L4}, &\qquad
y_{R1} &= y_{R4}, &\quad
c_d &= 0, \nonumber
\end{alignat}
\vspace{-0.7cm}
\begin{alignat}{2}
g_{\rho_L} &= g_{\rho_R}, &\qquad
a_{\rho_L} &= a_{\rho_R} = \frac{1}{\sqrt{2}}\, ,
\end{alignat}
\vspace{-0.7cm}
\begin{alignat}{2}
\beta_{2_L} &= \beta_{2_R} = 0, &\quad
c_L &= c_R = -1. \nonumber
\end{alignat}

Note that the above constraints imply $M_{\rho_L} = M_{\rho_R}$. 
The Lagrangian obtained after imposing the above relations is equivalent to a model characterised by the global symmetry breaking pattern 
$SO(5)_L \otimes SO(5)_R \rightarrow SO(5)_V$ in which the EW gauge bosons and the vectorial resonances appear as gauge fields 
arising from gauging part of the $SO(5)_L$ and  the $SO(5)_R$ groups.  As a characteristic feature of this construction, the EWSB can 
only be achieved if the global symmetry in both the left and the right sites are explicitly broken. As shown in Ref. \cite{Panico2site1}, a consequence 
of this fact is that the degree of divergence of the 1PI contributions to any EWSB-related quantity is lowered. In particular, the fermionic contribution 
to the Higgs potential is changed from quadratically to logarithmically divergent, and the contributions to the $\hat{S}$ and $\hat{T}$ 
parameters become finite. Non-1PI divergent contributions can still be present because of the presence of spin-1 resonances. As pointed out in 
Ref. \cite{ContinoSalvarezza}, for the $\hat{S}$ parameter such divergences can be reabsorbed by the renormalization of the parameters of 
$\mathcal{L}_\rho$, making it fully calculable at one-loop, while this is not the case for the $\hat{T}$ parameter because an additional operator 
is needed to remove the non-1PI subdivergences. 
The coefficient of such operator gives an additional contribution to the mass splitting between neutral and charged vector resonances, hence 
the $\hat{T}$ parameter becomes calculable after fixing the splitting. Similarly, it is found that the Higgs potential is calculable by fixing a 
single counterterm to reproduce the Higgs VEV.

Another special case is obtained by $c_d = \pm 1$. In this case, as discussed in ref.~\citep{PanicoGrojean}, in the composite fermion Lagrangian 
the goldstone fields can be shown to appear only in the mass terms. 
Hence the degree of divergence in any EWSB related quantity is decreased. As a result of this, fermionic contributions to the $\hat{S}$ 
parameters become finite in the case $c_d = \pm 1$.

The global $SO(5)$ symmetry is explicitly broken by two different sources: the EW gauging and the fermion mass mixings. The breaking can be parameterised 
by the couplings $g_{\text{el}}$, $g^{\prime}_{\text{el}}$ and the four $y_{L/R}$ couplings. The EW observables $\hat{T}$ and $\delta g_L$ are sensitive to these 
symmetry-breaking couplings. Hence it is useful to analyse their effect with a spurion method. In particular, it is relevant to analyse the number of multiplicative 
symmetry breaking couplings that is required to generate $\hat{T}$ and $\delta g_L$.

As mentioned earlier, the breaking of $SO(5)$ proceeds through the EW gauging and the fermion mass mixing Lagrangian \eqref{eq:Lmix}. 
In particular, the terms involving the $y_L$ couplings break $SU(2)_R^\prime$, while the $y_R$ couplings preserve the full $SO(4)^\prime$. The terms 
involving the gauge couplings $g_{\text{el}}$ and $g^\prime_{\text{el}}$ respectively preserve $SO(4)^\prime$ and the gauge 
group $SU(2)_L^\prime \otimes U(1)_Y$.  Hence the custodial symmetry is broken by the couplings $y_L$ and $g^{\prime}_{\text{el}}$.
On the other hand, as $t_R$ is an $SO(4)$ singlet the $y_R$ couplings do not break the $P_{LR}$ parity, which is only broken by the $y_L$ 
and the gauge couplings.  
Moreover, since in our calculation we neglect $\mathcal{O}\rpar{g_{\text{el}},g^\prime_{\text{el}}}$ contributions to $\delta g_L^{(b)}$, we will be 
only interested in the $y_L$ mediated contributions to $\delta g_L^{(b)}$ and $\hat{T}$, and $g^\prime_{\text{el}}$ mediated contributions 
to $\hat{T}$. 
As far as these couplings are concerned, the global $SO(5)$ can be restored by promoting them to fields with suitable transformation rules.
However, for our purpose we find it more suitable to promote the associated matrices $\Delta_L$ and $T^{R\;3}_\theta$. This is because in the fermionic 
sector there are four mass mixing couplings but only two possible mixing-related structures. For the EW gauging part we promote 
$T^{R\;3}_\theta$ to the spurion $\widetilde{T}^{R\;3}_\theta$ transforming as
\vspace{0.15cm}
\beq
\widetilde{T}^{R\;3}_\theta \rightarrow g \widetilde{T}^{R\;3}_\theta g^\dagger, 
\eeq
while for the fermion mixing, we introduce the $\widetilde{\Delta}_L$ spurions transforming as 
\beq
\widetilde{\Delta}_L \rightarrow g \widetilde{\Delta}_L k^\dagger, 
\eeq
where $k=\exp\rpar{i\;\alpha_{Li} \; \sigma^i/2}$ is a $2\times 2$ $SU(2)_L^\prime$ representation.

Using $\tilde{T}^{R\;3}_\theta$ we can form the $SO(4)$-covariant matrix
\beq
\tau = U^\dagger \widetilde{T}^{R\;3}_\theta U,
\hspace{1cm}
\tau \rightarrow h \tau h^\dagger.
\eeq

Two structures, a matrix and a 5-component vector, can also be constructed by $\tilde{\Delta}_L$:
\begin{alignat}{2}
\chi_L &= U^\dagger \Delta_L q_L, &\qquad
\chi_L &\rightarrow h\chi_L, \\
\eta_L &= U^\dagger \Delta_L \Delta_L^\dagger U, &\qquad
\eta_L &\rightarrow h \eta_L h^\dagger.
\end{alignat}

The structures $\tau$, $\chi_L$ and $\eta_{L}$ can be further deconstructed as fourplets and singlets of $SO(4)$ as

\begin{align}
\rpar{\tau}_{IJ} &= \; \tau_{ij} + \tau_{i5} + \tau_{5j} + \tau_{55}, \\
\rpar{\chi_L}_I &= \; \chi_{L\,i} + \chi_{L\,5}, \\
\rpar{\eta_L}_{IJ} &= \; \eta_{L\,ij} + \eta_{L\,i5} + \eta_{L\,5j} + \eta_{L\,55}.
\end{align}
where, the range of the indices are given by $I,J=1\dots 5$ and  $i,j=1\dots 4$.

Using only the $\tau$ field one can form a custodial symmetry breaking operator which contributes to the $\hat{T}$ parameter \cite{ContinoSalvarezza},
\beq
O_T^{(\tau)} = \text{Tr}\spar{d_\mu \tau}^2.
\label{eq:OTtau}
\eeq

Similarly, using the $\eta$ field we can build
\beq
O_T^{(\eta)} = \text{Tr}\spar{d_\mu \eta}^2 \, .
\label{eq:OTtau}
\eeq
It is now clear that in order to generate the $\hat{T}$ parameter one either needs two powers of the coupling $g^\prime_{\text{el}}$ or four 
powers of $y_{L(1/4)}$.\footnote{Other invariant operators with two powers of $\tau$ or $\eta$ are not independent and differ only by 
custodial symmetry invariant terms.}

As shown in ref.~\cite{PanicoGrojean}, in the effective theory with only SM particles and at the lowest order in the mass mixing spurions, one can form 
the following local operator contributing to $\delta g_{L}^{(b)}$, 
\begin{align}
O_\delta^{\rpar{qq}} &= \overline{\chi}_L \gamma^\mu \chi_L \text{Tr}\spar{\eta_L d_\mu}\label{Odeltaqq} \, .
\end{align}

The above operator contains four powers of the spurion $\Delta_L$, and hence at least four powers of $y_{L1/4}$ are needed to generate 
$\delta g_L^{(b)}$. 
Notice that the vectorial sector also introduces a genuine breaking of $P_{LR}$ without breaking $SO(5)\times U(1)$.

\section{Calculation of Electroweak Precision Observables}
\label{sec:calculations}

In this section we will discuss the details of our computation of the electroweak precision observables. Some of our results were computed for the first 
time in ref.~\cite{PanicoGrojean,ContinoSalvarezza}. The $\epsilon$ parameters
\cite{Altarelli:1990zd,Altarelli:1991fk} that are relevant for us are
\begin{align}
\epsilon_1 &= e_1 - e_5 + \text{non-oblique leptonic terms} , 
\\
\epsilon_3 &= e_3 + c^2 e_4 - c^2 e_5 + \text{non-oblique leptonic terms} \, , 
\end{align}
where $c$ and $s$ denote the sine and cosine of the weak mixing angle respectively, and the $e_i$ are related to gauge bosons vacuum 
polarisation amplitudes in the following way, 
\beq
\begin{aligned}
e_1 &= \frac{1}{M_W^2}\rpar{A_{33}(0) - A_{11}(0)}, &
e_3 &= \frac{c}{s} F_{3B}\rpar{M_Z^2}, \\
e_4 &= F_{\gamma\gamma}(0) - F_{\gamma\gamma}\rpar{M_Z^2}, &
e_5 &= M_Z^2 F^\prime_{ZZ}\rpar{M_Z^2} \, .
\end{aligned}
\eeq
The $A_{ij}$ and the $F_{ij}$ functions are defined through the vacuum polarization amplitudes of 
the EW gauge bosons,
\beq
\Pi^{\mu\nu}_{ij}(q) = -i g^{\mu\nu} \Pi_{ij}(q^2) + q^\mu q^\nu {\rm \, terms} \, = -i g^{\mu\nu}
\left( A_{ij}(0) +  F_{ij}(q^2) q^2 \right) + q^\mu q^\nu \rm \, terms \, .
\eeq
As we have neglected couplings between the composite sector and leptons, non-oblique leptonic terms vanish at one-loop. For our analysis we will 
need to compute the deviations from the SM predictions, namely $\Delta\epsilon_i = \epsilon_i - \epsilon_i^{(\rm SM)}$.

As discussed in reference \citep{ContinoSalvarezza}, the calculation of the oblique observables can be interpreted as a matching procedure 
at the scale of the heavy resonances followed by renormalisation group (RG) evolution down to the weak scale and an additional calculation 
for the light physics contributions (i.e. threshold corrections from the composite Higgs and the top quark at the EW scale). 
The first step corresponds to the calculation of the effects coming from the heavy resonances, and  by neglecting terms of 
$\mathcal{O}\rpar{M_Z^2/M_*^2}$ (where $*$ stands for any resonance) this becomes equivalent to the calculation of  $\hat{S}$ 
and $\hat{T}$ parameters,
\begin{align}
\hat{S} &= \frac{c}{s} \Pi_{3B}^\prime (0) \, ,\\
\hat{T} &= \frac{1}{M_W^2} \rpar{ \Pi_{33}(0) - \Pi_{11}(0) } \, .
\end{align}

On the other hand, both the RG evolution and the threshold corrections arise from the non-standard dynamics 
of the Higgs boson and also the top quark (if it mixes with the heavy fermion resonances).

Another quantity which is relevant for us is the  parameter $g_L^{(b)}$ which is defined through the non-universal corrections to the effective coupling
\beq
g_{\rm eff}^{Z \, \bar{b}_L b_L}  \equiv \dfrac{e}{sc} \, \Gamma_L^{(b)}(q^2) \, \bar{b}_L(p_1) \, \gamma^\mu \,  b_L(p_2) \, Z_\mu (q=p_1+p_2)\, .
\label{eq:EWobs}
\eeq
The quantity $\delta g_L^{(b)}$ is defined as the NP contribution to $\Gamma_L^{(b)}(q^2)$ at $q^2=0$.

We could in principle also consider the parameters $\epsilon_2$ and $g_R^{(b)}$. In general, in composite Higgs models the resonance contributions 
to the first one is suppressed by powers of $g^2/g_*^2$ (where $g_*$ is any strong coupling) and the IR contribution from the composite Higgs is 
finite and small. These suppressions are often strong enough to make the impact of $\epsilon_2$ negligible in any strongly coupled model and 
can be safely neglected (see, for example \cite{ContinoSalvarezza}). 
On the other hand, the coupling $g_R^{(b)}$ is in general expected to be produced with similar size as $g_L^{(b)}$. However, it is much less 
constrained than the left coupling (see Fig.~\ref{fig:comparison}), hence it will not have much effect on our fit and we will neglect 
it for simplicity.

We will use dimensional regularisation with the $\overline{\rm MS}$ scheme and fix the renormalisation scale to the cutoff $\Lambda$, which will be 
fixed to the somewhat arbitrary value of three times the mass of the heaviest particle in the particular model\footnote{A naive alternative 
would be to use $\Lambda = g_* \, f $ with some fixed value of $g_*$ . 
If $g_* = 4 \pi$ is chosen, in the region of small $\xi$, $\Lambda$ can be easily in the 
range 10 - 20 TeV. In this case, when the $\rho$ is in the interesting mass region 
(say 1-3 TeV), the ratio $\Lambda/M_\rho$ can be large. As the $\beta$-function of $g_\rho$ is 
generally big,  it changes a lot in the window between $\Lambda$ and $M_\rho$, and often becomes 
non-perturbative. As, perturbativity in the whole range ($\Lambda$ to $M_\rho$) is required for 
our calculations to make sense, we used [$\Lambda = 3 \, \times$ largest mass] in order to keep 
the running under control. 
Also note that, for $\Lambda = 4 \pi f$ and small $\xi$, the $\rho$ in the interesting mass region 
(again say 1-3 TeV) is too light compared to the cut-off.}.

The various contributions to the EWPO can be divided into four categories: 
\begin{itemize}
\item The purely spin-1 resonance contributions where heavy vectors propagate alone or together with light particles (NGBs, Higgs or EW gauge bosons).
\item The purely heavy fermion contributions arising from diagrams where at least one heavy fermionic resonance propagates in the loop.
\item The mixed heavy fermion-vector contribution comprising of diagrams where both type of resonances propagate.
\item The light physics contribution from the loops of only light particles. Since we are interested in the deviations with respect to the SM predictions, 
only light particles with non standard dynamics will contribute, i.e. the Higgs boson and the top quark.
\end{itemize}

Note that part of the above contributions have already been calculated in the literature. While we perform 
all the calculations independently, we will explicitly point out to the original literature in all such cases.

Logarithmically divergent terms in all four categories account for RG evolutions, while the finite terms of the first three categories represent the 
short-distance effects at the $M_*$ threshold. This is the part for which we can neglect $\mathcal{O}\rpar{M_Z^2/M_*^2}$ terms and calculate 
contributions to $\Delta\hat{T}$ and $\Delta\hat{S}$ instead of $\Delta \epsilon_{1,3}$. On the other hand, when calculating the light physics 
contributions (which must also undergo a subtraction of the SM result), we have instead to consider the full $\epsilon$ parameters, thus probing 
the full IR structure of the models: for this reason, the contribution coming from this last part to the oblique $\epsilon_{1,3}$ parameters will contain 
a complicated finite part. In all our calculations we neglect the effects coming from the propagation of 
the EW gauge bosons inside loops except for the pure vectorial contribution to the $\hat{T}$ parameter, as it is the leading contribution of the purely vectorial part.

We stress that in the class of models we are interested in (or in general, in any effective theory), EWPO also get additional unknown contributions from 
local operators that are generated from physics above the cut-off. Since they depend 
on the details of the UV theory, we are completely ignorant about these contributions. We will assume them to vanish at the cut-off scale 
(at lower scales they will in general be generated by loops of resonances and SM particles). 
It is important to stress that these contributions can be non-negligible in particular, if non-decoupling effects are present. 
Thus, without a concrete UV completion, this always leads to some (unknown) uncertainties in the predictions.

Both the fermionic and the spin-1 resonances mix with each other, hence the interaction eigenstates are in general not the mass eigenstates. 
In the fermion sector, when both the 4-plet and the singlet are present,  in the limit $\xi = 0$ there are two degenerate $SU(2)_L$ doublets 
$\rpar{T,B}$ and $\rpar{X^{5/3},X^{2/3}}$ with masses $\sqrt{M_4^2 + y_{L4}^2 f^2}$ and $M_4$ respectively and a singlet $\tilde{T}$. 
The mass degeneracy is lifted by non-zero $\xi$. The same happens for the $SU(2)_L$ triplet of vector resonances, while the mass degeneracy 
of the $SU(2)_R$ triplet resonances is already lifted  at $\xi=0$ by the coupling $g^\prime_{\text{el}}$. Elementary states are also involved in the 
mixings: the top quark and the EW vector bosons acquire mass for $\xi \neq 0$, while the bottom quark remains exactly massless, as its 
compositness has been neglected. See appendix \ref{app:mixings} for more details on mass mixings. 

Mixings in the vector boson sector are regulated by powers of $\rpar{g_{\text{el}},g_{\text{el}}^\prime}/g_{\rho_r}$, and as long as $\beta_{2\,r} \lesssim 1$ 
they can be safely treated in perturbation theory. Things are different for fermionic mixings, as they are associated with powers of 
$y_i M_j / f = y_i/g_j$ which in general may not be a suitable expansion parameter. Hence calculations have been carried out by resumming 
power of mass insertions in the fermionic sector by numerically diagonalizing fermionic mass matrices. 

In the previous few paragraphs we sketched the details of our calculation for the EWPO that will be used in the numerical fits of section \ref{sec:results}. 
We will now present simplified analytic expressions for the various contributions so that (at least) some broad features of the results can be understood 
analytically. In order to do so, at first we will neglect the threshold corrections at the EW scale coming from top quark loops. Moreover, for the fermionic resonance 
contributions we will only keep the terms that are leading order in $\xi$ and also in the mass mixings $y_i$.  In the fermionic 5-plet scenario, additionally we 
will set $y_{L1} = y_{L4} = y_L$ and $y_{R1} = y_{R4} = y_R$. Finally, the mixed fermion-vector contributions will be expanded in $M_4^2/M_{\rho_r}^2$ and 
only the leading order terms will be kept.

\subsection{Contributions to the $\hat{T}$ parameter}
\label{ssec:T}

\begin{figure}[t]
\centering
\begin{tabular}{c c c}
\includegraphics[scale=0.28]{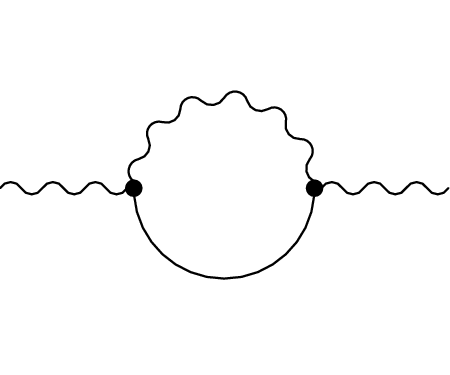} &
\includegraphics[scale=0.28]{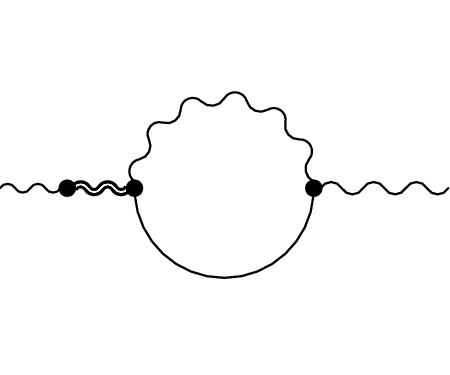} &
\includegraphics[scale=0.28]{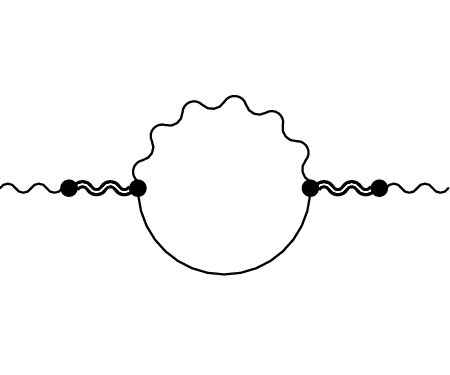} 
\\
\includegraphics[scale=0.28]{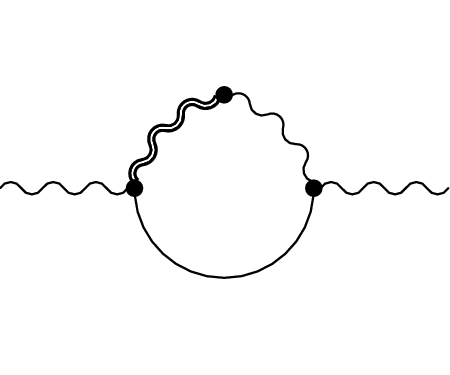} &
\includegraphics[scale=0.28]{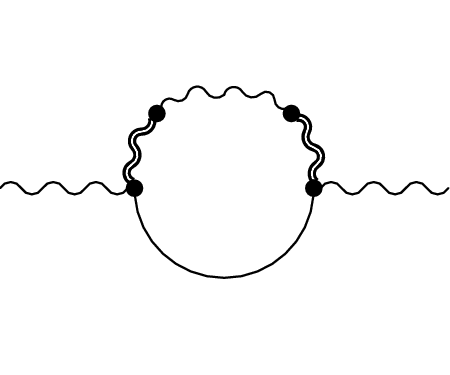} &
\includegraphics[scale=0.28]{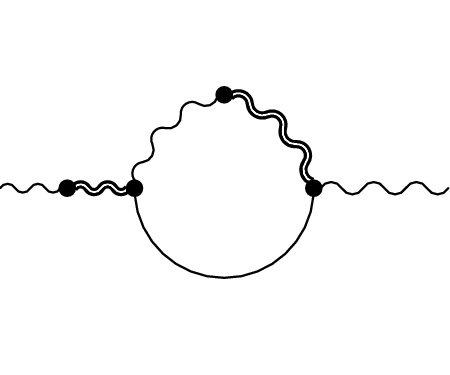}
\end{tabular}
\begin{tabular}{c c}
\includegraphics[scale=0.28]{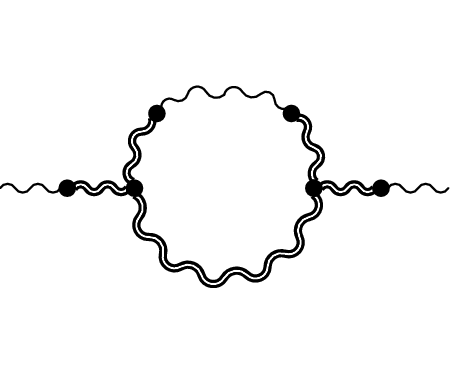} &
\includegraphics[scale=0.28]{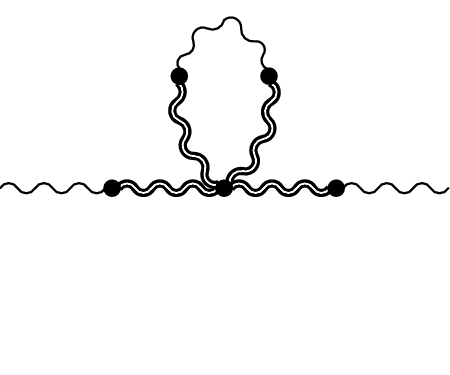} 
\end{tabular}
\caption{Pure vectorial contributions to the $\hat{T}$ parameter. Smooth single lines represent all the NGBs including the Higgs. 
External (internal) single wavy lines stand for a $W_\mu^i$ ($B_\mu$) field. Double wavy lines represent either a $\rho_L$ or a $\rho_R$. 
Although in the first diagram there is no actual resonance propagation, it gets contributions from the $\rpar{E_\mu}^2$ term of 
Eq.~\eqref{eq:rho Lagrangian}. Hence it must  be included in the resonance contribution. Note that, due to a mistake in ref.~\cite{ContinoSalvarezza},
the set of diagrams listed there (Figs.~12 and 13 therein) is slightly different from the above set. However, the mistake was only done in showing the 
list of diagrams, and in both calculations all the correct contributions have been included.}
\label{fig: diagrams T vect}
\end{figure}

Every contribution to the $\hat{T}$ parameter depends on sources of breaking of the custodial symmetry. In the absence of EW gauging, 
the sector of vectorial composite resonances is $SO(5)$ symmetric, hence insertions of the hypercharge coupling (at least two insertions, see 
section \ref{ssec:symmetries}) are needed to generate 
a contribution as shown in figure \ref{fig: diagrams T vect}. Pure vector contributions to the $\hat{T}$ parameter have been firstly calculated 
in~\cite{ContinoSalvarezza}. As pointed out in the same reference, the contribution from a single $\rho_L$ has 
the same form as that coming from a single $\rho_R$. Below, we report the expressions for the single resonance case as well as for the scenario when both 
$\rho_L$ and $\rho_R$ are present.
\begin{align}
\hat{T}\big|_{\rho_{L/R}} &= \frac{3g_{\text{el}}^{\prime\,2}}{32\pi^2} \frac{3}{4} \xi a_{\rho_{L/R}}^2\spar{\rpar{1-\frac{8}{3}\beta_{2_{L/R}}^2}\log\rpar{\frac{\Lambda}{M_{\rho_{L/R}}}}+\frac{3}{4} - \frac{4}{3}\beta_{2_{L/R}} + \frac{2}{9} \beta_{2_{L/R}}^2}, 
\label{eq:T 1rho}\\
\hat{T} \big|_{\rho_L + \rho_R} &= \hat{T}\big|_{\rho_L}+\hat{T}\big|_{\rho_R} + \hat{T} \big|_{\rho_{LR}}, 
\label{eq:T 2rho}
\end{align}
where
\beq
\begin{split}
\hat{T} \big|_{\rho_{LR}} &= -\frac{g^{\prime\,2}_{\text{el}}}{32\pi^2}\frac{3}{2}\xi a_{\rho_L}^2 a_{\rho_R}^2 \rpar{1-2\beta_{2L}-2\beta_{2R}+
4\beta_{2L}\beta_{2R}} \times \\
&\spar{ \log\rpar{\frac{\Lambda}{M_{\rho_L}}} + \log\rpar{\frac{\Lambda}{M_{\rho_R}}} - \frac{M_{\rho_L}^2 + M_{\rho_R}^2}{M_{\rho_L}^2 - 
M_{\rho_R}^2} \log\rpar{\frac{M_{\rho_L}}{M_{\rho_R}}} + \frac{5}{6} }.
\end{split}
\eeq
Notice that the contributions coming from $\beta_2$ are different from the ones presented in 
ref.~\cite{ContinoSalvarezza}. This is because we used  different basis for the four-derivative operators of the spin-1 resonances, 
see eq.~\eqref{eq:rho Lagrangian} and the text following it. 

If $|\beta_{2_r}| \lesssim 1$, it can be easily seen that the contribution coming from a single resonance can compensate for the negative shift coming from 
the composite Higgs boson (see Eqs.~\ref{eq:Deps1 Higgs} and \ref{eq:Deps3 Higgs}). When both the resonances are present, the contribution can still 
be positive however, the interplay is much more complicated and involves also 
the values of $a_{\rho_r}$.

The fermionic sector provides instead its own source of breaking, i.e. the mass mixings with elementary fermions, thus fermionic self-energies will contribute to 
the $\hat{T}$ parameter. The contributions have been firstly calcuated in ref.~\cite{PanicoGrojean}. As already reported in the same reference, 
a spurion analysis (see section \ref{ssec:symmetries}) shows that the contributions 
must appear at least at $\mathcal{O}\rpar{y_L^4}$, and are therefore finite by power counting. By including the top contributions and subtracting the SM result 
we find the following approximate results for $\Delta\hat{T} = \hat{T} - \hat{T}^{(SM)}$ in the singlet, fourplet and fiveplet cases (which are defined 
as the scenarios in which we only include a $\bf{1}$, $\bf{4}$ and $\bf{4 \oplus 1}$ of $SO(4)$ respectively):
\begin{alignat}{3}
\Delta \hat{T} \big|_{\Psi_1} &\approx \frac{3y_{L1}^4}{64\pi^2g_1^2} \xi, \label{eq:T psi1} \\
\Delta \hat{T} \big|_{\Psi_4} &\approx - \frac{y_{L4}^4}{32\pi^2g_4^2} \xi,  \\
\Delta \hat{T} \big|_{\Psi_5} &\approx \frac{y_L^4}{32\pi^2g_1^2} \xi \frac{x^2}{\rpar{1-x^2}^2} \Bigg[  -3 c_d^2 \rpar{2x^3-5x^2+18x-5} \nonumber \\
&  + 3c_d \rpar{x^3+7x^2-x-5+\frac{2}{x}} - x^4-\frac{5}{2}x^2-4+\frac{3}{2x^2} \nonumber \\
& + 6\frac{\log(x)}{1-x^2} \rpar{c_d^2 \rpar{x^4-12x^3+8x^2-8x+1} + c_d\rpar{x^4+2x^2+x-2}-2x^4+x^2-1 } \Bigg].
\end{alignat}

In the above and also the rest of this section 
$x \equiv M_1/M_4 = g_1/g_4 $.

The signs for the singlet and fourplet cases are fixed in the above expressions. This is tipically true also in the full numerical calculations, where all powers of 
$\xi$ and $y_i$ are resummed. On the other hand, in the fiveplet case there is no preference for positive or negative sign. Due to the finiteness of the 
above contributions, inverse powers of $g_{1,4}$ arise, resulting in good decoupling properties for both the fourplet and the singlet.

If one introduces the operators in eq. \eqref{eq:rhopsi Lagrangian}, new diagrams with the external propagation of heavy spin-1 resonances  appear
(as shown in Fig.~\ref{fig:diagrams S ferm+vect}), yielding however 
no contribution to $\hat{T}$.  This can be easily seen by the fact that substituting the equations of motion, namely
\beq
\rho_\mu^{r\,a} = E_\mu^{r\,a} - \frac{c_{r} g_{\rho_r}^2}{M_{\rho_{r}}^2} \overline{\Psi}_4 \gamma_\mu T^{(r)\,a} \Psi_4 - \frac{1}
{M_{\rho_r}^2} \nabla^\alpha E^{r\,a}_{\alpha\mu} + \mathcal{O}(p^5), \hspace{1cm} r= L,\,R \, .
\label{eq:eqmotion}
\eeq
in the UV Lagrangian of Eqs.~\eqref{eq:rho Lagrangian} and \eqref{eq:rhopsi Lagrangian} does not produce any local operator contributing to $\hat{T}$.  
In other words, the tree level exchanges of vector resonances cannot generate shifts to the couplings of gauge bosons to fermions at $q^2 = 0$, where 
$q_\mu$ is the momentum of the gauge boson. 
From a diagrammatic point of view, the absence of these contributions is due to a cancellation between two diagrams. The operators in 
Eq.~\eqref{eq:rhopsi Lagrangian} generate additional tree level contributions to the vertices between a gauge boson and two fermions 
proportional to $c_r$, which cancels out (at $q^2 = 0$) with the diagram containing a gauge boson-$\rho$ mixing.

\subsection{Contributions to the $\hat{S}$ parameter}
\label{ssec:S}

\begin{figure}[t]
\centering
\begin{tabular}{c c c}
\includegraphics[scale=0.28]{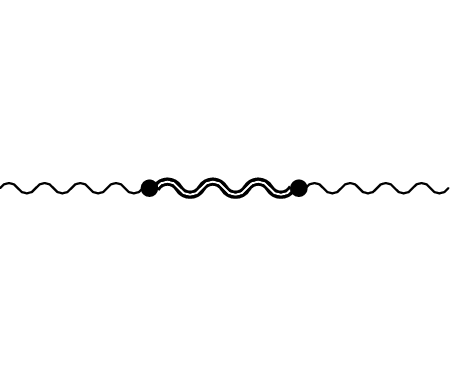} & 
\includegraphics[scale=0.28]{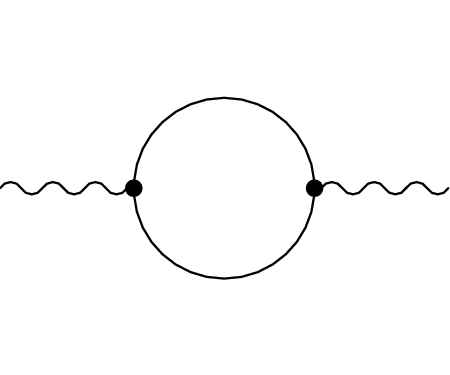} &
\includegraphics[scale=0.28]{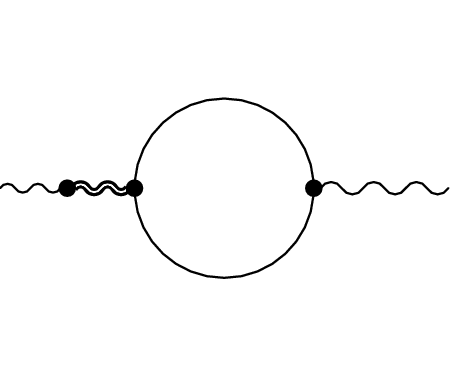} \\
\includegraphics[scale=0.28] {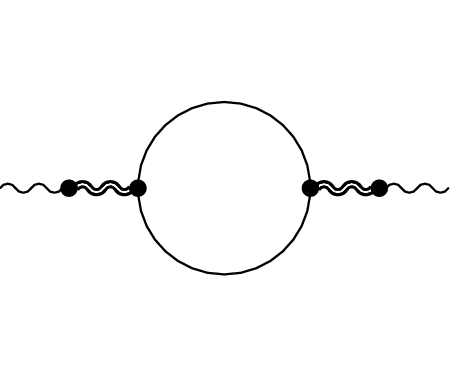} &
\includegraphics[scale=0.28]{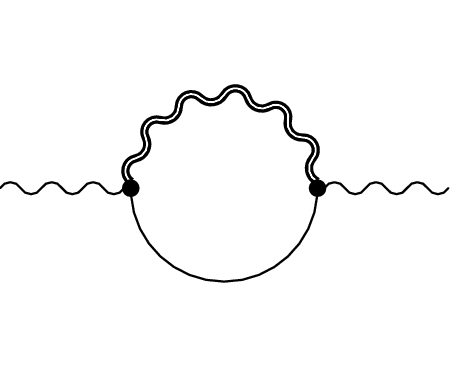} & 
\includegraphics[scale=0.28]{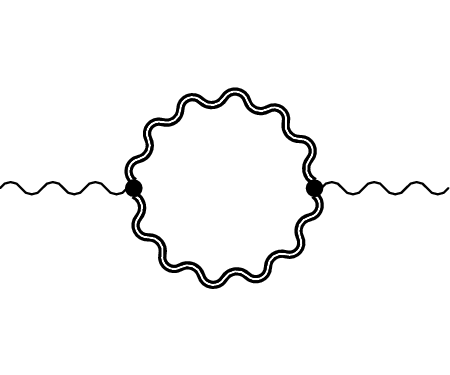} \\
\includegraphics[scale=0.28] {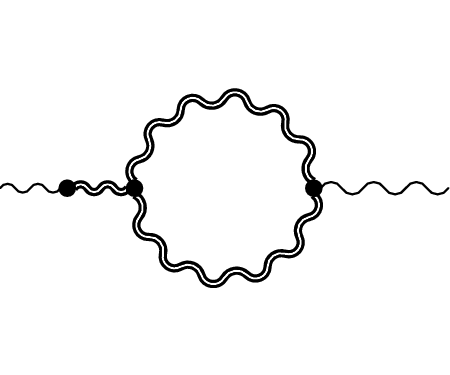} &
\includegraphics[scale=0.28] {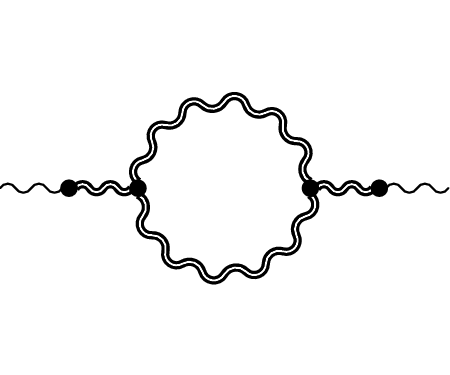} &
\includegraphics[scale=0.28] {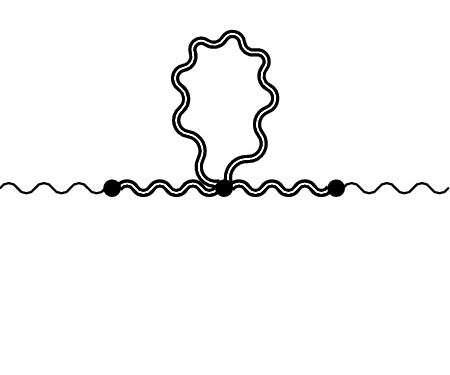}
\end{tabular}
\caption{Pure vectorial contribution to the $\hat{S}$ parameter. Smooth single lines represent all the NGBs including the Higgs. Single wavy 
lines on the left (right) represent a $W^3_\mu$ ($B_\mu$) field. Double wavy lines represent either a $\rho_L$ or a $\rho_R$. Although in 
the first diagram there is no actual resonance propagation, it gets contributions from the $\rpar{E_\mu}^2$ term of 
Eq.~\eqref{eq:rho Lagrangian}. Hence it must  be included in the resonance contribution.}
\label{fig:diagrams S vect}
\end{figure}

The $\hat{S}$ parameter gets an important contribution at tree level from the exchange of vector resonances.
The pure vectorial contribution at one-loop was calculated in ref. \citep{ContinoSalvarezza} and is shown in figure \ref{fig:diagrams S vect}. 
In analogy with the $\hat{T}$ parameter, the single resonance case is $L/R$ symmetric. We report the contributions:\footnote{In this case the 
change of operator basis from the one used in ref \citep{ContinoSalvarezza} has no effect (see the beginning of sec.~\ref{ssec:T}).}
\begin{align}
\hat{S} \big|_{\rho_{L/R}} &= \frac{g^2}{4g_{\rho_{L/R}}^2\rpar{\Lambda}} \xi \rpar{1-4\beta_{2_{L/R}}\rpar{\Lambda}} \nonumber \\
&- \frac{g^2}{96 \pi^2} \xi \Bigg[ \frac{3}{4}\rpar{ a_{\rho_{L/R}}^2 + 28 + 24\beta_{2_{L/R}}\rpar{a_{\rho_{L/R}}^2\beta_{2_{L/R}}-
a_{\rho_{L/R}}^2-2}} \log\rpar{\frac{\Lambda}{M_{\rho_{L/R}}}} \nonumber \\
&+ 1 + \frac{41}{16}a_{\rho_{L/R}}^2 -\frac{3}{2}\beta_{2_{L/R}}\rpar{9 a_{\rho_{L/R}}^2-4}+ \frac{3}{2}\beta_{2_{L/R}}^2\rpar{9 a_{\rho_{L/R}}^2-8}\Bigg], 
\label{eq:S 1rho}\\
\hat{S} \big|_{\rho_L+\rho_R} &= \hat{S} \big|_{\rho_R} + \hat{S} \big|_{\rho_R}.
\label{eq:S 2rho}
\end{align}

The inclusion of $\beta_{2_r}$ can partly or even fully cancel  the tree level contribution within its considered range (see Eq.~\eqref{arho-beta2} and the 
text following it), and in principle can also make it negative.\footnote{It was pointed out in \cite{RychkovSO(5),Contino:2015gdp} that the enforcement of 
a good UV 
behaviour of the $SU(2)_L$ and $SU(2)_R$ spectral functions within the low-energy effective theory would produce a non-negative tree level 
contribution to $\hat{S}$.}
On the other hand, the one-loop term can be negative even when when $\beta_{2_r}$ is small, hence reducing the total contribution to $\hat{S}$. 
Note that the explicit dependence on the RG scale (here, fixed to $\Lambda$) in the tree level term (first line of Eq.~\ref{eq:S 1rho}) ensures 
the RG independence of the full result till one loop order. 
See again ref. \citep{ContinoSalvarezza} for a detailed discussion about the running of the parameters in the vectorial sector.

Fermionic self-energies will also contribute to the $\hat{S}$ parameter. For the fourplet and fiveplet case the contributions start at 
$\mathcal{O}\rpar{y^0}$, while in the singlet case the leading order is $\mathcal{O}\rpar{y^2}$, as it must provide a finite contribution to 
$\hat{S}$ (see ref. \citep{PanicoGrojean}). By including the top contribution and subtracting the corresponding SM result, we find the approximate expressions:
\begin{align}
\Delta \hat{S} \big|_{\Psi_1}  \approx & \, - \frac{g^2}{96\pi^2} \frac{y_{L1}^2}{g_1^2} \xi \spar{ \log\rpar{\frac{f^4\xi y_{L1}^2 y_{R1}^2}{2M_1^4}} + \frac{5}{2} }, \\
\Delta \hat{S} \big|_{\Psi_4} \approx & \, \frac{g^2}{4 \pi^2} \xi \log\rpar{\frac{\Lambda}{M_4}}, \\
\Delta \hat{S} \big|_{\Psi_5} \approx & \, \frac{g^2}{4 \pi^2} \xi \Bigg[ \rpar{1-c_d^2} \log\rpar{\frac{\Lambda}{M_4}} + 
\log(x) c_d^2 \frac{x^3(x^3-3x+3)}{\rpar{1-x^2}^3} - \nonumber \\
&c_d^2 \frac{2x^4+9x^3-16x^2+9x+2}{12(1-x^2)^2} \Bigg].
\end{align}
\begin{figure}[t]
\centering
\begin{tabular}{c c c}
\includegraphics[scale=0.28] {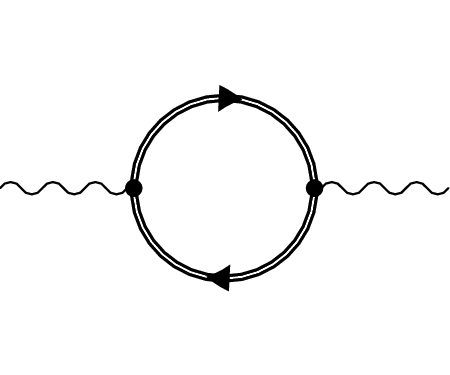}
\includegraphics[scale=0.28] {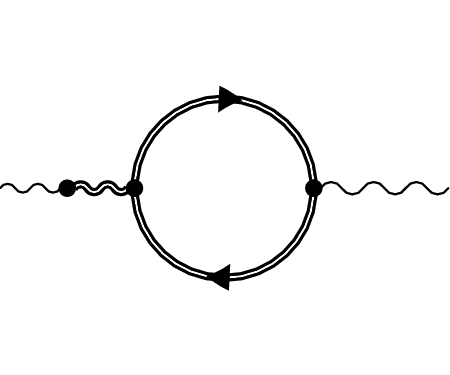} &
\includegraphics[scale=0.28] {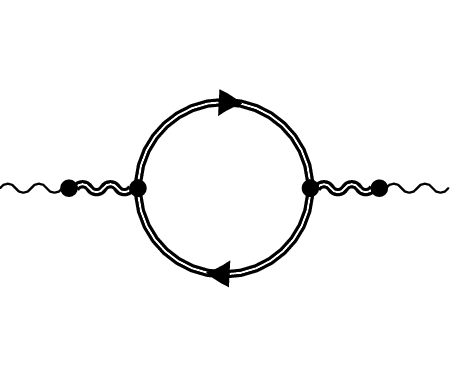}
\end{tabular}
\caption{Mixed fermion-vector contribution to the $\hat{S}$ parameter. Fermionic lines in the above diagrams stand for any fermionic 
resonance or light quark. Double wavy lines represent either a $\rho_L$ or a $\rho_R$. Although in the first diagram there is no actual $\rho$ 
propagation, it gets contributions from the $\overline{\Psi}\slashed{E}\Psi$ term of Lagrangian \eqref{eq:rhopsi Lagrangian}, which must be 
included in the mixed fermion-vector part.}
\label{fig:diagrams S ferm+vect}
\end{figure}
Part of these contributions were previously computed in \cite{PanicoGrojean,Azatov:2013ura}.
The contributions in the fourplet and fiveplet cases are tipically big also in the full numerical calculation. In the fourplet and singlet cases the full contributions 
are mostly positive, while in the fiveplet it can have both signs.

With the introduction of the operator in Eq.~\eqref{eq:rhopsi Lagrangian} vector resonances can couple to fermions and thus contribute to 
the $\hat{S}$ parameter forming vector-fermion diagrams shown in figure \ref{fig:diagrams S ferm+vect}. Such contributions exist only if a 
composite fourplet is present in the spectrum. 
They start at $\mathcal{O}\rpar{y^2}$ and, more interestingly, at order $\mathcal{O}\rpar{M_4^2/M_\rho^2}$. Their approximate expressions yield:
\begin{align}
\hat{S} \big|_{\rho_{L/R} + \Psi_4} &\approx \frac{3g^2}{32\pi^2} \frac{y_{L4}^2 - 2y_{R4}^2}{g_4^2} \frac{M_4^2}{M_{\rho_{L/R}}^2} 
\xi \rpar{1-2\beta_{2_{L/R}}} c_{L/R}\spar{\log\rpar{\frac{\Lambda}{M_4}} + \frac{1}{4} },  \label{eq:S4plet1rho}\\
\hat{S} \big|_{\rho_{L/R} + \Psi_5} &\approx \frac{3g^2}{32\pi^2} \frac{y_L^2 - 2y_R^2}{g_4^2} \frac{M_4^2}{M_{\rho_{L/R}}^2} 
\xi \rpar{1-\beta_{2_{L/R}}} c_{L/R} \times \nn \\
&\spar{ \rpar{1+c_d}\log\rpar{\frac{\Lambda}{M_4}} + c_d \log(x) \frac{(2-x)x^2}{(1-x)^2(1+x)}+\frac{1}{4}c_d\frac{3-x}{1-x}+\frac{1}{4} },  \\
\hat{S} \big|_{\rho_L + \rho_R + \Psi_4} &\approx \hat{S} \big|_{\rho_L + \Psi_4} + \hat{S} \big|_{\rho_R + \Psi_4}, \\
\hat{S} \big|_{\rho_{L} + \rho_{R} + \Psi_5} &\approx \hat{S} \big|_{\rho_L + \Psi_5} + \hat{S} \big|_{\rho_R + \Psi_5}.
\end{align}

Since lower mass bounds on vectors are sizeably higher than the ones on fermions (both from direct and indirect indications), a factor $M_4^2/M_\rho^2$ 
can be a strong suppression. Just like the absence of analogous terms in the $\hat{T}$ parameter, this suppression can be also explained using the 
solution \eqref{eq:eqmotion} to the equations of motion of vectorial resonances. Using such solution it is found that an interaction of the form 
$c_r \rpar{q^2/M_{\rho_r}^2} \overline{\Psi} \gamma_\mu \Psi A_\mu$ is generated, which contributes to $\hat{S}$ when used to form a fermionic loop 
together with a standard $\overline{\Psi} \gamma_\mu \Psi A_\mu$ gauge-like interaction. The explicit $q^2$ appearing in the vertex forces us to extract a mass term proportional to $M_4^2$ from the loop, thus generating a factor $M_4^2/M_{\rho_r}^2$.

\subsection{Contributions to $\delta g_L^{(b)}$}
\label{ssec:Zbb}

Tree level contributions $\delta g_L^{(b)}$ from the 
fermionic sector are proportional to the degree of compositeness of the bottom quark, and although in some models they can be comparable with the ones 
coming from fermion loops (see ref. \citep{PanicoGrojean}) they are generally expected to be small. As we neglect the bottom quark mass, this contribution 
is absent in our case. 
Moreover, tree level contributions proportional to $c_r$ due to the presence of spin-1 resonances vanish for the same reason discussed 
at the end of sec.~\ref{ssec:T} for the $\hat{T}$ parameter. Hence, the contributions to  $\delta g_L^{(b)}$ start at the one loop level.

\begin{figure}[t]
\centering
\begin{tabular}{c c}
\includegraphics[scale=0.25]{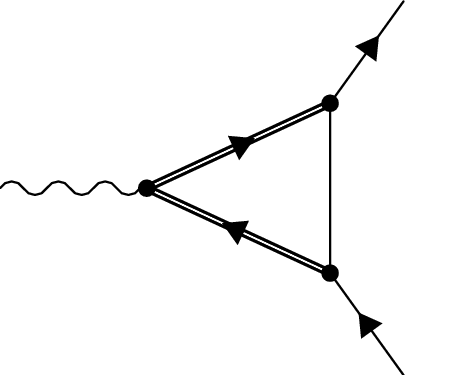} &
\includegraphics[scale=0.25]{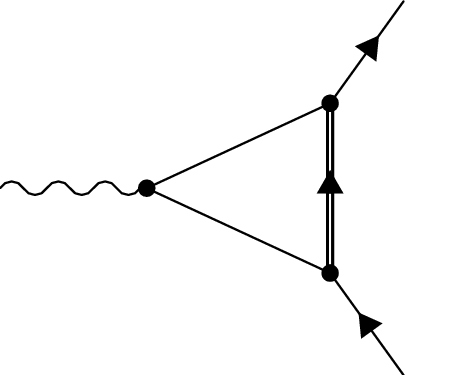} \\
\hspace{0.4 cm}\includegraphics[scale=0.28]{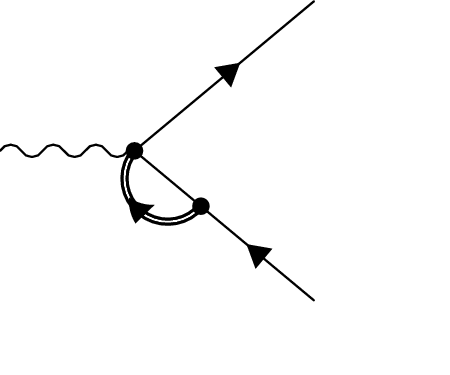} &
\hspace{0.4 cm}
\includegraphics[scale=0.28]{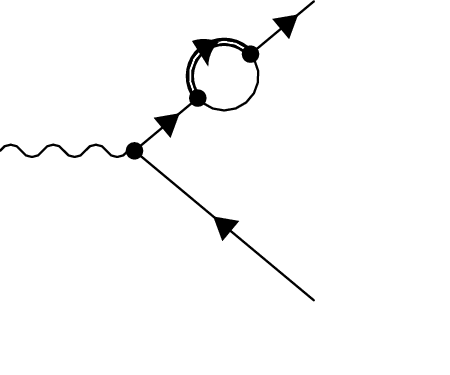}
\end{tabular}
\caption{Pure fermionic contribution to $\delta g_L^{(b)}$. Fermionic lines in the above diagrams stand for any resonance or light quark, while smooth continous lines denote any NG boson including the Higgs. Note that the wave function renormalization of the $Z$ leg must not be included in the calculation.}
\label{fig:glb pure fermionic}
\end{figure}

In our models the shift to the $Z\overline{b}_Lb_L$ coupling is linked to the breaking of the $P_{LR}$ parity. 
The spurion analysis of section \ref{ssec:symmetries} shows that the total pure fermionic contributions must appear at least at $\mathcal{O}\rpar{y_L^4}$, see 
the operator $O_\delta^{(qq)}$ in equation \eqref{Odeltaqq}. In the singlet and the fourplet cases 
the insertions of four mass mixings is enough to generate finite pure fermionic contributions. In the fiveplet case we may also make use of the 
couplings proportional to $c_d$, which contain a derivative and can therefore increase the degree of divergence. However, as explained in 
ref. \citep{PanicoGrojean}, the contributions are finite also in this case. Purely spin-1 resonance contributions to $\delta g_L^{(b)}$ are absent, as their effects 
only appear when coupled to the composite fermions through the operators in Eq.~\ref{eq:rhopsi Lagrangian}.

On the pure fermionic side, the various contributions (firstly calculated in ref.~\cite{PanicoGrojean}) are shown in figure 
\ref{fig:glb pure fermionic}, and the approximate expressions read:
\begin{align}
\delta g_L^{(b)} \big|_{\Psi_1+t} &\approx \frac{y_{L1}^4}{64\pi^2g_1^2} \xi, \\
\delta g_L^{(b)} \big|_{\Psi_4+t} &\approx \frac{y_{L4}^4}{32\pi^2g_4^2}\frac{y_{R4}^2}{g_4^2}\xi\spar{\log\rpar{\frac{2g_4^4}{\xi y_{L4}^2y_{R4}^2}}+2}, \\
\delta g_L^{(b)} \big|_{\Psi_5+t} &\approx -\frac{y_L^4}{64\pi^2g_1^2}\xi \frac{x^2}{1-x^2} \times \nonumber \\ &\Bigg[\frac{\log(x)}{1-x^2}\rpar{4\rpar{1-x^2}+c_d\rpar{2-4x^3}-4c_d^2\rpar{1-x+x^2}-2c_d^3x\rpar{2-x}} \nonumber \\
& +1-\frac{1}{x^2}+c_d\rpar{1+2x-\frac{4}{x}}+c_d^2\rpar{x^2+2x-5}+c_d^3x\rpar{x-2} \Bigg].
\end{align}

Differently from the oblique parameters, $\delta g_L^{(b)}$ is loosely correlated with the other EW observables and almost equally accepts positive and negative values (see section \ref{ssec:data}), so the sign of these contributions is not a relevant information.

\begin{figure}
\centering
\begin{tabular}{c c c}
\includegraphics[scale=0.28]{figures/diagrams/gl/fermions/png/loop9} &
\includegraphics[scale=0.28]{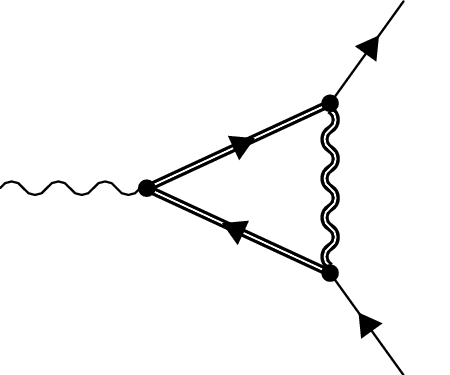} &
\includegraphics[scale=0.28]{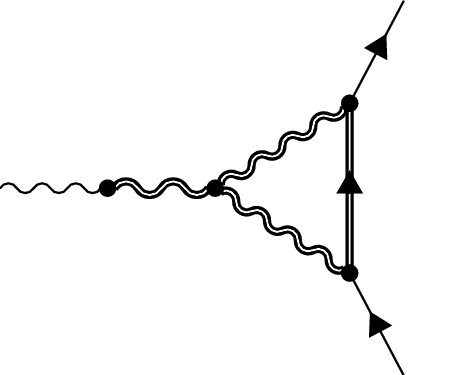} \\
\includegraphics[scale=0.28]{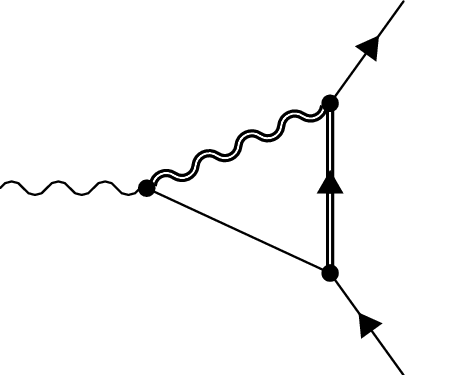} &
\includegraphics[scale=0.28]{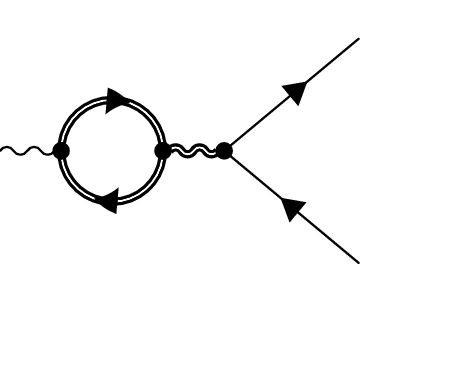} &
\includegraphics[scale=0.28]{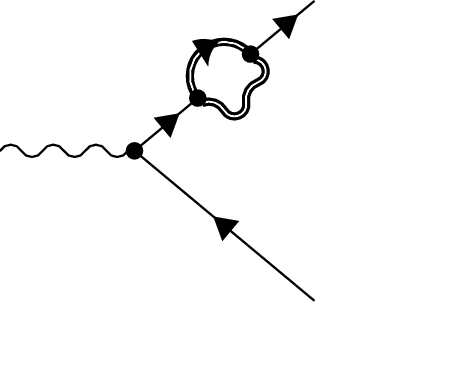}
\end{tabular}
\caption{Mixed vector-fermion contributions to $\delta g_L^{(b)}$. Fermionic lines stand for any resonance or light quark, while smooth continous lines denote any NG boson including the Higgs. Double wavy lines represent either a $\rho_L$ or a $\rho_R$. Although in the first diagram there is no actual $\rho$ propagation, it gets a contribution proportional to $c_r$ from the operator \eqref{eq:rhopsi Lagrangian}, which must be included in the mixed fermion-vector part. Note that the wave function renormalization of the $Z$ leg must not be included, and mixings on the same leg followed by vertices proportional to $c_r$ do not contribute (see text).}
\label{fig:glb vect+fermion}
\end{figure}

The vectorial sector is also able to break the $P_{LR}$ parity. 
The  vectors couple to the fermionic sector through the parameters $c_r$ introduced in Eq.~\eqref{eq:rhopsi Lagrangian}. As discussed in the beginning 
of this section, their effect starts at the one-loop level. Even in this case, the same cancellation discussed for the tree level contributions takes place, so 
that one-loop contributions are only generated by diagrams with the propagation of spin-1 resonances inside loops. The approximate expressions are given by,


\begin{align}
\delta g^{(b)}_L \big|_{\rho_L+\Psi_4+t} \approx&
-\frac{3y_{L4}^4}{64\pi^2g_4^2}\xi\frac{c_L^2}{a_{\rho_L}^2}
\spar{\log\rpar{\frac{\Lambda}{M_4}}+\frac{1}{12}} + \phi^{(4)}_L,  \label{cL} \\
\delta g^{(b)}_L \big|_{\rho_R+\Psi_4+t} \approx&\,
\frac{3y_{L4}^4}{64\pi^2g_4^2}\xi\frac{c_R^2}{a_{\rho_R}^2}
\spar{\log\rpar{\frac{\Lambda}{M_4}}+\frac{1}{4}} - \phi^{(4)}_R, \label{cR} \\
\delta g^{(b)}_L \big|_{\rho_L+\Psi_5+t} \approx&
-\frac{y_{L}^4}{64\pi^2g_4^2}\xi\frac{c_L^2}{a_{\rho_L}^2} \times \nn \\
&\left[ 3\rpar{1+c_d}\log\rpar{\frac{\Lambda}{M_4}} -  c_d \frac{x^2(x-2)(3x+1)}{\rpar{x^2-1}^2}\log(x) + \right. \nn \\
& \hspace{4cm} \left. c_d \frac{3x^2 - 10x - 1}{4\rpar{x^2-1}} + \frac{1}{4} \right] + \phi^{(5)}_L, \\
\delta g^{(b)}_L \big|_{\rho_R+\Psi_5+t} \approx&
\frac{y_{L}^4}{64\pi^2g_4^2}\xi\frac{c_R^2}{a_{\rho_R}^2}
 \nn \\
&\left[ 3\rpar{1+c_d}\log\rpar{\frac{\Lambda}{M_4}} - 3c_d \frac{x^2(x-2)}{(x-1)^2(x+1)}\log(x) + \right. \nn \\
& \hspace{4cm} \left. 3c_d \frac{x-3}{4(x-1)} + \frac{3}{4} \right] - \phi^{(5)}_R, \\
\delta g_L^{(b)} \big|_{\rho_L+\rho_R+\Psi_4+t} &\approx \delta g_L^{(b)} \big|_{\rho_L+\Psi_4+t}+\delta g_L^{(b)} \big|_{\rho_R+\Psi_4+t} \;, \\
\delta g_L^{(b)} \big|_{\rho_L+\rho_R+\Psi_5+t} &\approx \delta g_L^{(b)} \big|_{\rho_L+\Psi_5+t}+\delta g_L^{(b)} \big|_{\rho_R+\Psi_5+t} \; .
\end{align}

where the $\phi$ functions are given by ($r=L,\,R$):

\begin{align}
\phi_r^{(4)} =& \, \frac{3y_{L4}^2y_{R4}^2}{64\pi^2g_4^2}\xi 
\left\{c_r \spar{\log\rpar{\frac{\Lambda}{M_{\rho_r}}}+\frac{5}{12}} + \right. \nonumber \\
& \hspace{1.8cm} \left. \frac{c_r^2}{6a_{\rho_r}^2} 
\spar{ 11\log\rpar{\frac{\Lambda}{M_{\rho_r}}} - 8\log\rpar{\frac{M_4}{M_{\rho_r}}} + \frac{19}{4} } \right\}, \\
\phi_r^{(5)} =& \, \frac{y_{L}^2y_{R}^2}{64\pi^2g_4^2}\xi
\left\{ 3c_r(1+c_d) \spar{\log\rpar{\frac{\Lambda}{M_{\rho_r}}}+\frac{5}{12}} + \frac{c_r^2}{2a_{\rho_r}^2} \times \right. \nonumber \\
& \hspace{1.8cm} \left[ (11+8c_d)\log\rpar{\frac{\Lambda}{M_4}} + 3\log\rpar{\frac{M_4}{M_{\rho_r}}} + \frac{19}{4} -  \right. \nonumber \\
& \hspace{2.2cm} \left.\left. 4 c_d \frac{x^2(x-2)(2x+3)}{\rpar{x^2-1}^2}\log(x) + 2c_d \frac{x^2-2x-9}{2(x^2-1)} \right] \right\}.
\end{align}

The presence of contributions given by the $\phi$ functions, which do not scale as $y_{L}^4$, is a signal that the breaking of $P_{LR}$ is 
not coming from the fermionic sector only. 
Despite the appearance of four mass mixings the above results are divergent, as the massive spin-1 fields change the power counting. 
Notice that the $\phi$ functions enter with opposite sign in the contributions from $\rho_L$ and $\rho_R$. This happens because a 
spin-1 sector with equal $\rho_L$ and $\rho_R$ parameters is $P_{LR}$ symmetric, 
hence in this case any contribution to $\delta g_L^{(b)}$ with less than four powers of $y_L$ should cancel out in the final result. 

\section{Fit Procedure}
\label{sec:fit}

In order to fit the composite Higgs model parameters to the electroweak precision data, we follow a two step procedure. In the first step, we perform a relatively 
model independent fit that can be used for a broader class of models. The details of this step is described in the following subsection. In the second step we 
apply the results of the model independent fit to our composite Higgs scenarios. More details about the second step will be given in section \ref{ssec:parameters}.

\subsection{Data and fit interpretation}
\label{ssec:data}

\begin{figure}[t]
\begin{center}
\begin{tabular}{ll}
\hspace{-0mm}\includegraphics[scale=0.345]{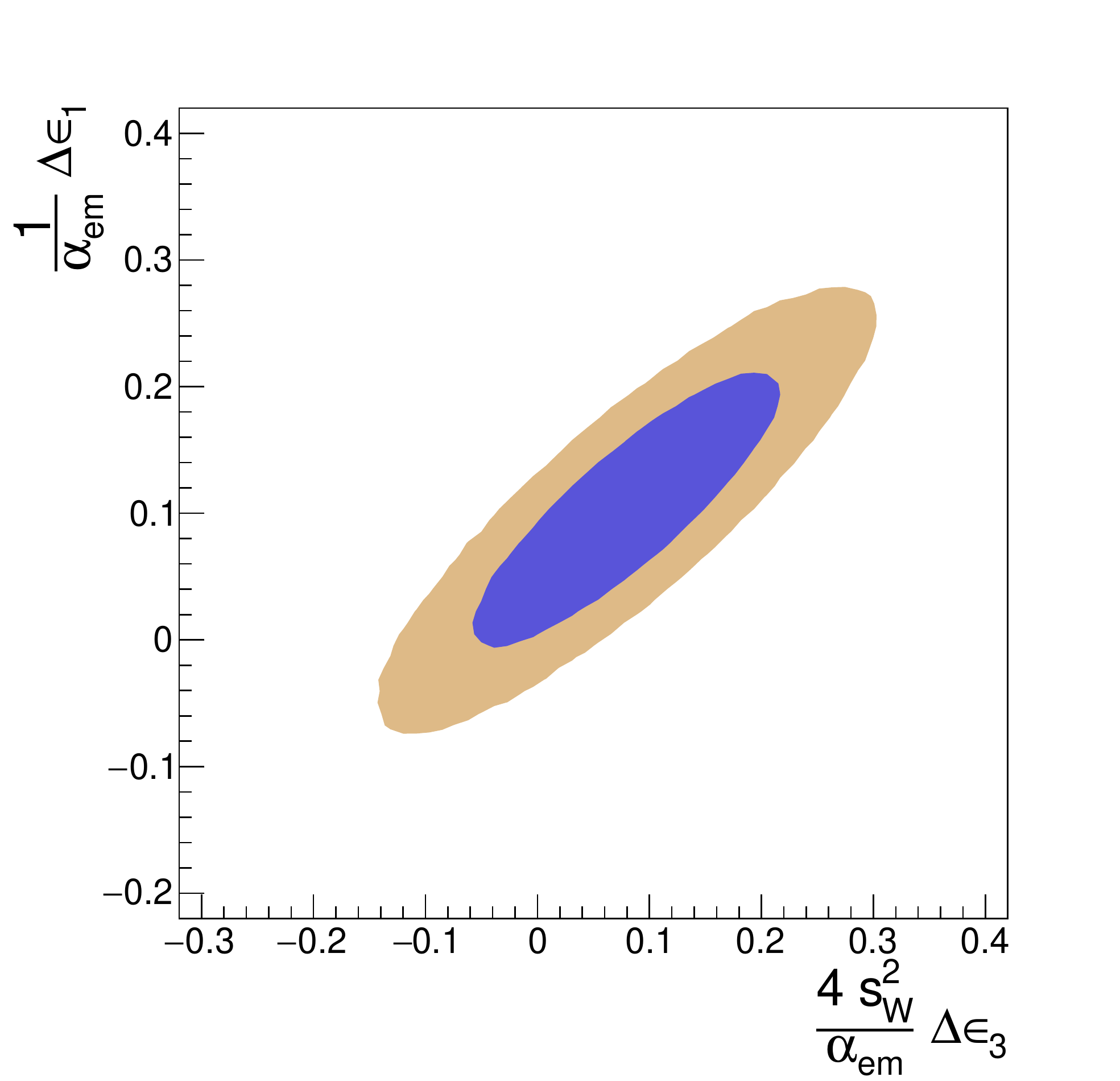} &
\hspace{-0mm}\includegraphics[scale=0.35]{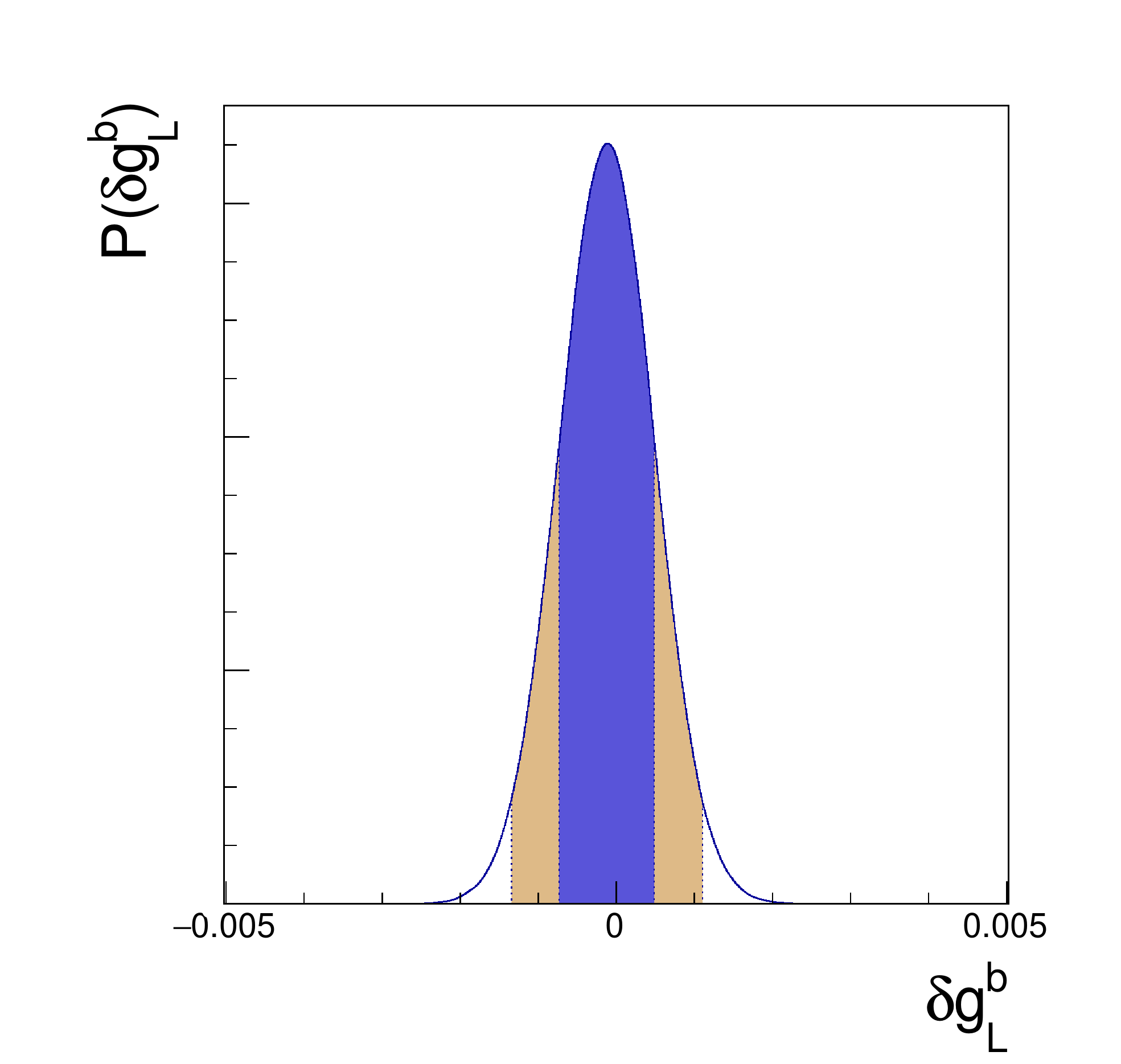} \\
\end{tabular}
\caption[]{The 68\% and 95\% marginalised probability regions of the epsilon parameters (left) and $\delta g_L^{(b)}$ (right). Since 
$\delta g_L^{(b)}$ is very weakly correlated to the other pseudo-observables (see table \ref{tab:corr8}), we show only its one dimensional 
probability distribution.}
\label{fig:BSM fit}
\end{center} 
\end{figure}

As discussed at length in section~\ref{sec:calculations},  the fit to the EWPO in the composite Higgs models of our interest can be performed 
by only using the two epsilon parameters $\Delta \epsilon_{1\, , 3}$  and the modification to the $Z \, b_L \, b_L$ vertex, 
namely $\delta g_L^{(b)}$.  Therefore, in our 
specific NP scenarios one can perform a complete EW fit  using $\Delta \epsilon_1$, $\Delta \epsilon_3$,  $\delta g_L^{(b)}$,  $\alpha_s(M_Z)$, 
$\Delta \alpha_{had}^{5}(M_Z)$,  $M_Z$, $m_t^{\rm pole}$ and $m_h$ as the input parameters
\footnote{For more details of the EWPO and the experimental data used for the fit, see \cite{SilvestriniEW}.} and keeping $\Delta\epsilon_2$ and 
$\delta g_R^{(b)}$ fixed to zero. The result of this fit, namely the central values, uncertainties and the correlation 
matrix for the 8 input parameters (we will call them pseudo-observables below) are shown in tables \ref{tab:obs8} and \ref{tab:corr8}. 
The marginalised 68\% and 95\%  probability regions in the $\Delta \epsilon_3 - \Delta \epsilon_1$ plane and for  $\delta g_L^{(b)}$ are also 
shown in Fig.~\ref{fig:BSM fit}. The numerical fit has been done using the {\tt HEPfit} package\footnote{The {\tt HEPfit} is available under the 
GNU General Public License (GPL) from \url{https/github.com/silvest/HEPfit}.} (formerly {\tt   SUSYfit}). 

\begin{figure}[t]
\begin{center}
\begin{tabular}{l}
\includegraphics[scale=0.4]{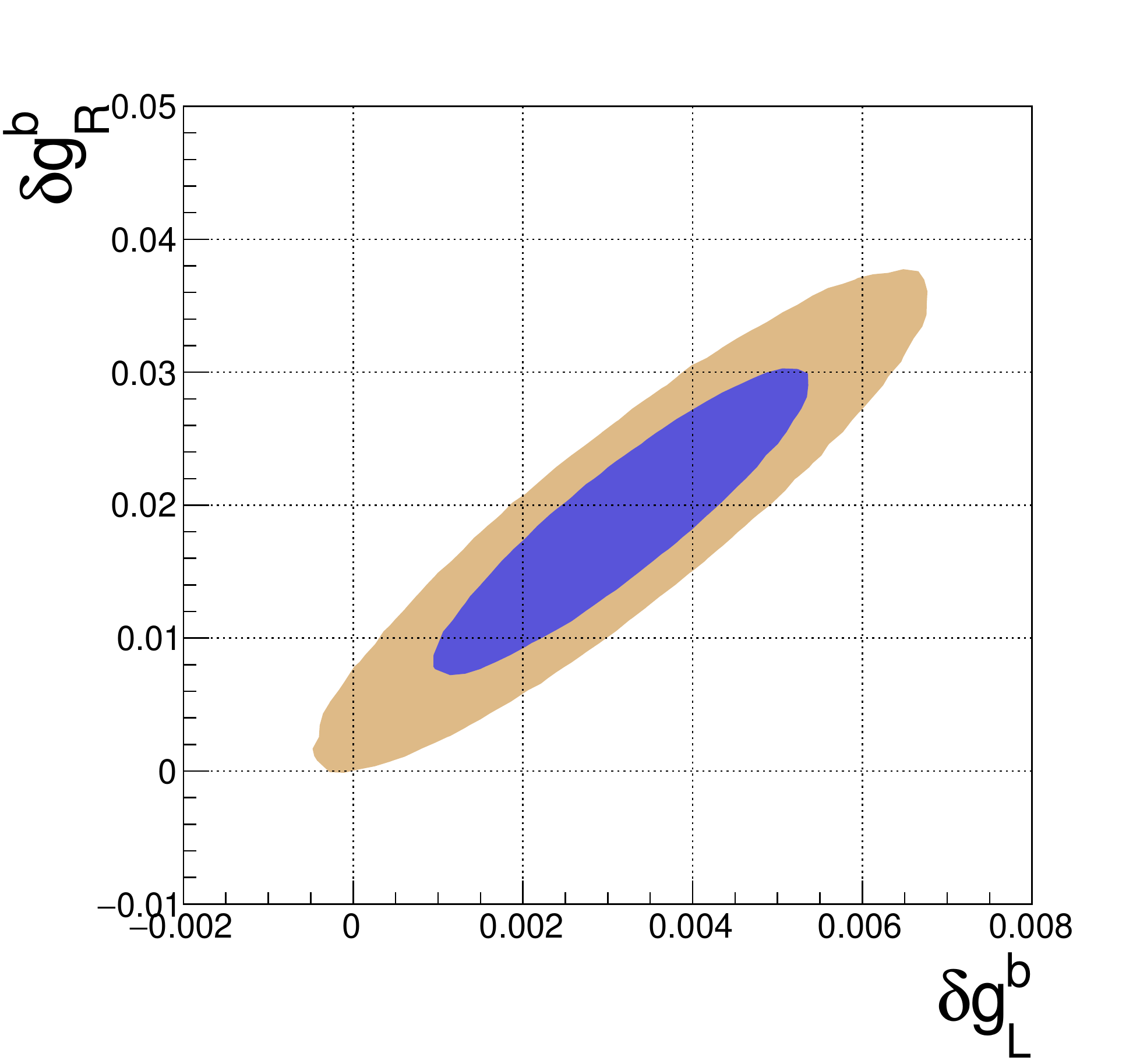}
\end{tabular}
\caption[]{The 68\% and 95\% marginalised probability regions in the $\delta g_L^{(b)} - \delta g_R^{(b)}$ plane.}
\label{fig:comparison}
\end{center} 
\end{figure}

Here we would like to emphasise 
that  $\delta g_L^{(b)} = 0$ is fully consistent with our fit (it is almost at the centre of the 68\% allowed region) which is slightly inconsistent 
with the results obtained in \cite{Batell:2012ca,Guadagnoli:2013mru} (which was also used in \cite{PanicoGrojean}). In order to directly compare their 
results with our fit, in Fig.~\ref{fig:comparison} we  show the 68\% and 95\% marginalised probability regions in the 
$\delta g_R^{(b)} - \delta g_L^{(b)}$ plane when $\delta g_R^{(b)}$ is also included in the input parameters of our fit. It can be seen that 
$\delta g_R^{(b)} = \delta g_L^{(b)} =0$ is now marginally consistent at the 95\% CL while in \cite{Batell:2012ca,Guadagnoli:2013mru} it was 
slightly outside the 95\% CL region. Our results are consistent with ref.~\cite{SilvestriniEWupdate}, and we believe that the difference with 
the earlier literatures is due to the new SM calculation of $R_b$ \cite{Freitas:2013dpa,Freitas:2014hra,Freitas:2014owa} and possibly also 
due to the different methodologies used to obtain the 68\% and 95\% probability regions.  
\begin{table}[h]
\small
\begin{center}
\tabulinesep=1.2mm
\begin{tabu}{|c|c|c|}
\hline
Quantity & Central value & Uncertainty \\
\hline
  $\Delta \epsilon_3$ & $0.62 \times 10^{-3}$ & $0.74 \times 10^{-3}$ \\
  $\Delta \epsilon_1$ & $0.71 \times 10^{-3}$ & $0.52 \times 10^{-3}$ \\
  $\delta g_L^{(b)}$  & $-0.13 \times 10^{-3}$ & $0.61 \times 10^{-3}$ \\
  $\alpha_s(M_Z)$   & $ 0.11850 $    &  $0.00050$  \\
  $\Delta \alpha_{\rm had}^{5}(M_Z)$  & $0.02750$  & $0.00033$ \\
  $M_Z$               & 91.1900 & $2.1 \times 10^{-3}$ \\
  $m_t$               & 173.30 & 0.76 \\ 
  $m_h$               &   125.50   &    0.30   \\
  \hline
\end{tabu}
\caption{ Central values and the uncertainties of the pseudo-observables.  \label{tab:obs8}}
\end{center}
\end{table}

\begin{table}[t]
\begin{center}
\begin{tabular}{c|rrrrrrrr} 
 & $\Delta \epsilon_3$ & $\Delta \epsilon_1$ & $\delta g_L^{(b)}$ & $\alpha_s$ & $\Delta \alpha_{\rm had}^{5}$ & $M_Z$ & $m_t$ & $m_h$ \\
 \hline
$\Delta \epsilon_3$ & $1$ & $0.864$ & $0.060$ & $-0.013$ & $-0.400$ & $-0.039$ & $-0.004$ & $-0.003$ \\
$\Delta \epsilon_1$ & $0.864$ & $1$ & $0.123$ & $-0.011$ & $-0.116$ & $-0.124$ & $-0.146$ & $0.002$ \\
$\delta g_L^{(b)}$ & $0.060$ & $0.123$ & $1$ & $0.110$ & $-0.033$ & $-0.003$ & $-0.064$ & $0.001$ \\
$\alpha_s(M_Z)$ & $-0.013$ & $-0.011$ & $0.110$ & $1$ & $0.003$ & $0.001$ & $0.001$ & $0.001$ \\
$\Delta \alpha_{\rm had}^{5}(M_Z)$ & $-0.400$ & $-0.116$ & $-0.033$ & $0.003$ & $1$ & $0.004$ & $-0.002$ & $-0.001$ \\
$M_Z$ & $-0.039$ & $-0.124$ & $-0.003$ & $0.001$ & $0.004$ & $1$ & $0.000$ & $ 0.000$ \\
$m_t$ & $-0.004$ & $-0.146$ & $-0.064$ & $0.001$ & $-0.002$ & $0.000$ & $1$ & $0.000$ \\
$m_h$ & $-0.003$ & $0.002$ & $0.001$ & $0.001$ & $-0.001$ & $0.000$ & $0.000$ & $1$ \\
\hline
\end{tabular}
\caption{Full correlation matrix of the 8 pseudo-observables.
\label{tab:corr8}}
\end{center}
\end{table}

The results of table \ref{tab:obs8} and \ref{tab:corr8} will be used as constraints in the fits of our NP models discussed in section~\ref{sec:results}.
 As $\alpha_s(M_Z)$ and $\Delta \alpha_{\rm had}^{5}(M_Z)$ do not get any NP contributions and their uncertainties are also very small, we fix 
them to their central values and remove them from the fit. Moreover, as the Higgs mass is not calculable in all the cases we consider in this paper 
(except the two-site model), it drops out from the list of constraints. Hence, we end up with only five pseudo-observables, namely 
$\Delta\epsilon_{1,3}$, $\delta g_L^{(b)}$, $M_Z$ and $m_t$ and a simpler correlation matrix of dimension 5. 
As the Higgs mass is calculable in the two-site model, we include it in that case. However, as the Higgs mass is very weakly correlated to all the other 
pseudo-observables, we treat it as an independent constraint.

In addition to the constraints coming from the pseudo-observables listed in table~\ref{tab:obs8}, we also take into account lower mass bounds on 
the masses of spin-1/2 and spin-1 resonances from direct searches. Including these direct search limits in a proper way is beyond the scope of 
this work, so we have chosen to impose approximate limits, namely of 800 GeV and 1.5 TeV on the lightest fermionic and vectorial resonances 
respectively. 
However, as we will see in the next section, the constraints from the EW fit are often stronger than the direct search limits, hence these lower 
limits do not have noticeable impact on our results. 

In order to perform the numerical fit, we use the data given in tables \ref{tab:obs8} and \ref{tab:corr8} and also the lower bounds. 
We follow a Bayesian statistical approach using the \texttt{BAT} library \cite{BAT} and present the 68\% and 95\% probability regions of the posterior  
distributions of the relevant model parameters. We do not make any effort to quantify the goodness-of-fit for the various composite Higgs 
scenarios. Hence, in general the fit results of two different scenarios can not be compared. However, as we know that in the limit $\xi \to 0$ one 
gets back the SM, for very small $\xi$ all the models are equally acceptable so that the posterior probability distributions of $\xi$ give a good qualitative indication of the goodness-of-fit. In other words, 
the posterior distributions actually carry the information of $\Delta \chi^2$ and not the absolute $\chi^2$, however, as all our models 
have a smooth SM limit (i.e., $\xi \to 0$) and the SM provides a very good fit to the EWPO, the shape and range of the probability distribution of $\xi$ can be taken as a qualitative measure of the goodness-of-fit. 

In particular, models that allow larger values of $\xi$ provide better fit to the EWPO.
Models with smaller upper bound on $\xi$ require more fine tuning than models with bigger upper bound. Note that the 
requirement of low fine tuning is not an input of our analysis. Our analysis only takes into account the electroweak precision 
data and checks how a limited set of composite Higgs scenarios compare with each other in view of these data.

\subsection{New physics parameters and physical masses}
\label{ssec:parameters}

In the presence of both, the fermionic 4-plet and the singlet as well as both the spin-1 resonances, the $\rho_L$ and $\rho_R$, there are in total 
20 free parameters in our fit which are listed below, 
\beq 
\left\{\xi, 
\, f,
\, g_{\text{el}},
\, g^{\prime}_{\text{el}},
\, g_{\rho_{L/R}},
\, a_{\rho_{L/R}}, 
\, \beta_{2_{L/R}},
\, g_{1/4}, 
\, y_{(L/R)(1/4)}, 
\, c_d, 
\, c_{L/R},
\, \Lambda \right\}.
\eeq

The global symmetry breaking scale $f$ can be related to $\xi$ using the Higgs VEV. In particular, we use the following tree-level relation,
\beq
G_F = \frac{1}{\sqrt{2}f^2\xi} ,
\eeq
where we take $G_F = 1.166371\times 10^{-5} \, \text{GeV}^{-2}$. Moreover, one of the vectorial couplings can be fixed  using the tree-level relation
\beq
\frac{1}{4\pi\alpha_{_{EM}}} = \frac{1}{g^{\prime 2}_{\text{el}}} + \frac{1}{g_{\text{el}}^2} + \frac{1}{g_{\rho_L}^2} + \frac{1}{g_{\rho_R}^2} ,
\label{eq:alphaEM constraint}
\eeq
where $\alpha_{EM} = 1/128.96$. We will use the above equation to fix $g_{\text{el}}$. The cut-off scale $\Lambda$, which only enters logarithmically in our formulas, will be set to three times the mass of the heaviest resonance. Thus, in the most general case, we end up with 17 free parameters:
\beq 
\left\{\xi, 
\, g^{\prime}_{\text{el}},
\, g_{\rho_{L/R}},
\, a_{\rho_{L/R}}, 
\, \beta_{2_{L/R}},
\, g_{1/4}, 
\, y_{(L/R)(1/4)}, 
\, c_d, 
\, c_{L/R} 
\right\}.
\eeq

Note that this choice is preferable because the decoupling limit of the composite sector (i.e., the SM limit) is achieved by setting  
$\xi \rightarrow 0$ with all the other parameters fixed to any value. In table \ref{tab:param range} we summarise the approximate 
ranges of the input parameters that were used in our fit. We use flat priors for all the parameters. 

\begin{table}[t]
\small
\begin{center}
\tabulinesep=1.2mm
\begin{tabu}{|c|c||c|c|}
\hline
Parameter & Range & Parameter & Range \\
\hline
  $\xi$                                & $\spar{0,1}$                     & $g_{1/4}$              & $\spar{0.1,7}$     \\
  $g^{\prime}_{\text{el}}$   & $\spar{0.34,0.4}$             & $y_{(L/R)(1/4)}$    & $\spar{0,5.5}$    \\
  $g_{\rho_{L/R}}$             & $\spar{1.5,5.5}$               & $|c_d|$                  & $\spar{-4,4}$      \\
  $a_{\rho_{L/R}}$             & $\spar{0.1,6}$                  & $|c_{L/R}|$            & $\spar{-3,3}$      \\
  $|\beta_{2_{L/R}}|$         & $\spar{-1,1.5}$    
&                     & \\
\hline
\end{tabu}
\caption{Ranges of the model parameters used in our fit. We use flat priors for all the parameters.
\label{tab:param range}}
\end{center}
\end{table}

Since in the one-loop expressions for the EWPO we fixed the renormalization scale to the cutoff $\Lambda$, all our input parameters are understood as 
$\overline{\rm MS}$ parameters at the same scale. Hence, also the tree-level masses obtained from these model parameters are understood as 
$\overline{\rm MS}$ masses. In order to compare with the top quark pole mass, we will use the SM QCD contribution to the running of the top mass 
from the cutoff to the EW scale which is given by
\beq  
m_t^{\overline{{\rm MS}}}(m_t) = m_t^{\overline{{\rm MS}}}(\Lambda) \rpar{ \frac{\alpha_s (\Lambda)}{\alpha_s (m_t)}  }^{\frac{\gamma_m^{(0)}}{2\beta_0}},  
\eeq
where,  $\gamma_m^{(0)} = 7$, $\beta_0 = 8$, $\alpha_s (m_t) = 0.1083$ and 
\beq
\alpha_s (\Lambda) = \frac{\alpha_s (m_t)}{1 + \beta_0 \dfrac{\alpha_s (m_t)}{2 \pi} \log\rpar{\Lambda / m_t}} \, . 
\eeq
In order to obtain the pole mass from the $\overline{{\rm MS}}$ mass we use the one loop conversion formula,
\beq
m_t^{\text{pole}} = m_t^{\overline{MS}}(m_t) \rpar{ 1 + \dfrac{4}{3} \dfrac{\alpha_s (m_t)}{\pi} } \, .
\eeq

Note that, besides SM QCD there are three more effects contributing to the running of the 
top mass: 1) Pure SM electroweak effects 2) loops of gluons and top partners 
3) loops of spin-1 resonances and top partners. While the first two effects are expected to be subdominant, 
the third contribution can in principle be comparable or even larger than the SM QCD contribution.
These effects (including the potentially large loops of resonances) have been neglected in this work and will 
be presented in a future publication.

\section{Results and discussion}
\label{sec:results}

In this section we present the results of our fits and discuss their qualitative features. As discussed in the previous section, we construct the likelihood function 
from the results of our tables \ref{tab:obs8} and \ref{tab:corr8} removing the entries corresponding to $\alpha_s$, $\Delta\alpha_{\text{had}}^5$ and $m_h$. 
Depending on the particular model under study, we also remove the entries related to $M_Z$ and/or $m_t$. For instance, if vectorial resonances are not 
considered, $M_Z$ does not receive NP contributions (at the tree level) and is automatically reproduced by setting $g_{\text{el}}^\prime = g^\prime$ 
(and $g_{\text{el}} = g$, as it would naturally follow from the constraint \eqref{eq:alphaEM constraint}). For the same reason, in these cases we will also 
remove the parameter $g_{\text{el}}^\prime$ from the fit. On the other hand, in the absence of fermionic resonances the particular mechanism which 
generates the top mass is not specified. In this case,  we will assume it to have the correct value and remove the corresponding constraint from the fit.

In all the different scenarios, the effect of the nonstandard Higgs dynamics on the $\Delta\epsilon_{1,3}$ parameters will always be present. These effects 
are completely dictated by the $SO(5)/SO(4)$ symmetry breaking pattern, thus being the same in all the cases. They were first computed in 
ref.~\cite{RychkovSO(5)} and are reported in Eq.~\eqref{eq:Deps1 Higgs}-\eqref{eq:Deps3 Higgs}.

\begin{figure}[t]
\begin{center}
\begin{tabular}{l}
\includegraphics[scale=0.4]{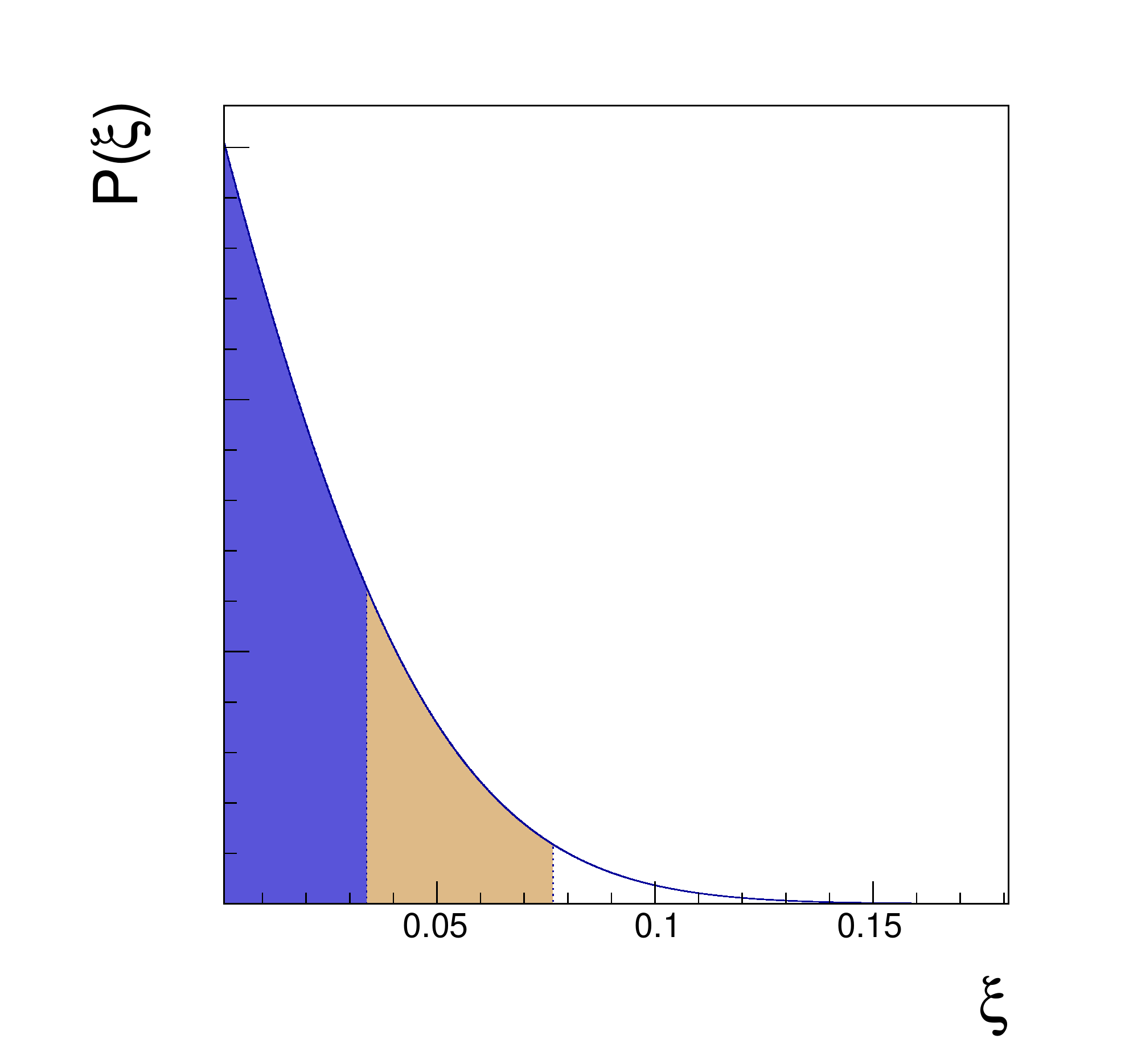}
\end{tabular}
\caption{
Posterior probability for $\xi$ including only the effects of the nonlinear higgs dynamics. The blue and the  grey colors respectively indicate the 
allowed regions at 68\% and 95\% probability. \label{fig:only-higgs}}
\end{center}
\end{figure}

As a warm-up and to compare with the literature, we first perform the fit taking into account only the composite Higgs contributions. 
In this case we set $\delta g_L^{(b)} = 0$ and $g^\prime_{\text{el}} = g^\prime$.
Once the cutoff is fixed to a value $\Lambda = 3\,\text{TeV}$\footnote{Another choice would be to use $\Lambda = g^\ast f$ and fix 
the value of $g^\ast$.  As expected, the result of the fit will be slightly different for different choices of $g^\ast$. For example, if $g^\ast= 2 \pi$ 
is used, the 95\% upper limit on $\xi$ reduces approximately by 10\% compared to the $\Lambda$ = 3 TeV case.}, the only remaining free 
parameter is the separation of scales $\xi$. 
Its posterior is presented in Fig.~\ref{fig:only-higgs}, which clearly shows the tight constraint, namely $\xi \lesssim 0.075$ at 95\% CL.

\begin{figure}[t]
\begin{center}
\includegraphics[scale=0.55]{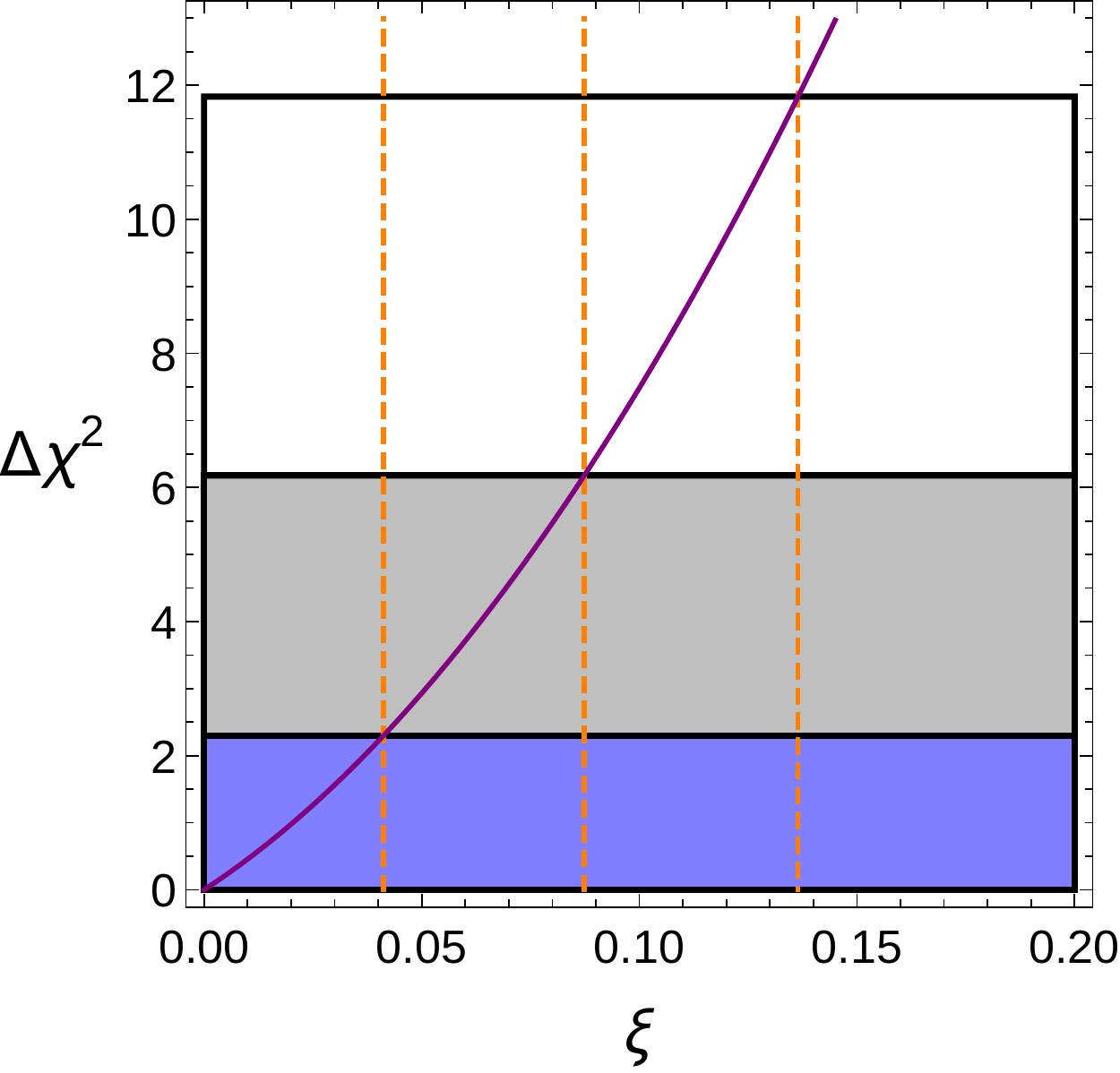}
\hspace{0.5cm}
\includegraphics[scale=0.53]{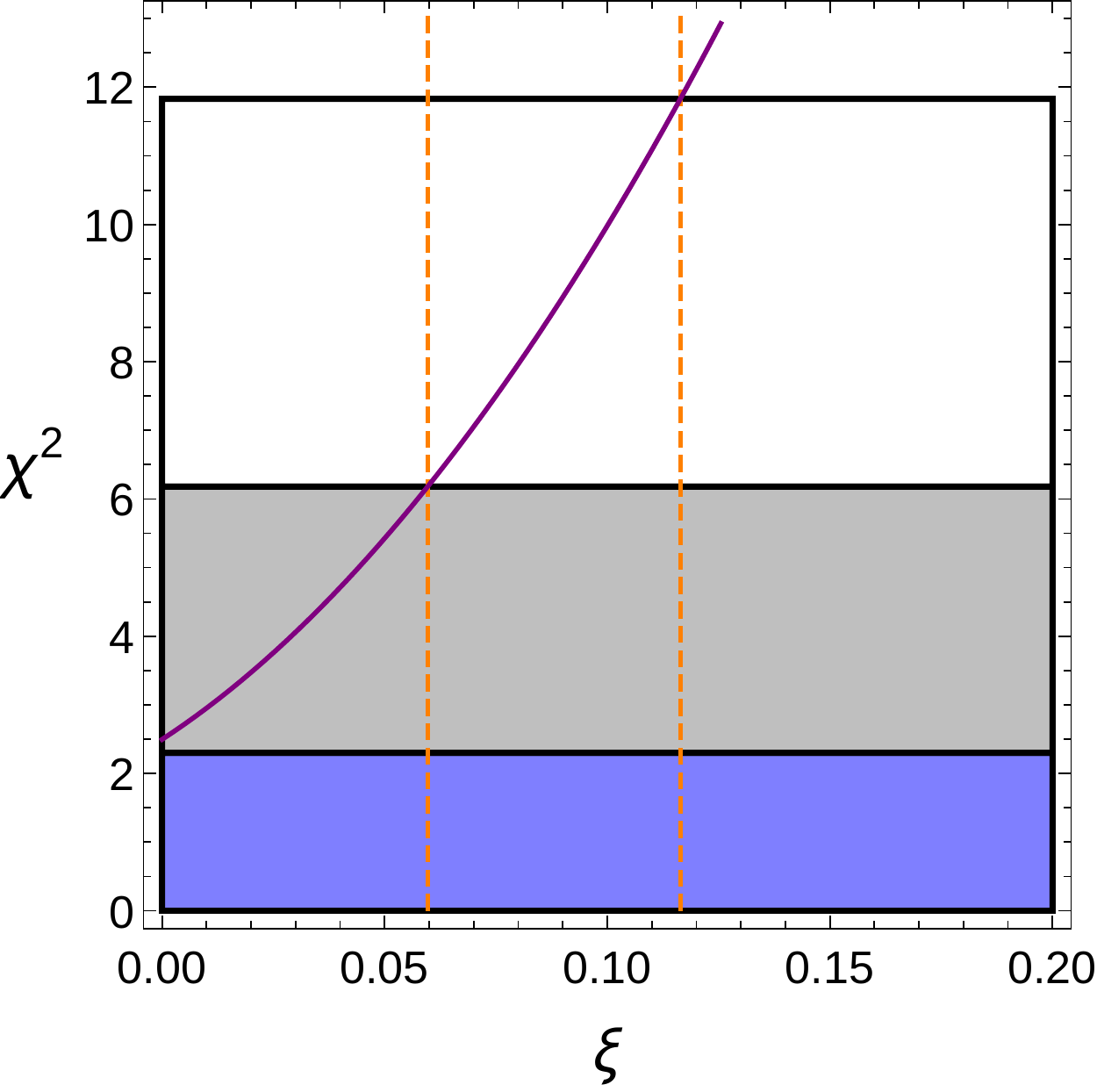}
\caption{The $\Delta \chi^2$ and the absolute $\chi^2$ as a function of $\xi$ in the only composite Higgs scenario. See text for more details.
\label{fig:frequentist}}
\end{center}
\end{figure}

In order to ease the comparison with the literature, in Fig.~\ref{fig:frequentist} we also plot the absolute $\chi^2$ and the $\Delta \chi^2$ as a function 
of $\xi$. In the left panel the blue, grey and transparent regions satisfy $\Delta \chi^2 < (2.3,\,6.18,\,11.83)$, which correspond to the 1-, 2- and 
3-$\sigma$ regions assuming 2 degrees of freedom and gaussian distributions of the observables as a function of $\xi$. Similarly, in 
the right panel we plot the absolute $\chi^2$ function. 

%
\subsection{Fermionic sector}
\label{ssec:fermionic sector}
%
In this section we investigate the constraints on the fermionic resonances. We start with the discussion of fermionic singlet 
for which a full analytical calculation is possible. The masses of the top quark and the heavy top partner $\widetilde{T}$ are given by
\beq
\begin{split}
m_{\widetilde{T},t}^2 &= \frac{1}{4} f^2 \, \bigg[2 g_1^2 +\xi y_{L1}^2 + 2 \rpar{ 1 - \xi } y_{R1}^2  \\
& \pm \sqrt{ 4 \rpar{ g_1^2 +\rpar{1-\xi}y_{R1}^2 }^2 + 4\xi y_{L1}^2 \rpar{g_1^2 - \rpar{1-\xi}y_{R1}^2} }\bigg]. 
\end{split}
\label{eq:masses singlet}
\eeq

The contributions to the EWPO $\delta g_L^{(b)}$, $\hat{T}$ and $\hat{S}$ coming from loops of $\widetilde{T}$ and the top quark 
are given by
\begin{align}
\delta g_L^{(b)} \big|_{\Psi_1+t} =& \; \frac{1}{3}\hat{T}\big|_{\Psi_1+t} \; , 
\\
\Delta\hat{T}\big|_{\Psi_1+t} =& \; \frac{3m_t^2}{16\pi^2v^2} \sin^2(2\phi) \Bigg[\frac{m_{\widetilde{T}}^2}{\Delta m^2}\log\rpar{\frac{m_{\widetilde{T}}}{m_t}}
+ \frac{1}{4}\rpar{\tan^2(\phi)\frac{m_{\widetilde{T}}^2}{m_t^2} - 1 - \frac{1}{\cos^2(\phi)}} \Bigg], \\
\Delta\hat{S}\big|_{\Psi_1+t} =& \; \frac{g^2}{96\pi^2} \sin^2(2\phi) \left[
\frac{1}{2}\rpar{6m_{\widetilde{T}}^4
\frac{m_{\widetilde{T}}^2 - 3 m_t^2 }{\Delta m^6}-3-\frac{1}{\cos^2(\phi)}} \log\rpar{\frac{m_{\widetilde{T}}}{m_t}} \right. \nn \\
& \left. \hspace{8.9cm} + 3 \frac{ m_{\widetilde{T}}^2m_t^2}{\Delta m^4} - \frac{5}{4} \; \right].
\end{align}

Where $\Delta m^2 = m_{\widetilde{T}}^2-m_t^2$, and the angle $\phi$ is related to the rotation matrix which defines the mass 
eigenstates $\rpar{t_L, \widetilde{T}_L}$
\beq
\cos^2(\phi) = \frac{2m_{\widetilde{T}}^2 - v^2 y_{L1}^2}{2\Delta m^2} \, .
\eeq

In order to obtain the contributions to $\Delta\epsilon_{1,3}$ parameters, the above contributions to $\Delta\hat{T}$ and $\Delta\hat{S}$ 
must be supplemented with the additional contributions coming from top loops. 
It is found that the fermionic resonance contributions to $\Delta\hat{S}$ and $\Delta\hat{T}$ are generically dominant  in the models under 
consideration, and the additional terms usually contribute with a smaller correction. In fact, we have checked that removing these terms from the 
fit does not produce sizeably different results. They can be obtained from equations \eqref{eq:eps1 top}, \eqref{eq:eps3 top} using the following 
expressions for the top couplings with EW gauge bosons (which, except for $g_{RZ}$, deviate from the SM values by a shift of ${\mathcal O} (\xi)$):
\begin{align}
g_{LZ} =& \; \frac{2}{3} \frac{g}{c}\rpar{s^2 - \frac{3}{4}\cos^2(\phi)}, &
g_\alpha =& \; \frac{1}{3} g^2 \frac{s}{c} \cos^2(\phi) \rpar{1 - \frac{3}{4} \cos^2(\phi) }, \\
g_{RZ} =& \; \frac{2}{3} \frac{g}{c} s^2, &
g_\beta =& \; \frac{1}{3} g^2 \frac{s}{c} \cos^2(\phi).
\end{align}

\begin{figure}[!h]
\begin{center}
\begin{tabular}{ll}
\hspace{-10mm}\includegraphics[scale=0.4]{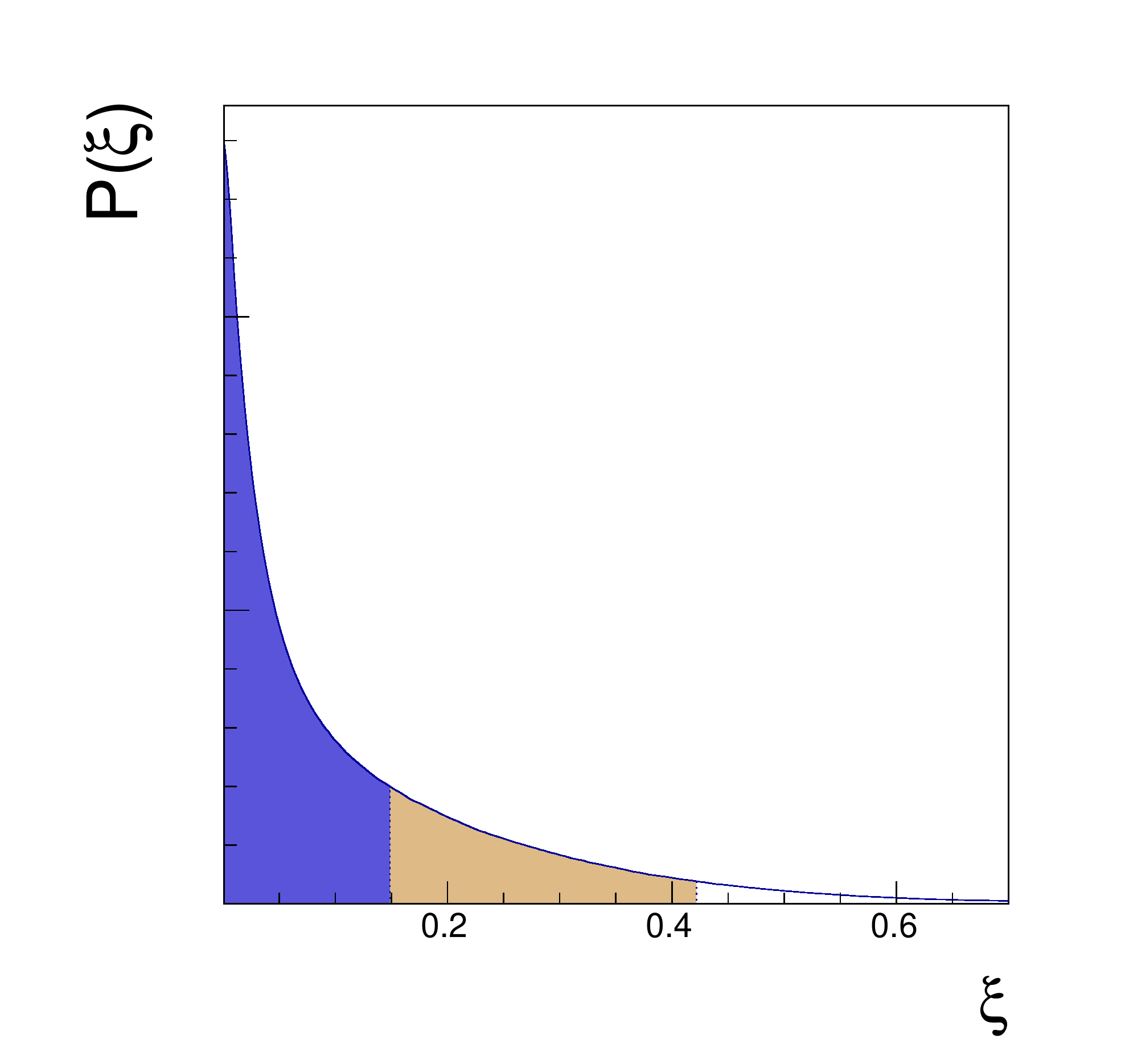} &
\hspace{-10mm}\includegraphics[scale=0.4]{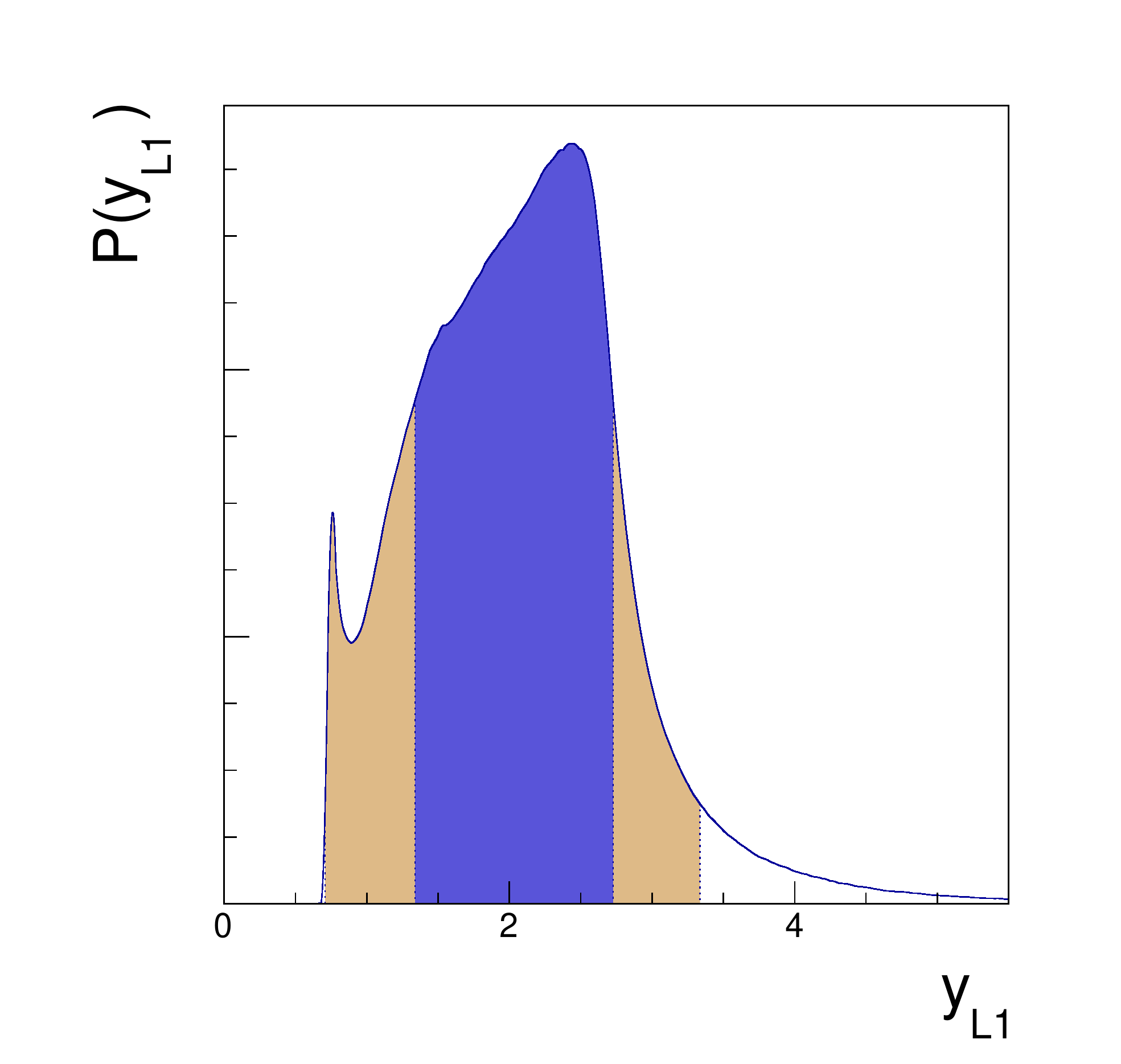} \\
\hspace{-10mm}\includegraphics[scale=0.4]{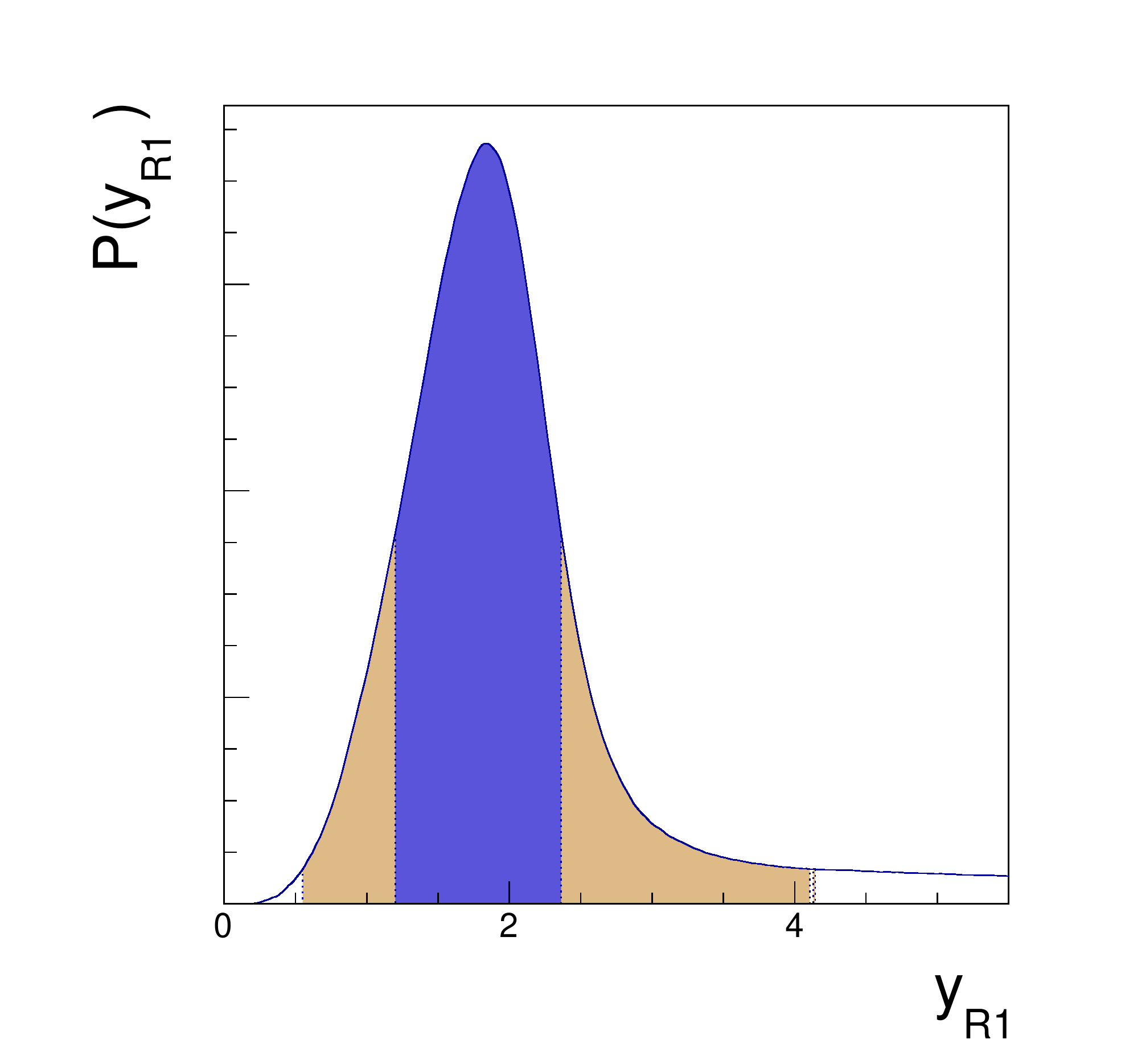} &
\hspace{-10mm}\includegraphics[scale=0.4]{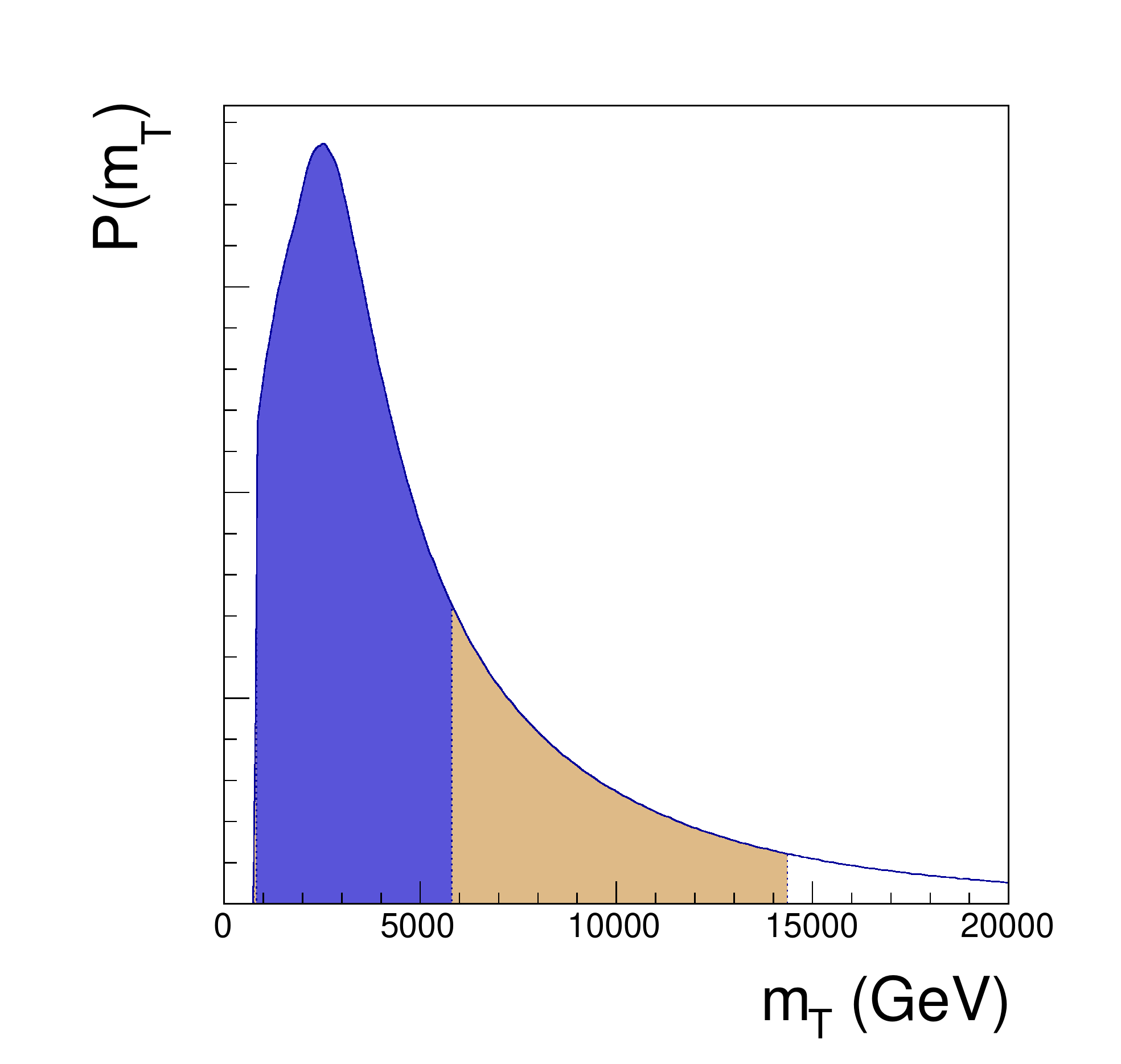} \\
\end{tabular}
\caption[]{Posterior probability density functions for the case of a fermionic singlet of $SO(4)$ (see main text). The sharp drop for small 
values of $y_{L1}$ is due to the top mass constraint.}
\label{fig:singlet}
\end{center} 
\end{figure}

The singlet contribution to the $\hat{T}$ parameter is often positive and big, while $\hat{S}$ can have both signs. 
The contribution to $\delta g_{L}^{(b)}$ is, on the other hand, always positive. Combination of these properties help 
the singlet scenario improve the overall fit to the data dramatically compared to the only composite Higgs scenario, as 
can be seen from the posterior of $\xi$ shown in Fig.~\ref{fig:singlet}. 
It can be seen that the 95\% CL allowed region in $\xi$ has now increased to $\sim 0.4$ as compared to $\sim 0.08$ in 
the only composite Higgs case (see Fig.~\ref{fig:only-higgs}). 
The posterior probability distributions for a few other relevant quantities are also shown in Fig.~\ref{fig:singlet}.

%
By expanding equations \eqref{eq:masses singlet} at leading order in $\xi$, it is found that scenarios in which the mass mixings $y_{(L/R)1}$ 
are both too big or too small are forbidden by the mass constraints. On the other hand, EWPO forbid them to have too different values. Hence 
the posteriors for both the couplings 
$y_{L1}$ and $y_{R1}$ are peaked around $\sim 2$. We are not showing the posterior distribution for the coupling $g_1$ as it is loosely constrained by the fit. 
As a consequence, the mass of the top partner $\widetilde{T}$ can also vary in a big range, as shown in the bottom right plot of Fig.~\ref{fig:singlet}.

We now move to the other cases, namely the $\bf 4$ and the $\bf{4}\oplus \bf{1}$ of $SO(4)$, for which fully analytical calculations are not
possible and a numerical approach is in order. We find that the fermionic 4-plet scenario is strongly disfavoured by the EWPO. This is because the 4-plet 
contributions to the oblique observables are rather large and have the same sign as the composite Higgs contributions. Our findings are qualitatively in 
agreement with \cite{PanicoGrojean}.  From the results of our fit we find a rather stringent 95\% CL upper bound on $\xi$, $\xi \lesssim 0.02$. Consequently, 
all the fermionic resonances are constrained to be rather heavy; with masses above 4 TeV at 95\% CL.

\begin{figure}[t]
\begin{center}
\begin{tabular}{ll}
\multicolumn{2}{c}{\topinset{\includegraphics[scale=0.18]{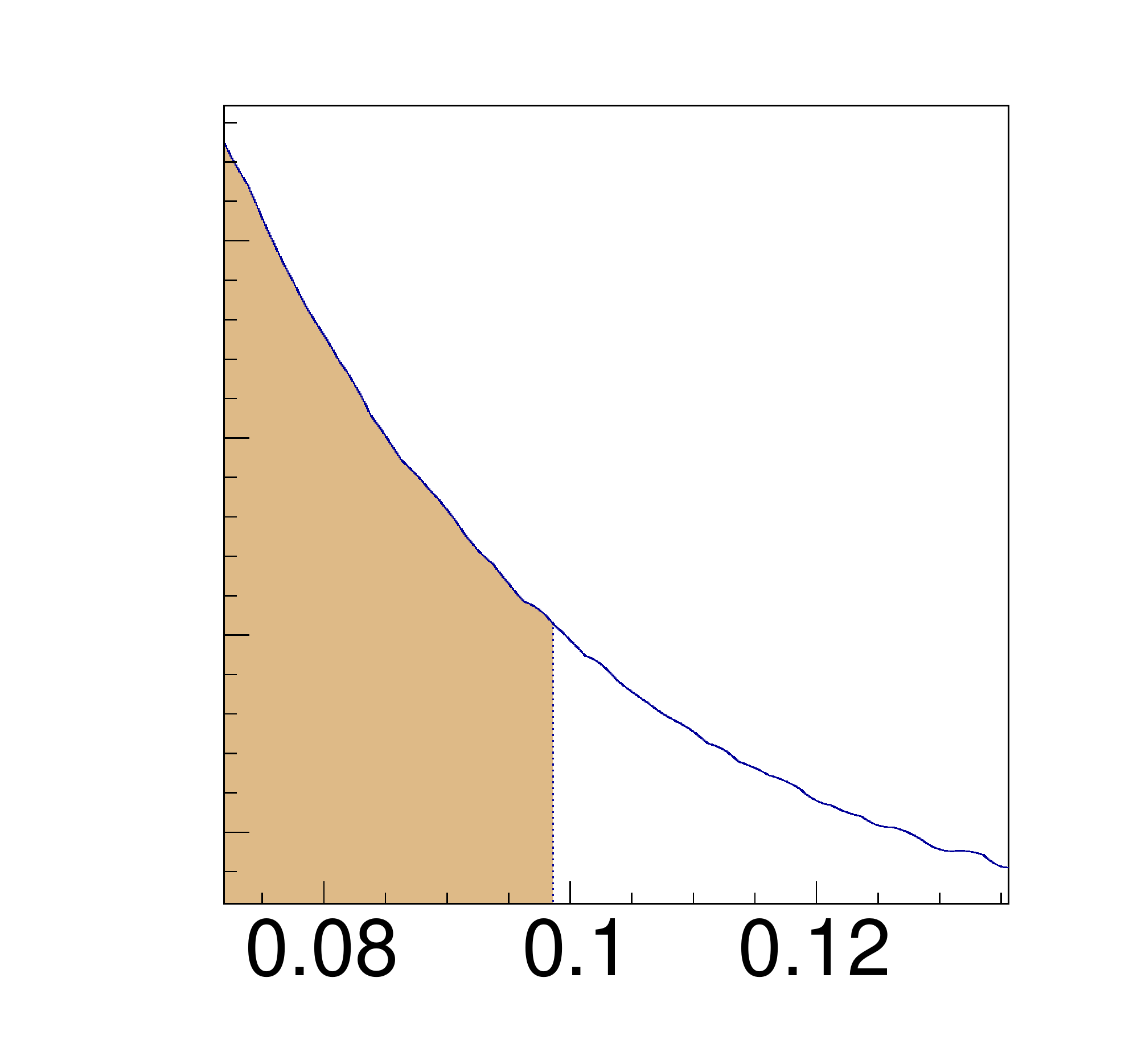}}{\includegraphics[scale=0.4]{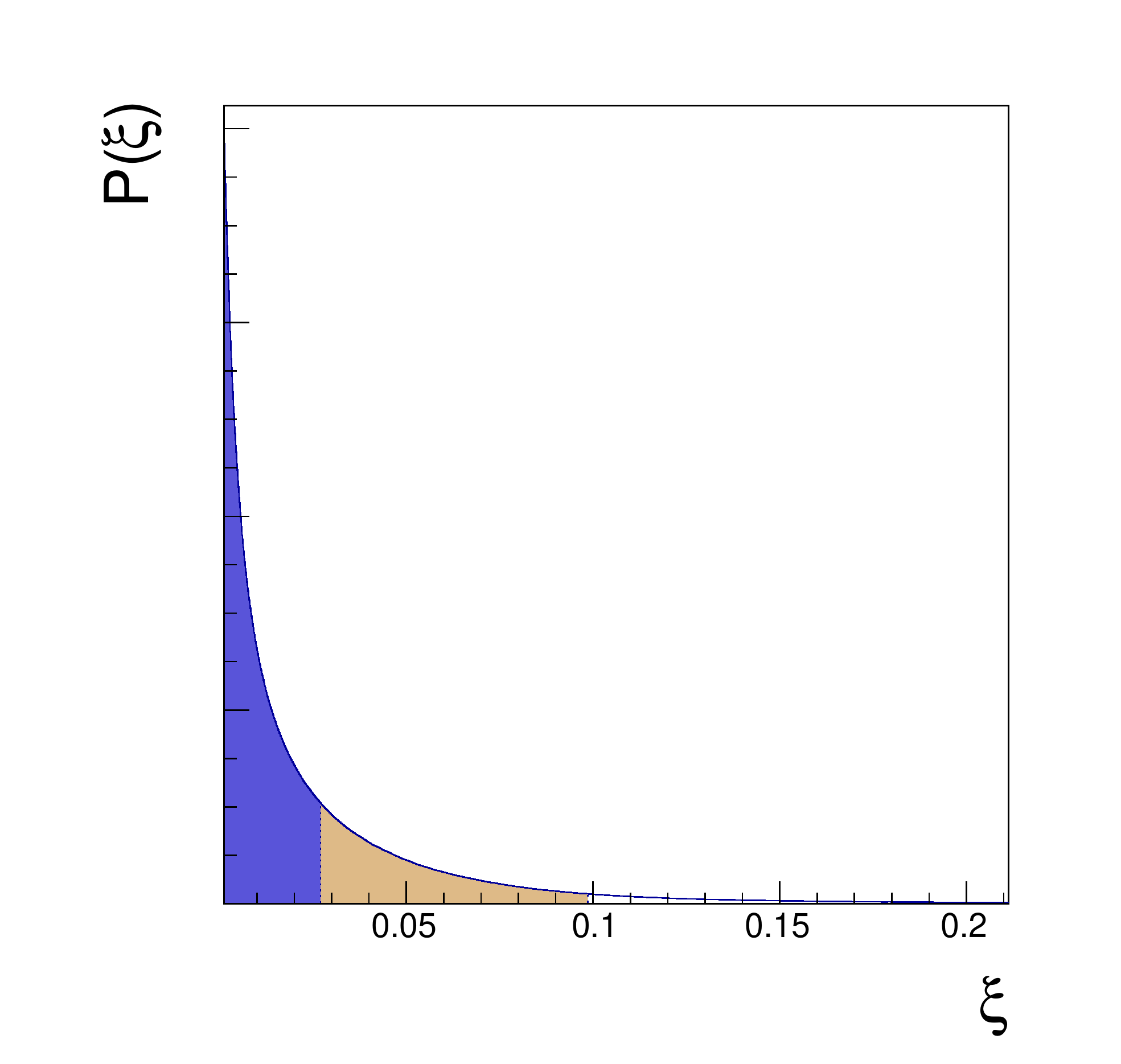}}{25pt}{35pt} }\\
\hspace{-10mm}\includegraphics[scale=0.4]{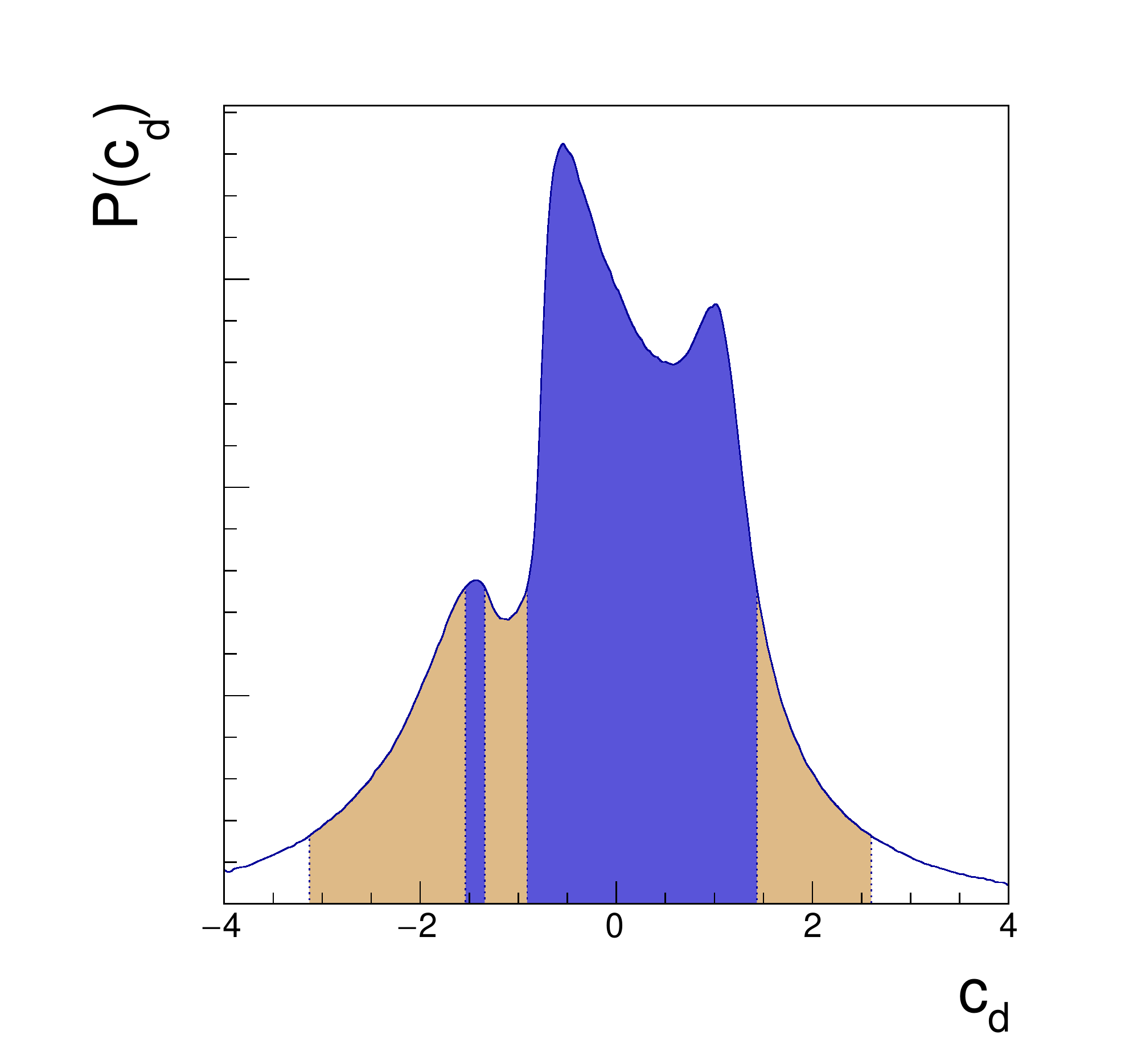} &
\hspace{-7mm}\includegraphics[scale=0.4]{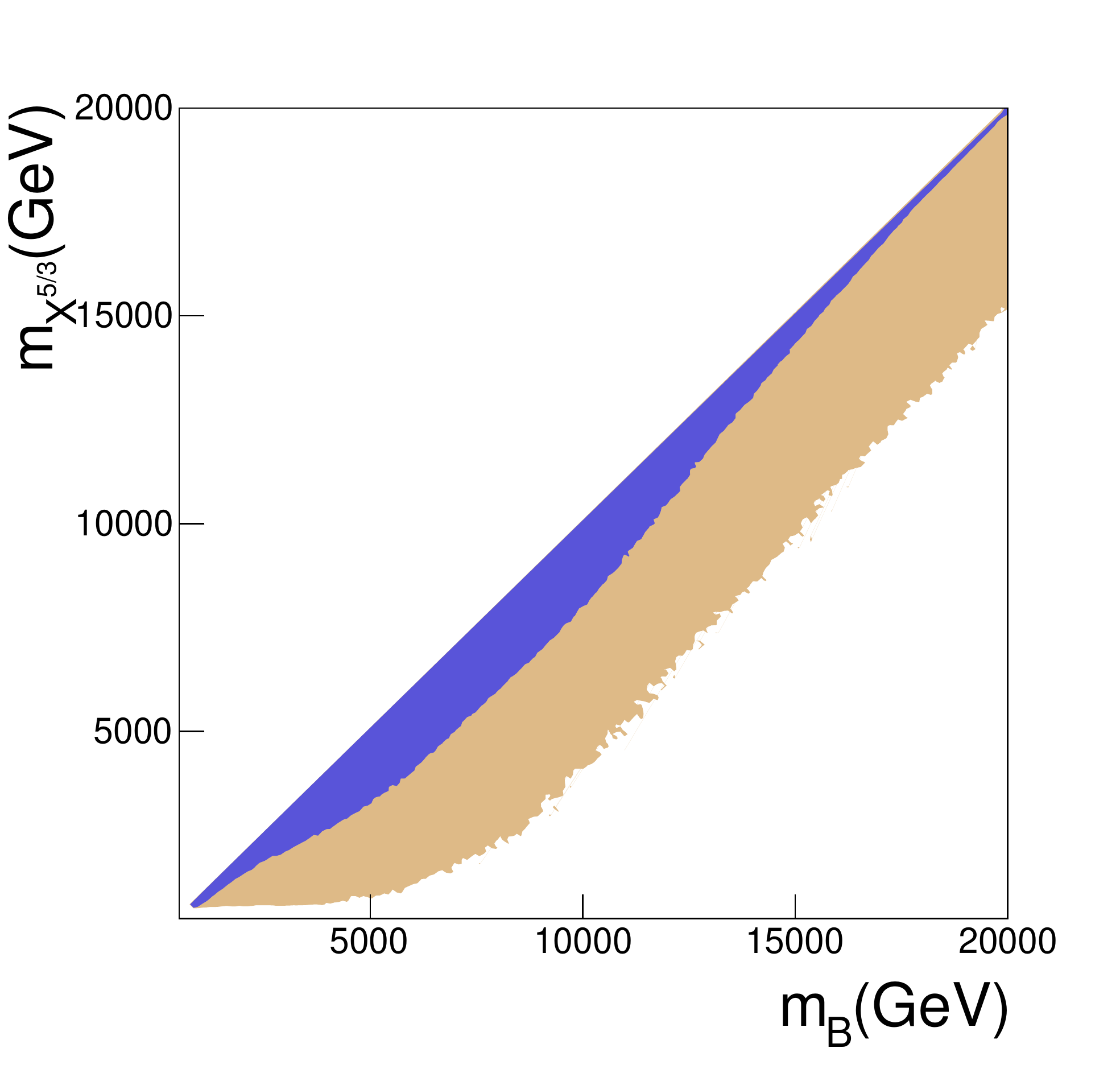}
\end{tabular}
\caption{Posterior probability density functions for the fermionic fiveplet (a $\bf{4}\oplus\bf{1}$ of $SO(4)$) case (see main text).}
\label{fig:5plet}
\end{center} 
\end{figure}

In the presence of both the fermionic singlet and the 4-plet (i.e., a full $SO(5)$ vector), the fit shows a mild improvement over the only Composite Higgs 
case, as shown in figure \ref{fig:5plet}. The interplay between the 4-plet and the singlet contributions to the observables is quite non-trivial. 
The parameters $y_{(L/R)4}$, $g_{1/4}$ are loosely constrained, while there is a preference for $y_{(L/R)1} > 1$, mainly in order to avoid the 
$\hat{T}$ parameter being too negative.
On the other hand, the $\hat{S}$ parameter contains a  big logarithmically divergent term proportional to $\rpar{1-c_d^2}$. In order to keep this term not 
so large, $|c_d|\sim \mathcal{O}(1)$ is preferred, as can be seen from  Fig.~\ref{fig:5plet}. All resonances are generally constrained to be rather heavy 
($\gtrsim 2$ TeV) with no clear mass hierarchy between the 4-plet and the singlet.

As discussed in Ref.~\cite{Panico2site2}, the amount of tuning coming from the requirement of correct EWSB for models with our fermionic sectors is 
estimated as $\mathcal{O}\rpar{\rpar{y_i/g_j}\xi}$, i.e. the naive estimate of $\mathcal{O}(\xi)$ is corrected by a factor $(y_i/g_j)$. From naive theoretical 
expectations (see for example, sec.~\ref{ssec:the model}) such factor is expected to be smaller than 1, thus increasing the amount of fine tuning. 
However, in our analysis we did not use any 
theoretical prejudice for the choice of the ranges of the input parameters, and from the results of our fit we do not see any preference for $(y_i/g_j)$ 
to populate values either lower or larger than 1.

%
\subsection{Spin-1 sector}
\label{ssec:spin1 sector}
%

In this section we will analyse models in which only vectorial resonances are present. In contrast to the contributions coming from fermionic 
resonances, in this case both a tree level and a one-loop term are present, which allows us to have more control on perturbation theory. 
As explained in section \ref{ssec:parameters}, our input parameters are taken at the scale $\Lambda$, but since important running effects 
of the tree level term arise, we have to ensure that the perturbative series controlled by the couplings $g_{\rho_r}$ holds also at the scale 
$M_\rho$, at which spin-1 resonances are integrated out: formally, this is the scale at which the effects of resonances is computed. 
Hence, in addition to the constraints mentioned in section \ref{ssec:data}, we also impose an upper bound on $g_{\rho_r}(M_\rho)$. In addition, 
couplings at both scales will be bounded from below, following the paradigm that the composite sector should remain reasonably strongly 
coupled.\footnote{This choice is also driven by the fact that we have neglected terms of $\mathcal{O}\rpar{g_{\text{el}}^2/g_{\rho_r}^2}$ in 
our calculations.} Thus, in order to be conservative, we have chosen a range of $\spar{1.5,5.5}$ for the couplings $g_{\rho_r}$ at both the 
scales $M_{\rho_r}$ and $\Lambda$. Their $\beta$ functions are calculated in ref. \citep{ContinoSalvarezza} and are given by:
\beq
\beta_{g_{\rho_r}} = g_{\rho_{r}}^3 \spar{ \frac{2a_{\rho_r}^4 - 85}{192\pi^2} - \beta_{2_r} \frac{a_{\rho_r}^4 - a_{\rho_r}^2 - 3}{24\pi^2} - 
\beta_{2_r}^2 \frac{a_{\rho_r}^4}{24\pi^2}},
\hspace{1cm}
r=L,\,R.
\eeq

They are generally negative for $a_{\rho_r}$ inside their PUVC bound, hence the couplings will increase in the Infra Red. However, as we 
will see in the next section, things can easily change when also including fermionic resonances.

Even after imposing these bounds, it is found that, at the mass scale of composite resonances, the one-loop contribution can still be of the 
same order of the tree level one. This happens in particular for big values of $a_{\rho_r}$, which enters quadratically in the one-loop expression 
of $\hat{S}$ (see eq. \eqref{eq:S 1rho}). Thus, in order to be as safe as possible, we also explicitly constrain the ratio between the one loop 
term and the tree level one to values smaller than $0.5$ at the mass scale of resonances. However, this upper bound is only implemented using 
$\beta_{2_r} = 0$, otherwise the reduction of the tree level term obtained for $\beta_{2_r} \sim 0.25$ would violate this constraint and be ruled out. 
Such possibility has nothing to do with the validity of perturbation theory, as it entails an accidental cancellation of the tree-level term only. 
Hence we will impose the bound
%
%
%
\beq
\rpar{
\frac
{\hat{S}^\text{(loop)}}
{\hat{S}^\text{(tree)}}
} 
\Bigg|_{\mu = M_{\rho_r},\,\beta_{2_r} = 0}
< 0.5 \, .
\eeq
%


It is relevant to note that, throughout this paper, we did not consider the corrections to $\Delta \epsilon_3$ coming from the parameters 
$W$ and $Y$ \cite{Barbieri:2004qk}. In fact they are suppressed by a factor $g^2/g_{\rho_r}^2$ with respect to $\hat{S}$, but since they 
also appear at tree level, their relative parametrical suppression with respect to the one loop contribution to $\hat{S}$ is only a factor 
$16\pi^2 g^2/g_{\rho_r}^4 \simeq \rpar{2.9/g_{\rho_r}}^4$. This factor can be easily 1, hence tree level contributions to $W$ and $Y$ 
must also be taken into account. They have been calculated in refs. \cite{ContinoSalvarezza,Contino:2015gdp}, and amount to:
\beq
W \big|_{\rho_L} = \frac{g^4}{4g_{\rho_L}^4} \xi \frac{\cos^4(\theta/2)}{a_{\rho_L}^2} (1-2\beta_{2_L})^2,
\hspace{1 cm}
W \big|_{\rho_R} = \tan^4\rpar{\frac{\theta}{2}} W\big|_{\rho_L},
\label{eq:Wpar}
\eeq
and the contributions to $Y$ are obtained by exchanging $\cos^4(\theta/2) \leftrightarrow \sin^4(\theta/2)$ (we recall that $\xi = \sin^2(\theta)$). These
$\sin^4(\theta/2)$ trigonometrical factors produce an additional $\xi$ suppression to $W\big|_{\rho_R}$ and $Y\big|_{\rho_L}$, which can be neglected.

As one could expect, the mass of the $Z$ boson exhibits a strong sensitivity to the coupling $g^\prime_{\text{el}}$, and in addition, aside from a mild 
dependence on $\beta_{2_r}$, no other parameter is sizeably relevant. As a consequence, the coupling $g_{\text{el}}^\prime$ always sits in a very 
narrow region centred around the physical coupling $g^\prime$, and the $M_Z$ constraint can be satisfied with a slight tuning of $g^\prime_{\text{el}}$ 
for any value of the other parameters. For this reason, it does not play an important role in the fit. 

We will start considering the case for which a single resonance is present. As far as the EWPO are concerned, it makes no difference wether we choose 
a $\rho_L$ or a $\rho_R$ vector, as their final expressions are exacty the same at our level of approximation (i.e., $g_{el} = g_{el}^\prime =0$). On the other 
hand, since our calculations for the masses of resonances are done by numerical diagonalization, they resum  terms of all orders in $g_{el}$ and $g_{el}^\prime$ making  
(as already observed, again, in the beginning of section \ref{sec:calculations}) the mass splittings between neutral and charged states slightly different for the left and 
the right vectors. However, the difference between the two scenarios is still very small, hence we will only discuss one single-resonance case: for instance, 
we will focus on a $\rho_L$.

The expressions for the oblique parameters are in this case given by equations \eqref{eq:T 1rho}, \eqref{eq:S 1rho}, \eqref{eq:Wpar} and the text following Eq.~\eqref{eq:Wpar}. 
This model is manifestly incomplete, as no mechanism for the generation of quark masses is 
explicitly present. In this case we can only test the EWPO, hence we remove the top mass 
constraint from the fit. We emphasise that here we are not assuming the Yukawa couplings to be SM like, rather 
we are just keeping the Yukawa sector unspecified and hence, no constraints are being applied on them.

As already commented, the positive tree level contribution to $\hat{S}$ can be reduced by tuning the parameter $\beta_{2_L}$, while the one-loop 
contribution is mostly negative. On the other hand, the contribution to $\hat{T}$ can be positive, and can sizeably compensate, or also overcome, 
the negative shift from the Higgs for $a_{\rho} \sim 1.5$ or more. Remarkably, all these effects can appear simultaneously, making a single vectorial 
resonance a good candidate to accomodate EWPO. 

\begin{figure}[t]
\begin{center}
\begin{tabular}{ll}
\hspace{-10mm}\includegraphics[scale=0.4]{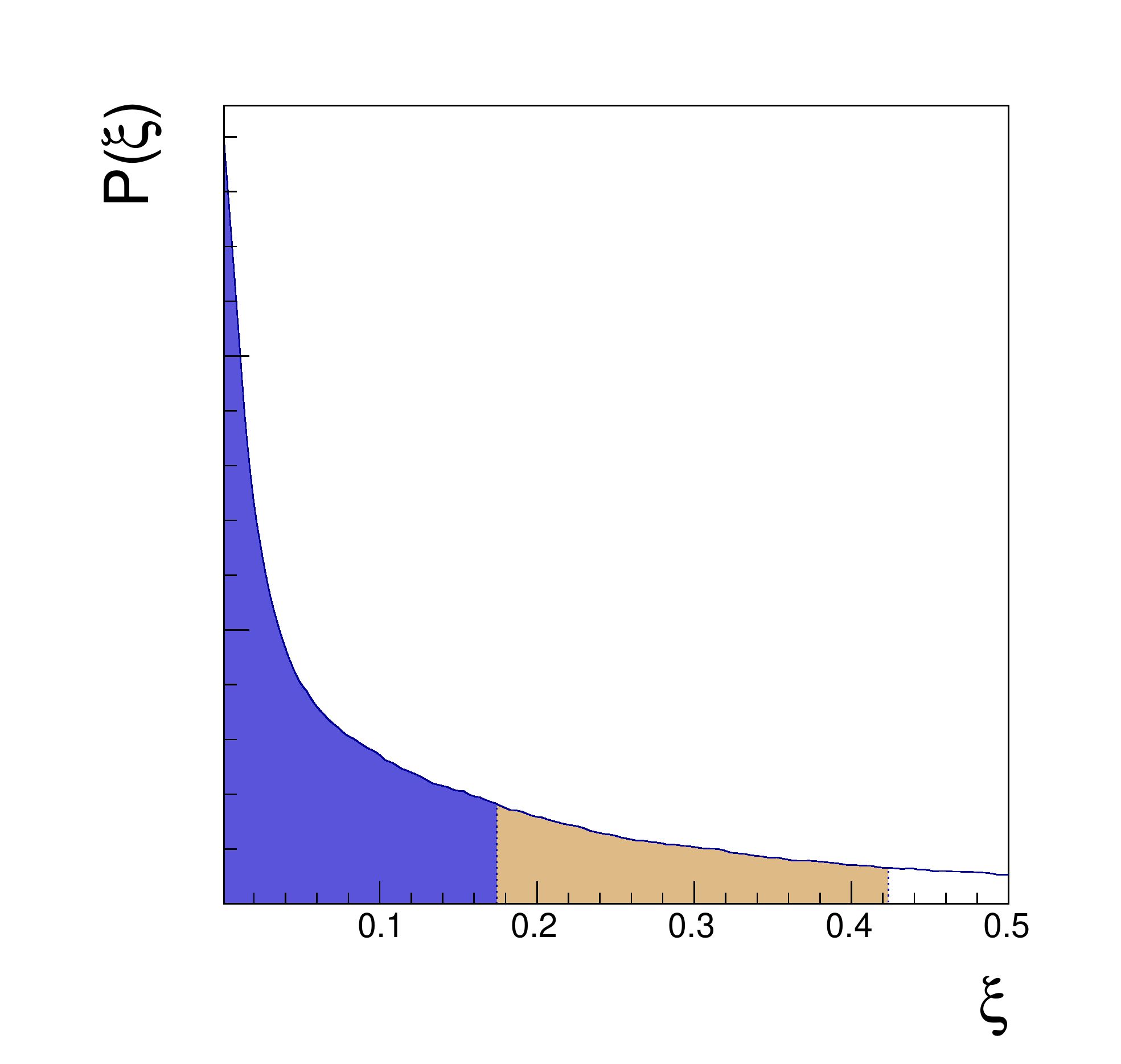} &
\hspace{-10mm}\includegraphics[scale=0.4]{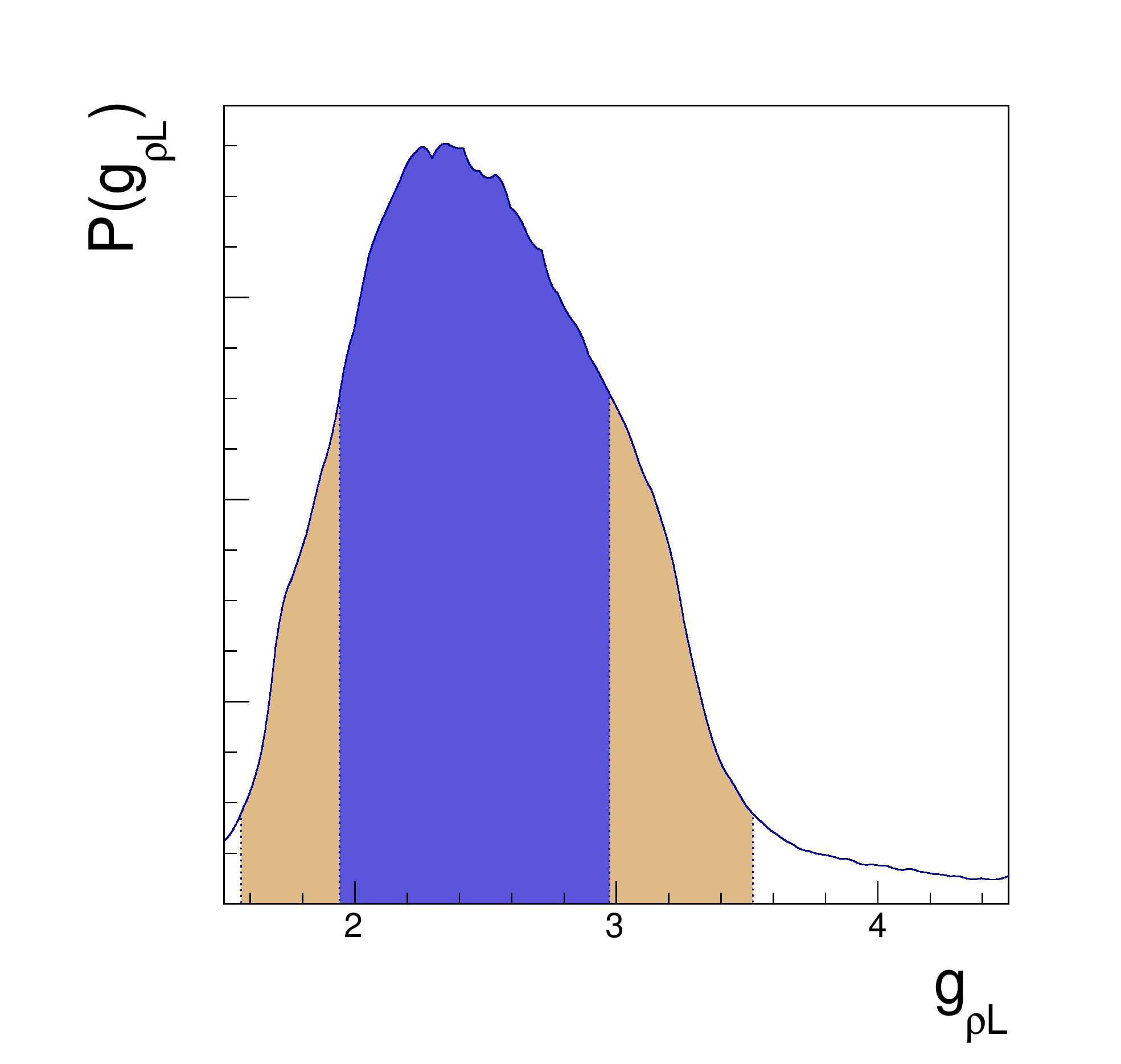} \\
\hspace{-10mm}\includegraphics[scale=0.4]{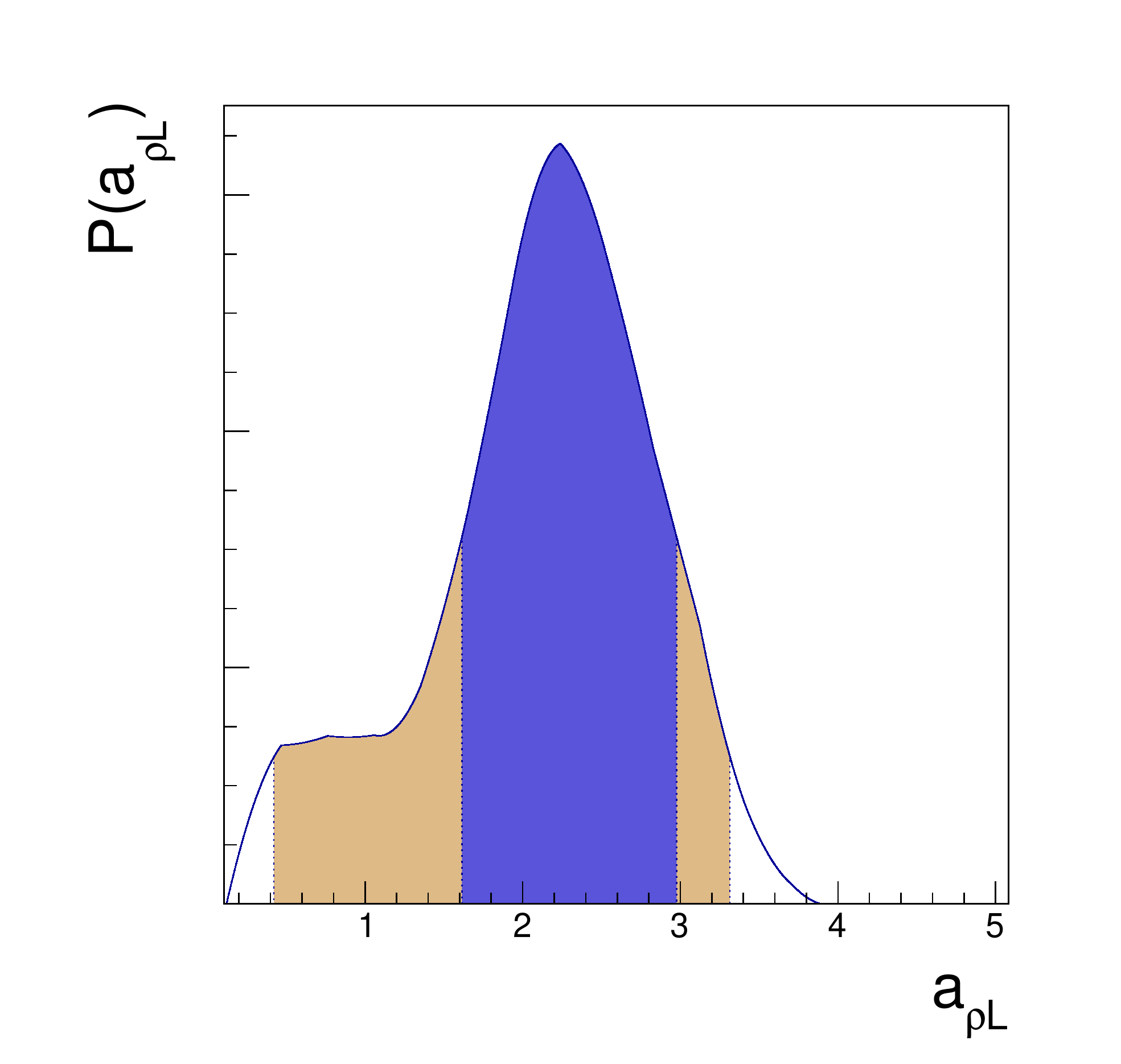} &
\hspace{-10mm}\includegraphics[scale=0.4]{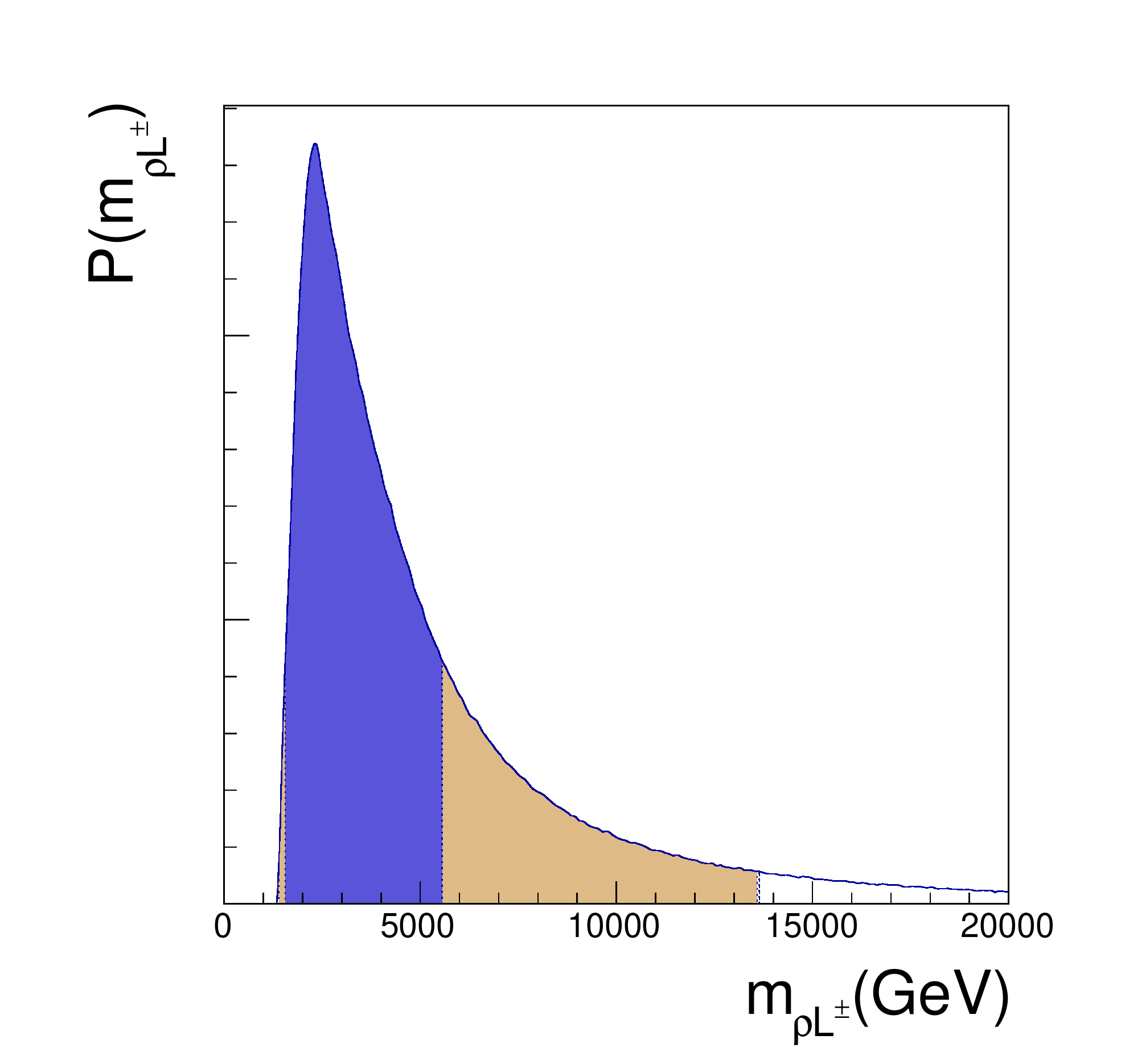} 
\end{tabular}
\caption[]{Posterior probability density functions for the case of an $SU(2)_L$ vectorial triplet (see main text). }
\label{fig:singleVector}
\end{center} 
\end{figure}

Indeed, as shown in Fig.~\ref{fig:singleVector} the bound on $\xi$ is weakened considerably and values up to $\sim 0.4$ are allowed at 95\% probability. 
The parameters $g_{\rho_L}$ and $a_{\rho_L}$ are mainly constrained, respectively, by the requests on the RG-evolved $g_{\rho_L}\rpar{M_{\rho_L}}$ 
and the ratio between the one-loop term and the tree-level term of $\hat{S}$ at the same scale. The parameter $\beta_{2_L}$, as expected, sits in a range 
$\sim \pm 1$ peaked around $\beta_{2_L} = 0.25$ \footnote{Notice that (according to our assumption $\Lambda=3 M_\rho$) in the central part of this 
range the estimate of the contributions coming from the local operators does not 
get the enhancement mentioned in sec.~\ref{ssec:the model}.}. Despite being parametrically sizeable, the contribution of $W$ to $\Delta \epsilon_3$ 
plays a minor role in the fit: this is due to a slight suppression $1/a_{\rho_L}^2$ and also accidental numerical factors. Resonances have mass in the multi-TeV 
range with a small splitting between neutral and charged states. Just like all the other cases, experimental lower mass bounds only play a small role.

We will now move to the case in which both resonances are present. In this scenario the observable $\delta g_L^{(b)}$ is still set to zero, and the contributions 
to the oblique observables are given by eqs. \eqref{eq:T 2rho}, \eqref{eq:S 2rho}. Since the contribution to $\hat{S}$ is just the sum of the two single-resonance 
contributions, the same qualitative statements of the single-resonance case also apply here, while on the quantitative side the contributions are of course 
typically bigger. On the other hand, in the $\hat{T}$ parameter an extra term (mostly negative and growing like $a_{\rho_r}^4$) arises, and the behaviour 
changes completely: in particular, while in the single-resonance case values of $a_{\rho_r}$ around $\sim 2$ provided an overcompensation of the Higgs 
contribution, in this case the extra term grows quickly and brings the overall resonance contribution to negative values. 

\begin{figure}[t]
\begin{center}
\begin{tabular}{ll}
\hspace{-10mm}\topinset{\includegraphics[scale=0.18]{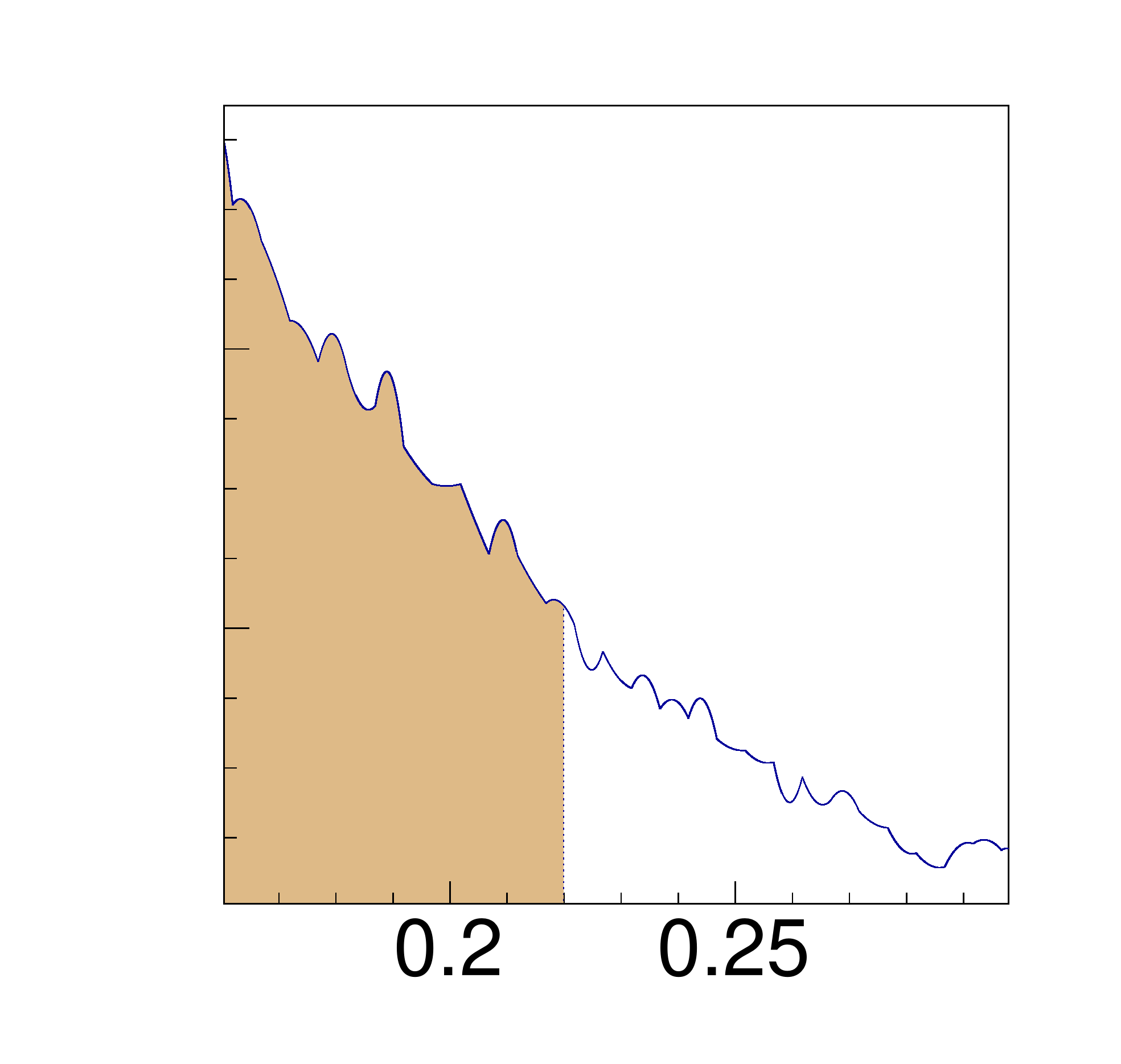}}{\includegraphics[scale=0.4]{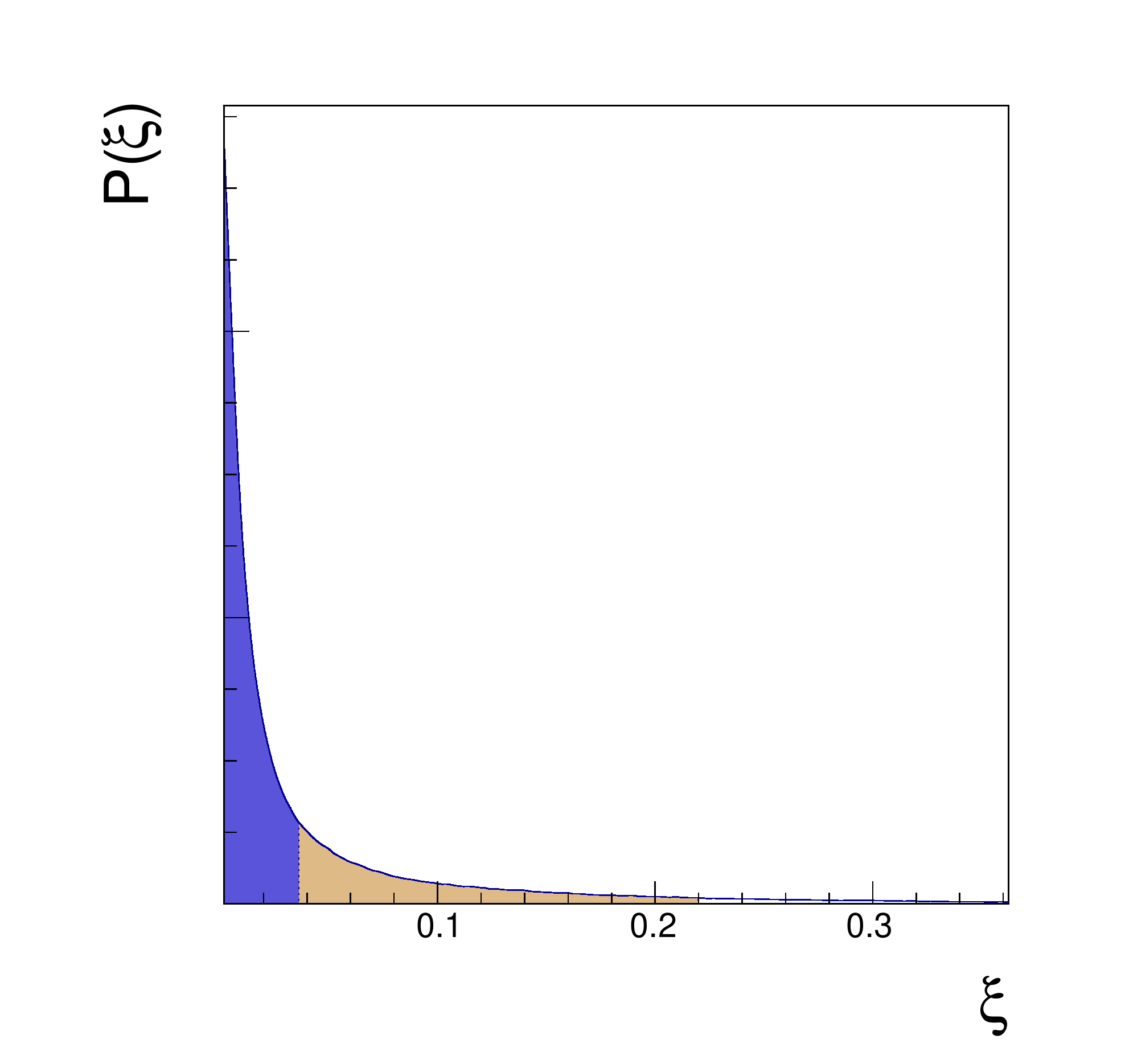}}{25pt}{35pt}&
\hspace{-10mm}\includegraphics[scale=0.4]{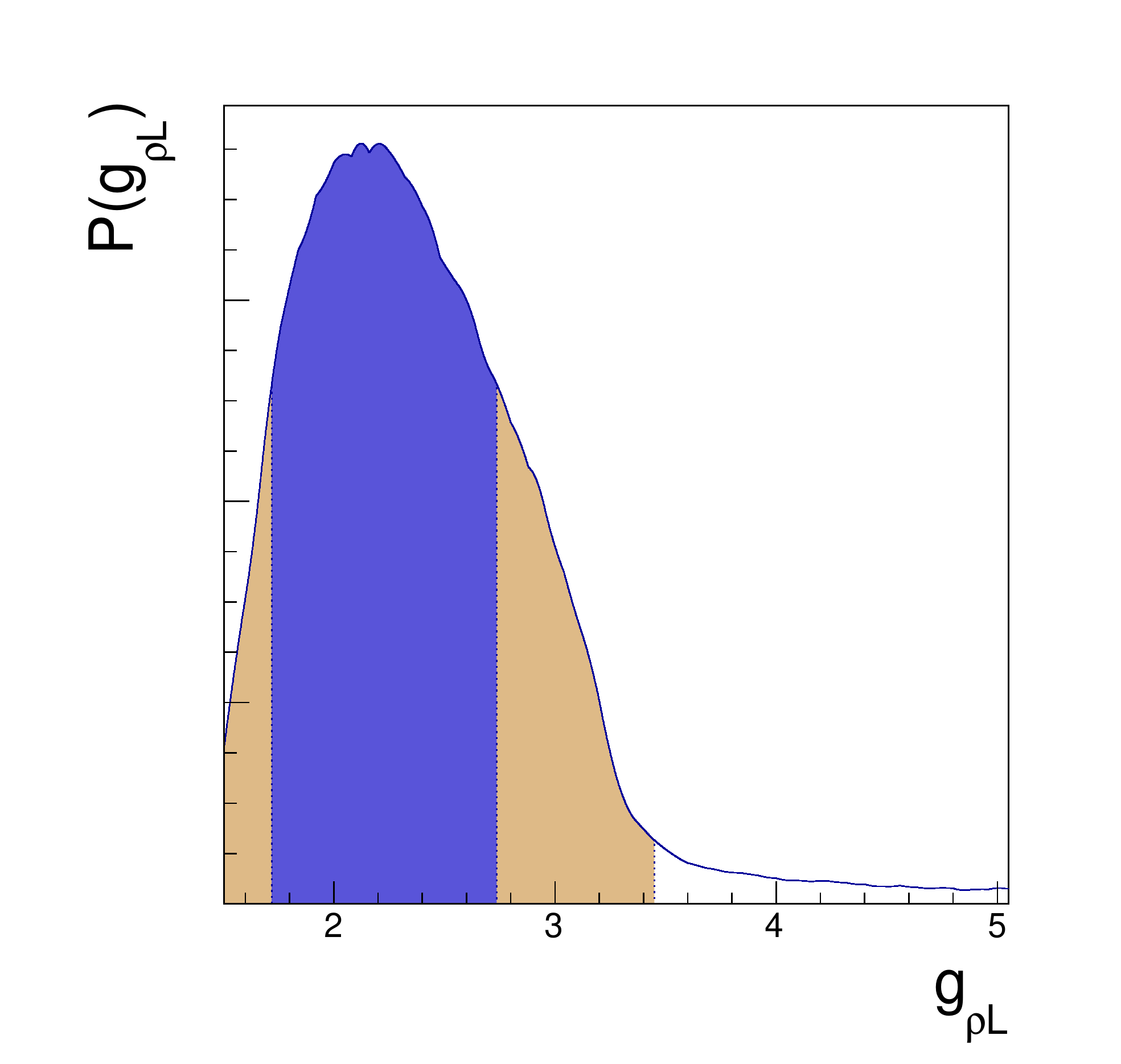} \\
\hspace{-10mm}\includegraphics[scale=0.4]{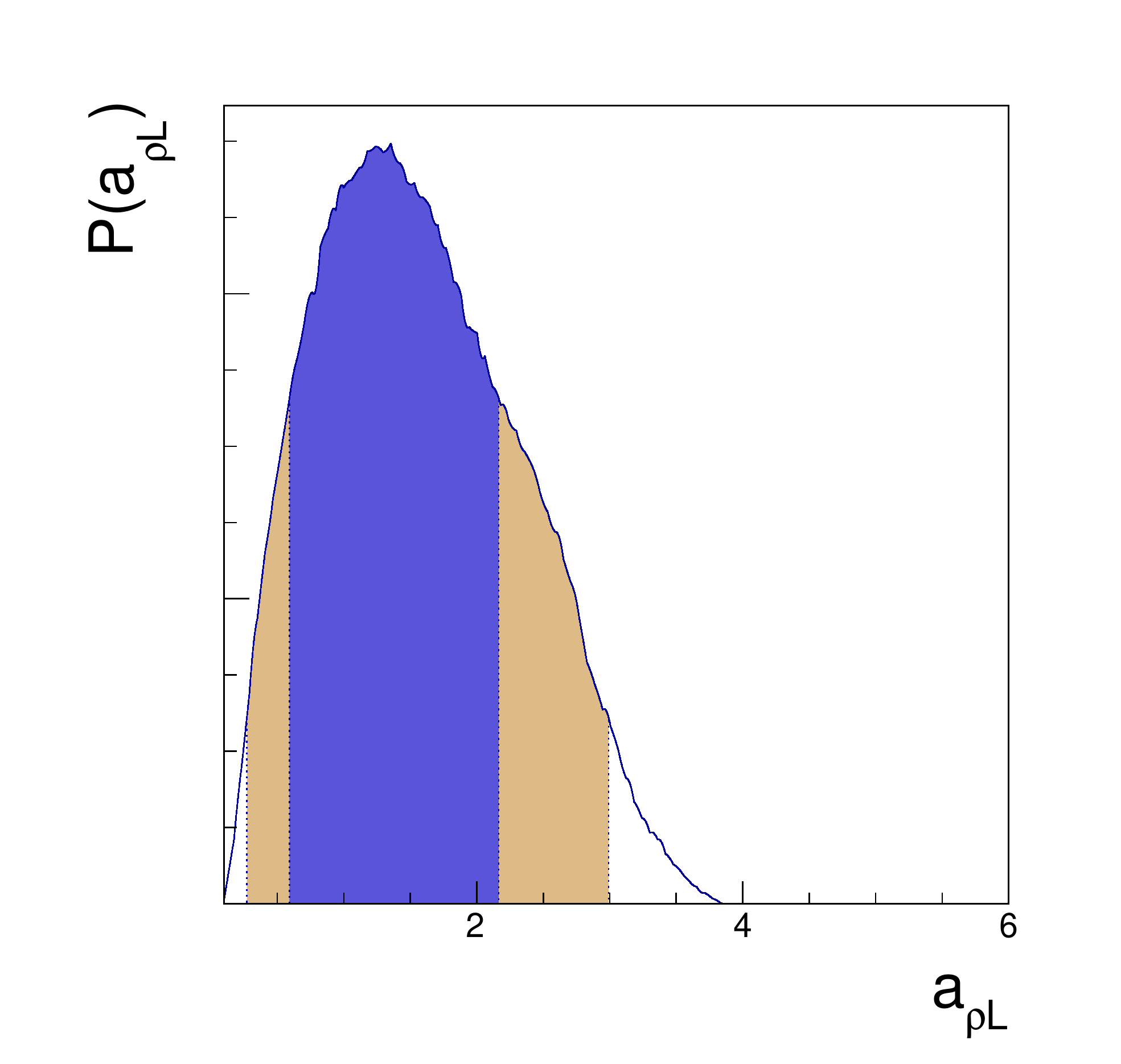} &
\hspace{-7mm}\includegraphics[width=220pt,height=220pt]{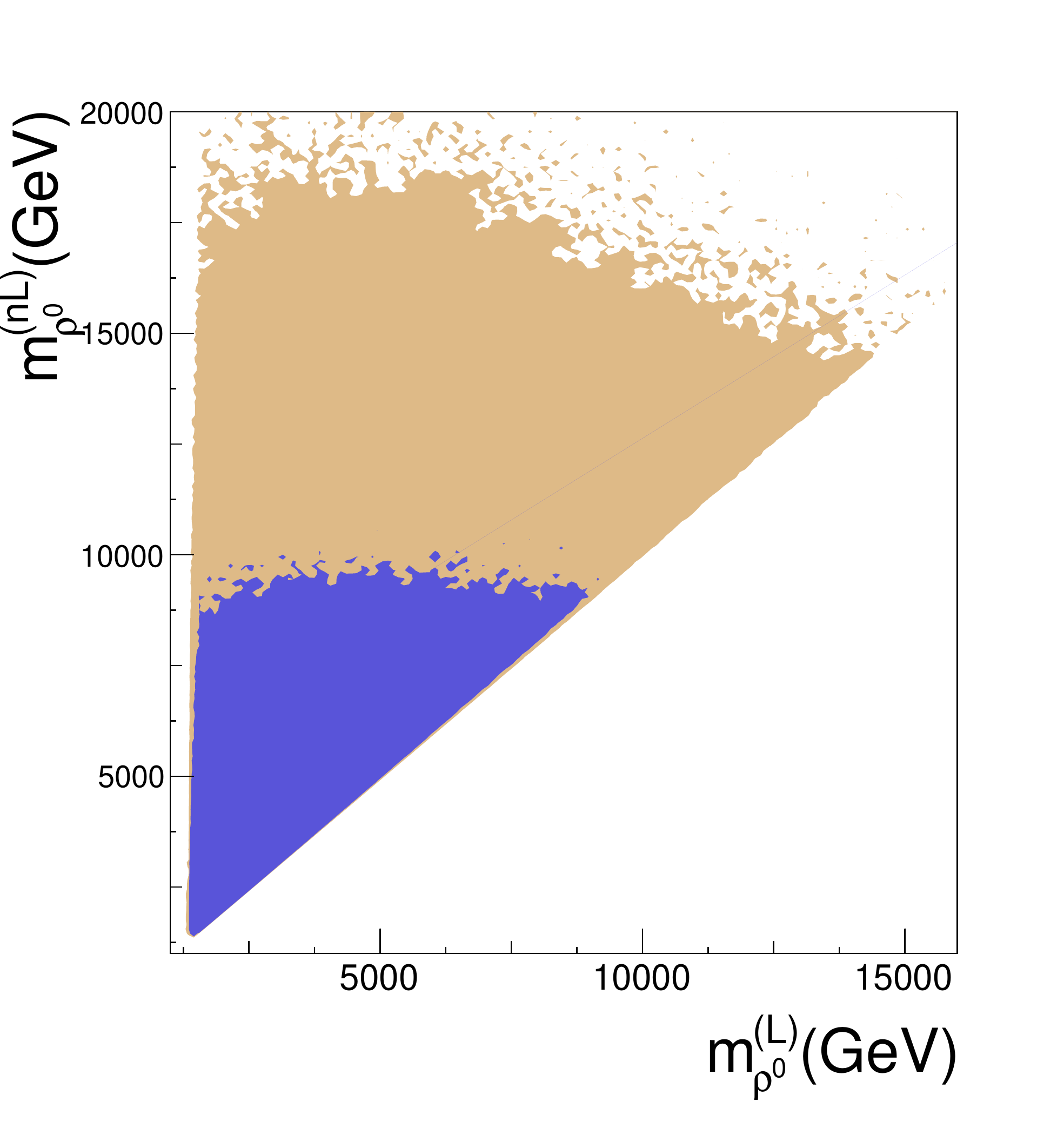} 
\end{tabular}
\caption[]{The posterior probability distributions for a few parameters and  masses in the scenario when both the $SU(2)_L$ and $SU(2)_R$ vectors 
are present. The superscripts (L) and (nL) correspond to the lightest and the next to the lightest resonances respectively. \label{fig:bothVector}}
\end{center} 
\end{figure}

Given these remarks, it turns out that the double-resonances case offers less room to accommodate the EWPT. Still, as shown in figure 
\ref{fig:bothVector}, an improvement of the picture offered by the Higgs is still obtained, with a small 95\% probability tail extending up to $\sim 0.2$.

\subsection{Combining spin-1/2 and spin-1 resonances}

In this section we will discuss a scenario where both the fermionic and vectorial resonances are present. In particular, we will investigate whether 
the presence of a single vectorial resonance can help relax the strong constraints on the fermionic 4-plet case. For definiteness we will consider the 
$SU(2)_L$ triplet of vectors. The fit results for the $SU(2)_R$ triplet do not show any sizeable difference compared to the $SU(2)_L$ one, except 
that the posterior of $c_R$ is similar to that of $-c_L$. This difference is driven by $\delta g_L^{(b)}$, as can be seen by the sign difference between 
Eq.~\eqref{cL} and Eq.~\eqref{cR} (see also the text below).

In this case we still apply all the additional requests on the spin-1 sector discussed in section \ref{ssec:spin1 sector}, namely the requests on 
$g_{\rho_L}\rpar{M_{\rho_L}}$ and the ratio between the one loop and the tree level term of the pure vectorial contribution. In this case the beta function 
of $g_{\rho_L}$ gets a big positive contribution by fermion loops proportional to $c_L^2$:
\beq
\beta_{g_{\rho_L}} = g_{\rho_L}^3 \spar{ \frac{2a_{\rho_L}^4 + 48c_L^2 - 85}{192\pi^2} - 
\beta_{2_L} \frac{a_{\rho_L}^4 - a_{\rho_L}^2 - 3}{24\pi^2} - \beta_{2_L}^2 \frac{a_{\rho_L}^4}{24\pi^2}},
\hspace{1cm}
r=L,\,R.
\eeq

Notice that, differently from the typical situation in the absence of fermionic resonances, the inclusion of such contribution can easily make the $\beta$ function positive.

\begin{figure}[t]
\begin{center}
\begin{tabular}{ll}
\hspace{-10mm}\topinset{\includegraphics[scale=0.18]{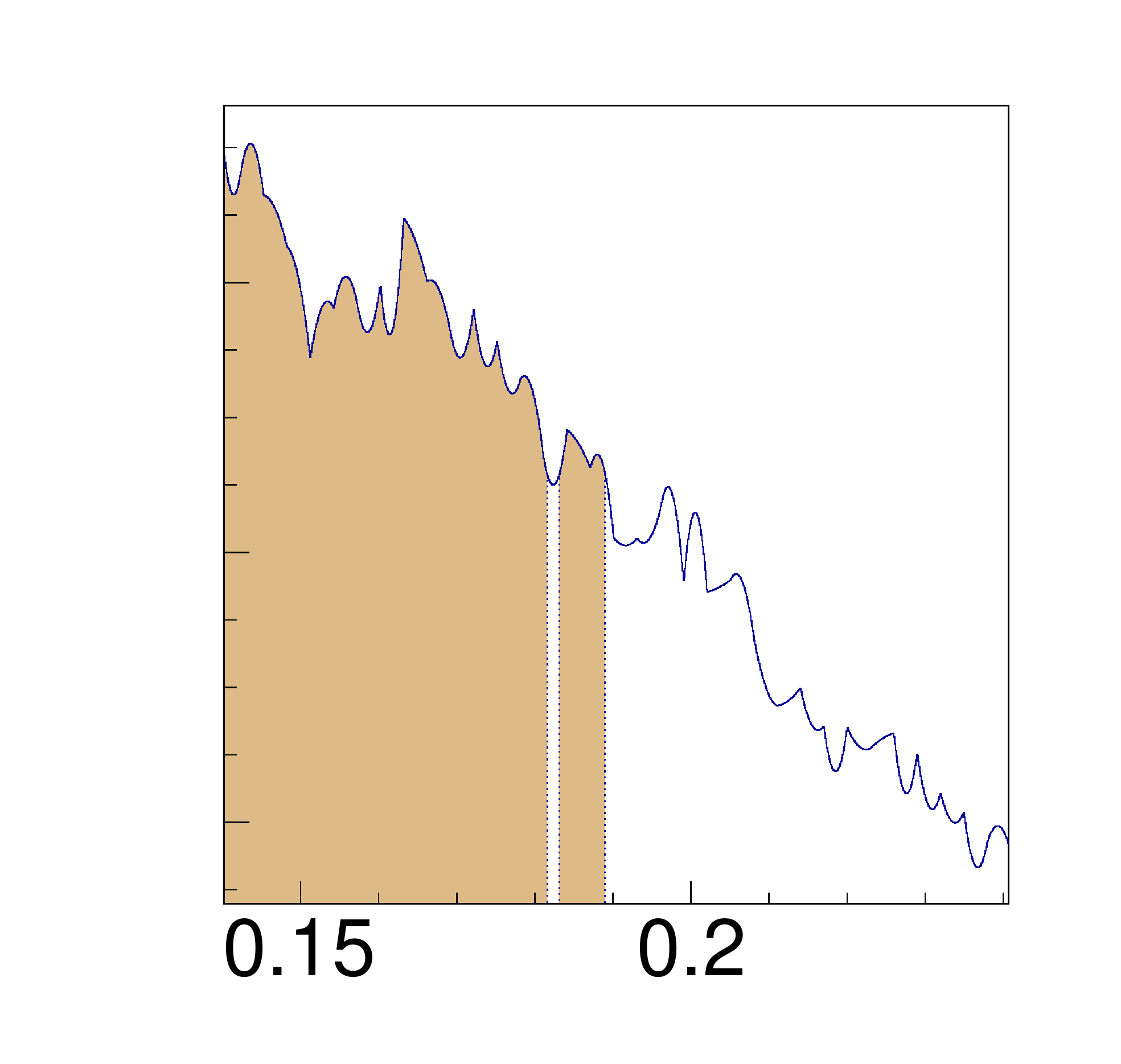}}{\includegraphics[scale=0.4]{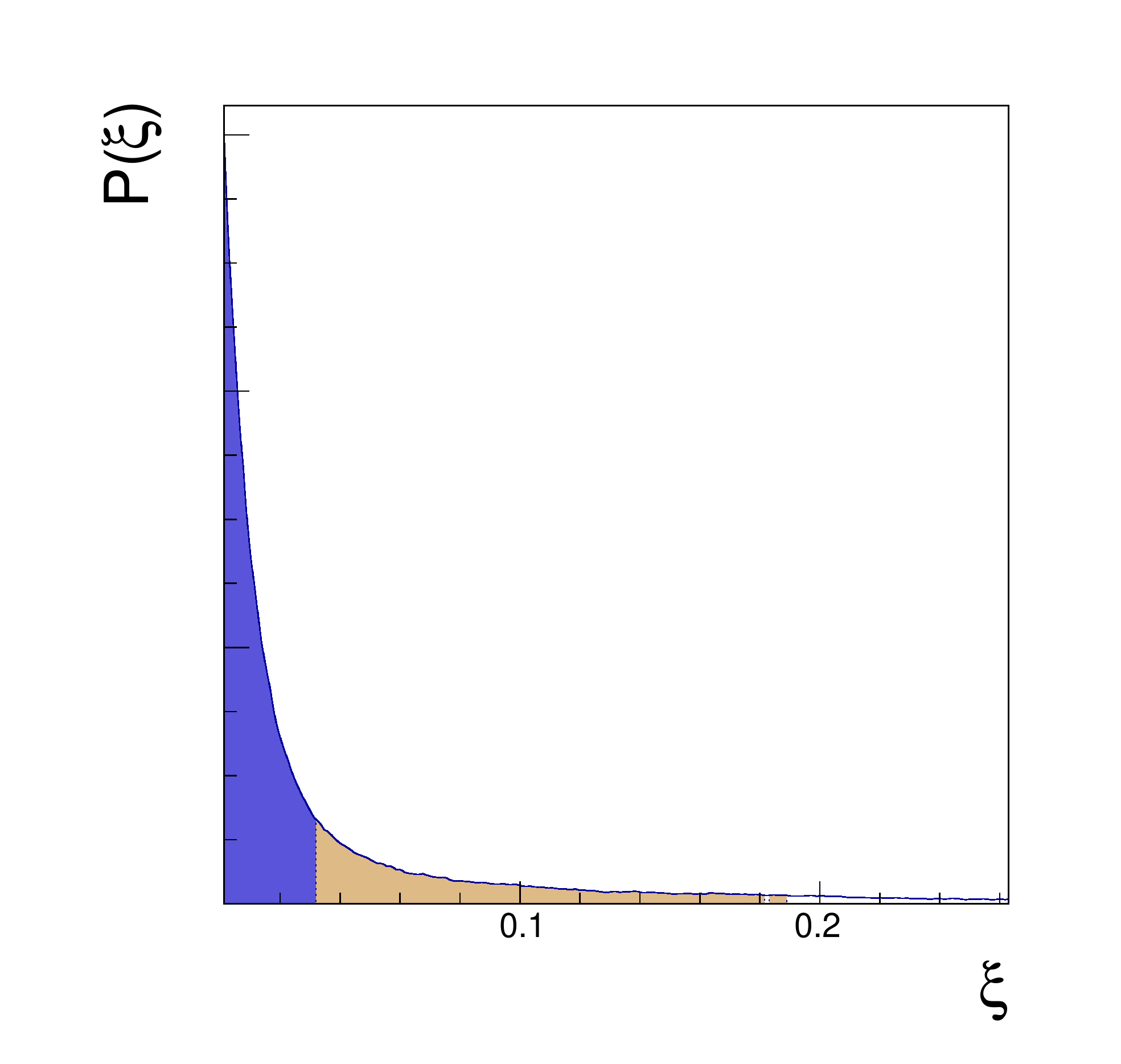}}{25pt}{35pt}&
\hspace{-10mm}\includegraphics[scale=0.4]{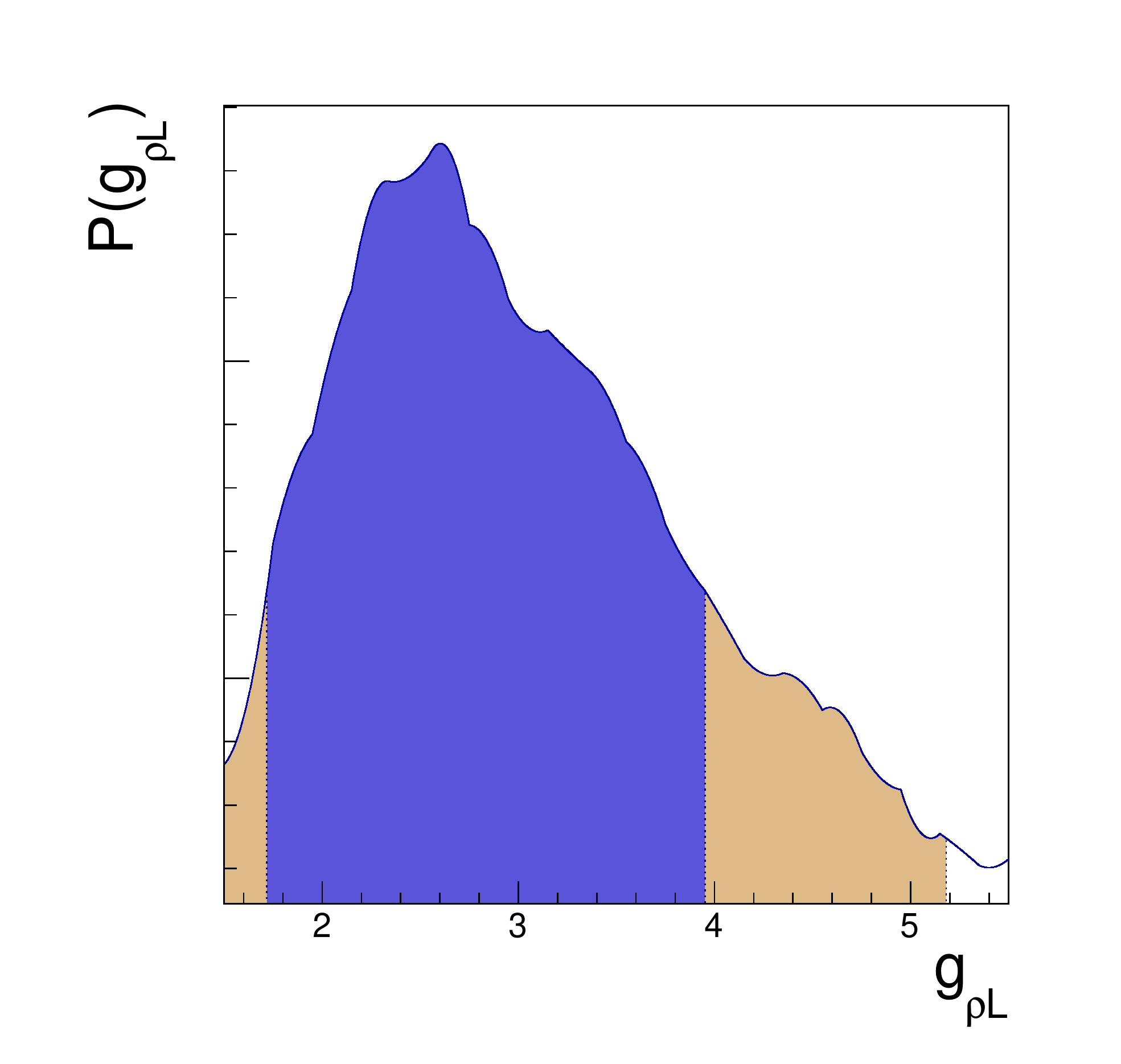} \\
\hspace{-10mm}\includegraphics[scale=0.4]{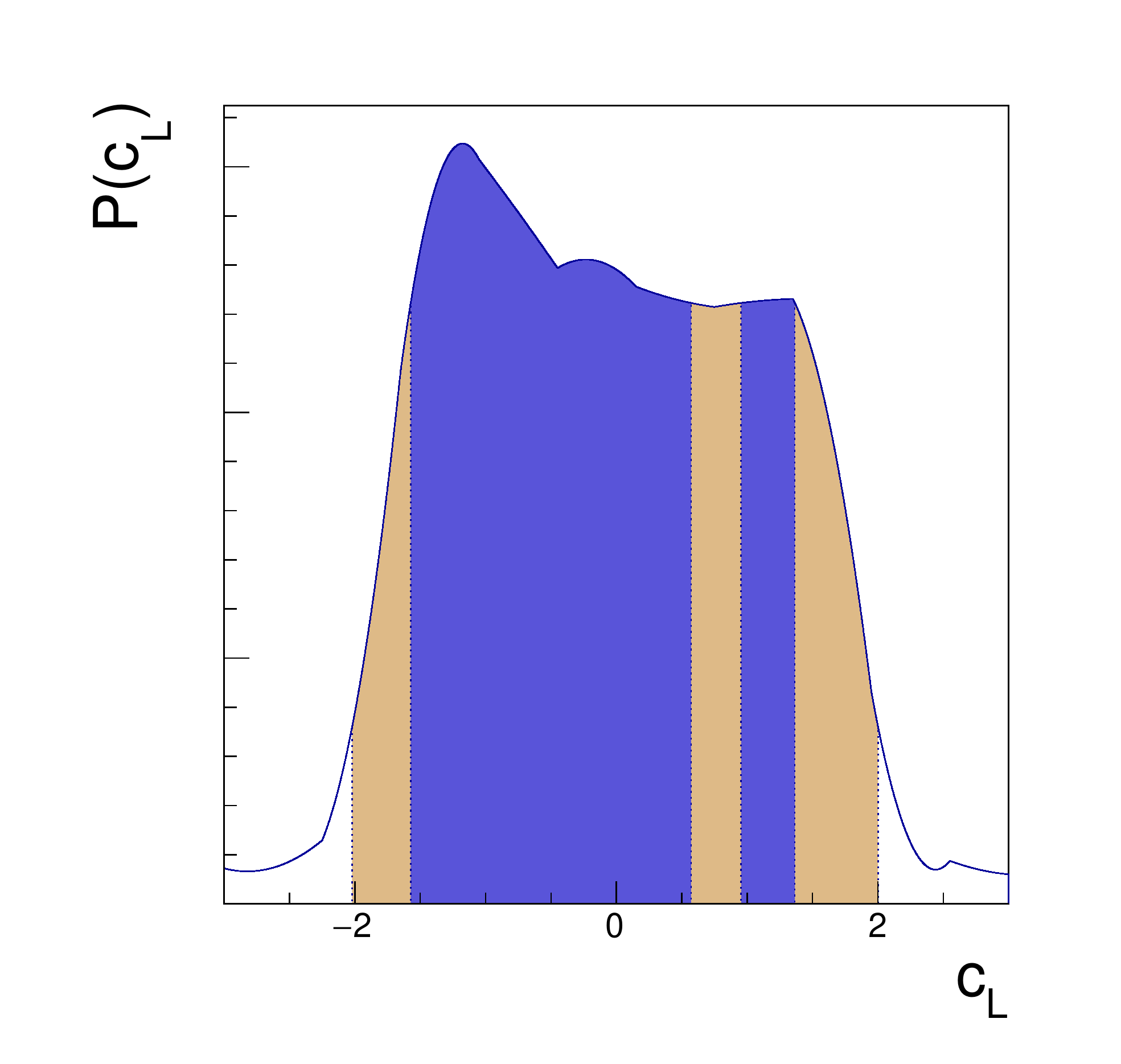} &
\hspace{-7mm}\includegraphics[width=220pt,height=220pt]{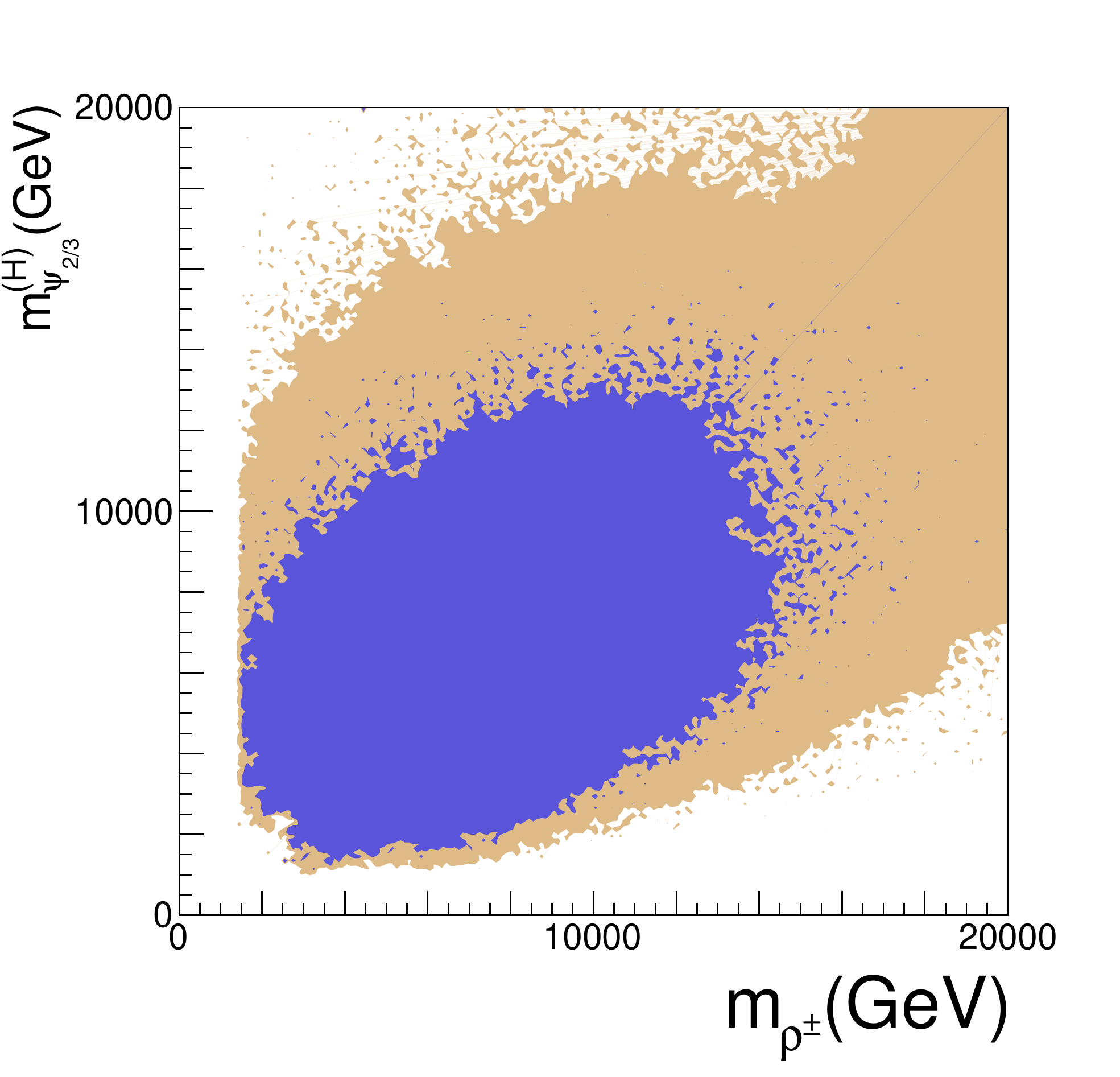} 
\end{tabular}
\caption[]{The posterior probability distributions for a few parameters and  masses in the scenario in which the fermionic 4-plet is present along with the 
the $SU(2)_L$ vectors. The superscript (H) refers to the heaviest charge 2/3 fermion. \label{fig:4plet+LVect}}
\end{center} 
\end{figure}

The results presented in Fig.~\ref{fig:4plet+LVect} show that a sizeable improvement is indeed present, with a 95\% probability upper bound 
relaxed from $\xi \lesssim 0.02$ to $\xi\lesssim 0.2$. We did not find any significant change in the posterior for the parameters $y_{L4}$, $y_{R4}$ 
and $g_{4}$ with respect to the pure fermionic case. After all, the purely fermionic contributions to the oblique observables cannot be too big, so the EWPO 
still require values of the fermionic parameteres that minimise them: in particular the large negative $\hat{T}$ can be partially tamed by choosing 
$y_{L4} < g_4$. On the other hand, the top mass constraint strictly forbids values of $y_{R4} \lesssim 1$, just like $y_{L1}$ for the singlet 
case (compare with fig. \ref{fig:singlet}).

The request $g_{\rho_L}\rpar{M_{\rho_L}} \in [1.5,5.5]$ forbids the beta function of $g_{\rho_L}$ from being too big, which is reflected in a posterior distribution 
showing $|c_L| \lesssim 2$. The presence of $c_L$ increases the possibility for some cancellation in the beta function, which results in a broader allowed range 
of $g_{\rho_L}$ compared to the purely vectorial case. For values of $c_L$ of $\mathcal{O}(1)$, the mixed fermion-vector contribution to $\hat{S}$ is typically 
subleading with respect to the pure vectorial and fermionic terms.
On the other hand, the contribution to $\delta g_L^{(b)}$ proportional to $c_L$ is typically dominant over the fermionic term. 
We have checked that for $c_L \sim -1$ $\delta g_L^{(b)}$ is typically smaller (and negative) than for $c_L \sim 1$, so that the former is slightly preferred by 
the fit.

Differently from the pure vectorial fit, in this case the posterior distribution for $\beta_{2_L}$ is peaked at the slightly negative value 
$\beta_{2_L} \sim -0.1$ instead of 0.25. With this choice it is possible to increase the positive shift to the $\hat{T}$ parameter from the pure vectorial 
contribution. On the other hand, the positive tree level contribution to the $\hat{S}$ parameter in Eq.~\ref{eq:S 2rho}  is increased by this choice. However, 
the one-loop contribution is also bigger and negative, and its relative effect to the tree level one is increased by the fact that bigger values of the 
coupling $g_{\rho_L}$ are now allowed by the fit. We stress the fact that $g_{\rho_L}$ is still a perturbative coupling in the window 
$\spar{M_{\rho_L},\Lambda}$ and we still constrain the ratio of the tree level to the one loop term of the pure vectorial contribution at the scale 
$M_{\rho_L}$ to be smaller than 0.5.

%
\subsection{The two-site model limit}
%
As discussed in sec.~\ref{ssec:symmetries}, the two-site model limits is obtained by enforcing the relations, 
\begin{alignat}{3}
y_{L1} &= y_{L4}, &\qquad
y_{R1} &= y_{R4}, &\quad
c_d &= 0, \nonumber
\end{alignat}
\vspace{-0.7cm}
\begin{alignat}{2}
g_{\rho_L} &= g_{\rho_R}, &\qquad
a_{\rho_L} &= a_{\rho_R} = \frac{1}{\sqrt{2}}\, ,
\end{alignat}
\vspace{-0.7cm}
\begin{alignat}{2}
\beta_{2_L} &= \beta_{2_R} = 0, &\quad
c_L &= c_R = -1. \nonumber
\end{alignat}

The two-site model is of phenomenological interest because the Higgs potential becomes calculable in terms of the Higgs VEV.  
The dominant contribution to the Higgs mass arises from the fermionic resonances, and an interesting approximate relation holds \citep{Panico2site2}:
\beq
\frac{m_h^2}{m_t^2} \simeq \frac{N_c}{\pi^2} \frac{m_T^2 m_{\widetilde{T}}^2}{f^2\rpar{m_T^2 - m_{\widetilde{T}}^2}} \log\rpar{\frac{m_T^2}{m_{\widetilde{T}}^2}}.
\label{eq:mh/mt}
\eeq

In the above formula $m_T$ and $m_{\widetilde{T}}$ are understood as the top partner masses before the EWSB, i.e. for $\xi = 0$. In the two-site model they are given by:
\beq
m_T = f \sqrt{y_L^2 + g_4^2}, \hspace{1cm}
m_{\widetilde{T}} = f \sqrt{y_R^2 + g_1^2}.
\eeq

Equation \eqref{eq:mh/mt} is a very good approximation of the exact numerical calculation of the fermionic contribution to the Higgs mass from the potential. 
Notice that, on substituting the explicit expressions for the masses in the right hand side of Eq.~\eqref{eq:mh/mt} it becomes independent of $\xi$. We also 
remind the reader that, according to our procedure, the top mass in the left hand side of Eq.~\eqref{eq:mh/mt} should be identified with 
the $\overline{\rm MS}$ top mass at the cut-off.

\begin{figure}
\centering
\begin{tabular}{c c}
\includegraphics[scale=0.5]{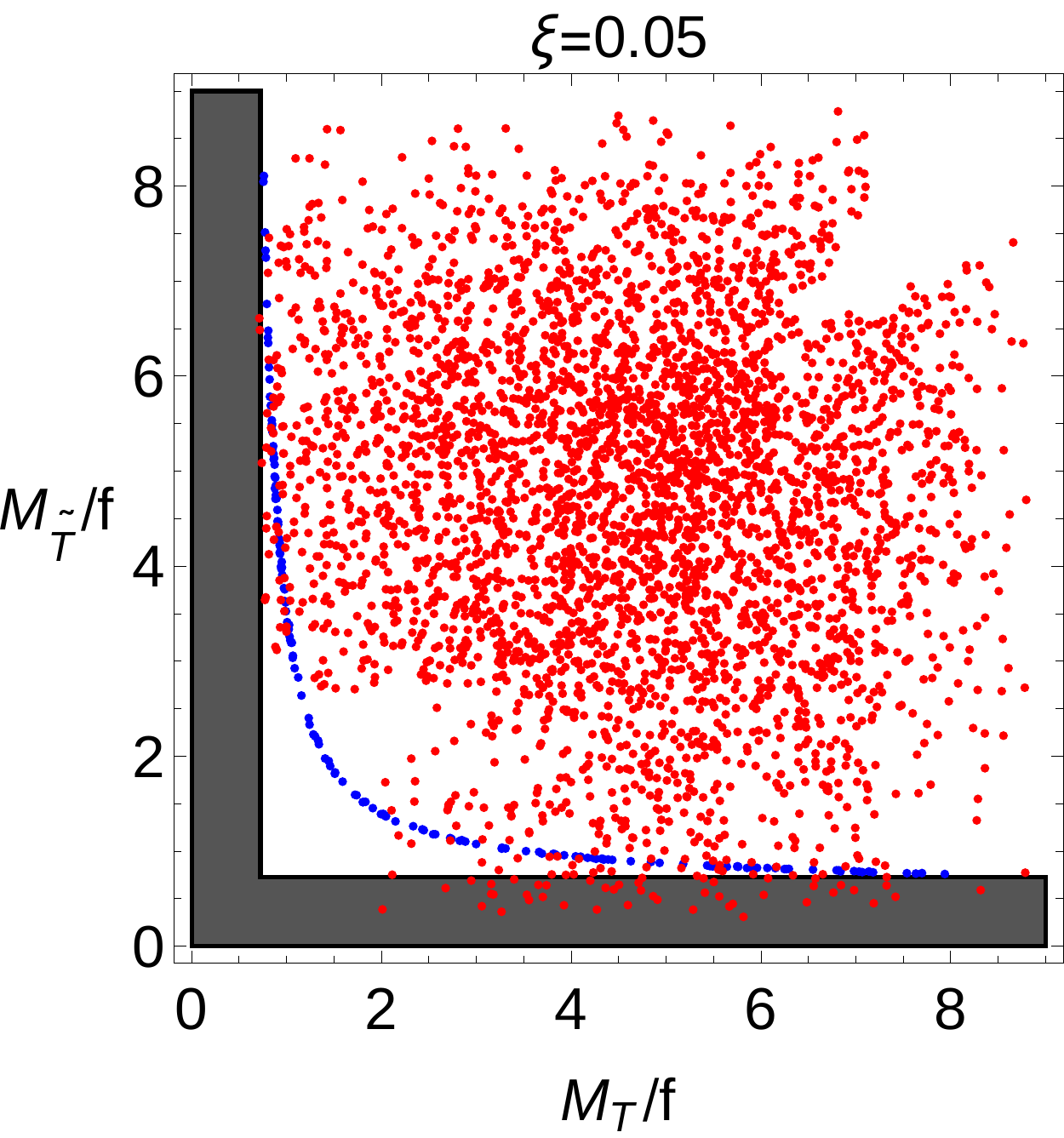} &
\hspace{0.5 cm}
\includegraphics[scale=0.5]{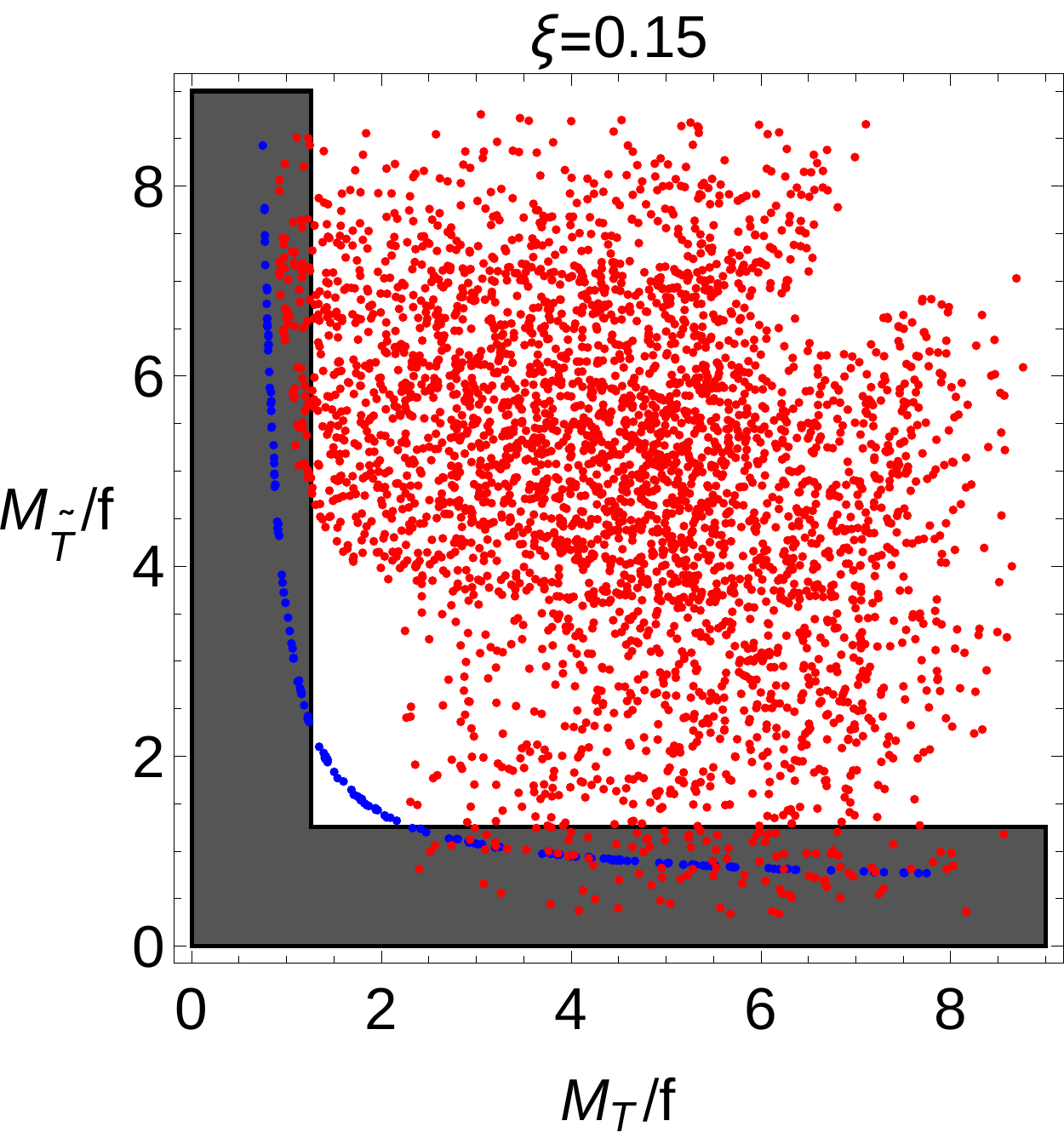}
\end{tabular}
\caption{Scattered plots showing the points allowed at 95\% probability by $m_t$ (in red) and $m_h/m_t$ (in blue) for two values of $\xi$, $\xi = 0.05$ 
(left panel) and $\xi = 0.15$ (right panel). The grey bands correspond to the area ruled out by lower bound on top  partner masses from direct searches. 
See text for more details.}
\label{fig:scatter plots}
\end{figure}

Differently from the benchmark models already discussed, in this case a (sizeable) tension with data already comes from mass constraints alone. 
It is already well known that sub-TeV top partners are needed to generate a light Higgs without large tuning of $\xi$, and a big part of the sub-TeV range 
is already ruled out by direct searches at the LHC.

In order to get a feeling of the different mass constraints,  we throw a large number of random points in the $\rpar{y_L,y_R,g_1,g_4}$ space and check 
how many of them satisfy the individual constraint from $m_t$ and $m_t/m_h$. The allowed points are plotted in Fig.~\ref{fig:scatter plots} in the plane  
of $\rpar{y_L^2 + g_4^2}^{1/2}$ and $\rpar{y_R^2 + g_1^2}^{1/2}$ (which correspond to $m_{T}/f$ and $m_{\widetilde{T}}/f$ at $\xi =0$ respectively) for two different 
values of $\xi$.\footnote{For the purpose of this analysis we have 
used the $\mathcal{O}\rpar{\xi^2}$ approximate expression for the top mass and fixed, for simplicity, $m_t\rpar{\Lambda} = 150 \, \text{GeV}$.} 
In this plane the points 
passing the $m_h/m_t$ constraint sit in a localised hyperbola-shaped region; the bigger the Higgs mass the more does this region move towards the 
upper-right corner of the plot. The red (blue) points refer to the points that are allowed at 95\% probability by the $m_t$ ($m_h/m_t$) constraint. The grey area 
correspond to the region ruled out by requiring the lightest top partner to have a mass greater than 800 GeV (for simplicity, here we neglected the 
$\xi$ dependent terms in the top partner masse). It can be seen from Fig.~\ref{fig:scatter plots} that already for $\xi \sim 0.15$ the lower bound on 
the top partner mass rules out almost all the points that satisfy the constraint from $m_h/m_t$. 

\begin{figure}[!ht]
\begin{center}
\begin{tabular}{cc}
\hspace{-10mm}\includegraphics[scale=0.4]{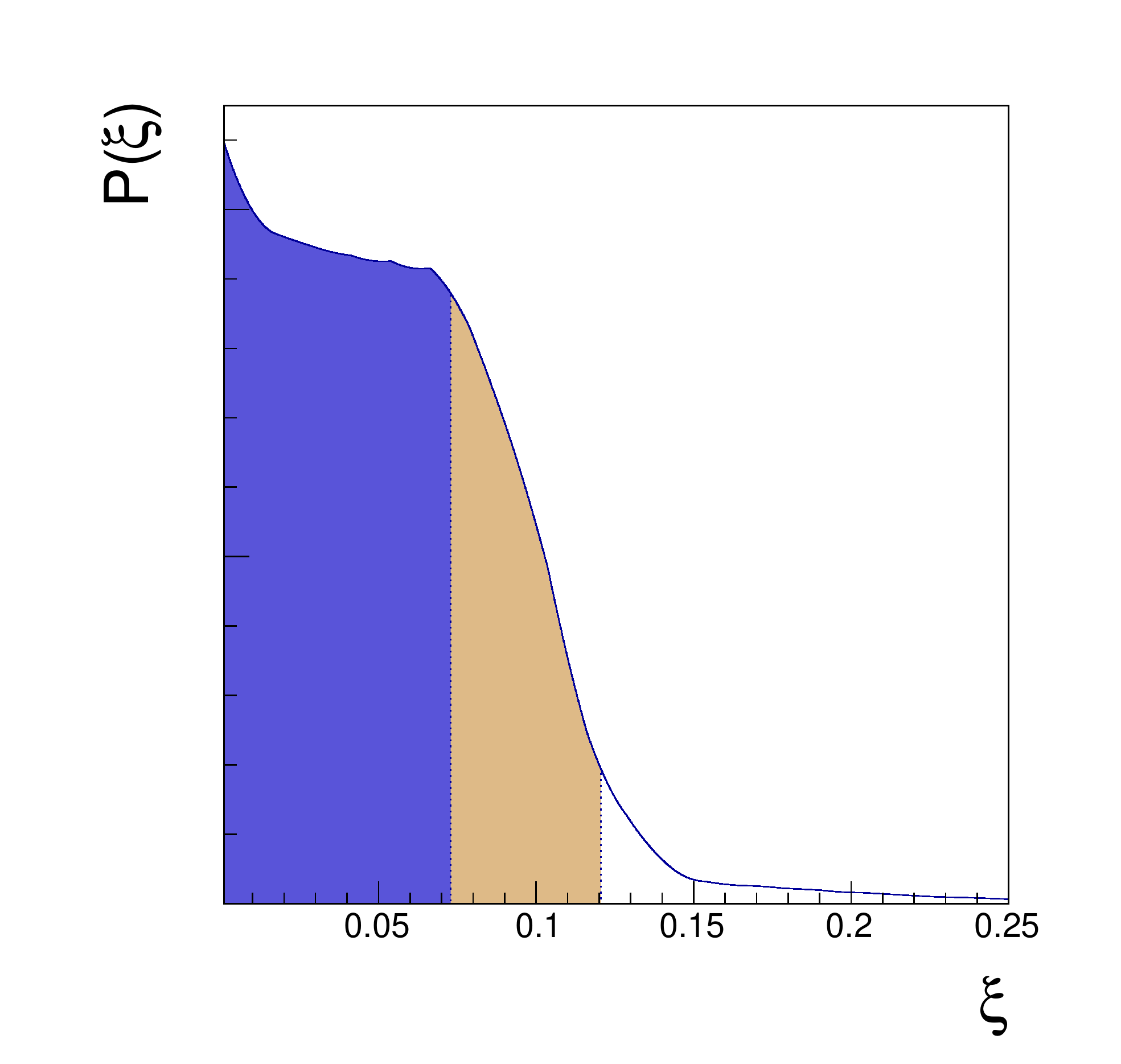} & 
\hspace{-10mm}\includegraphics[scale=0.4]{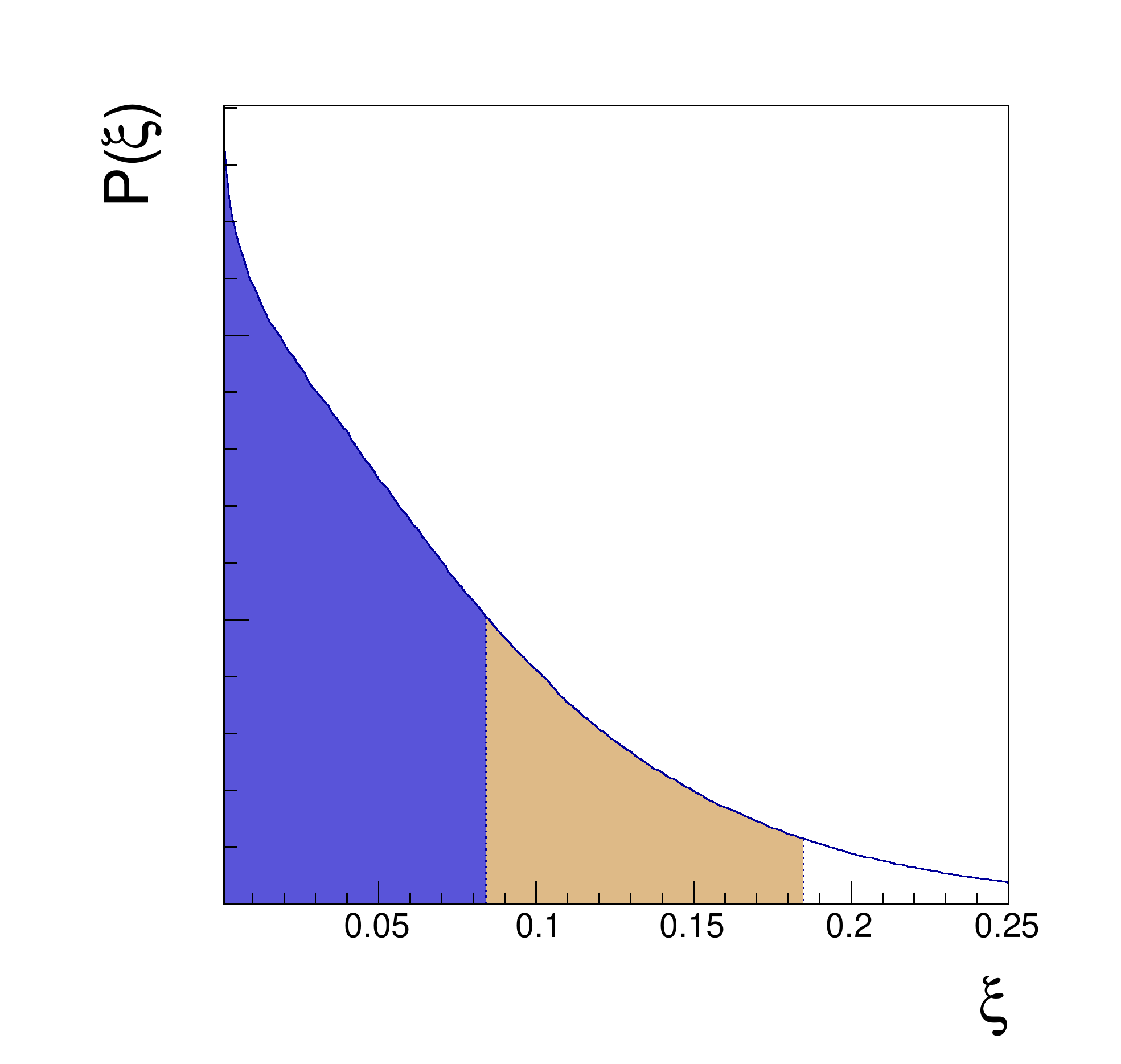} \\
\hspace{-10mm}\topinset{\includegraphics[scale=0.18]{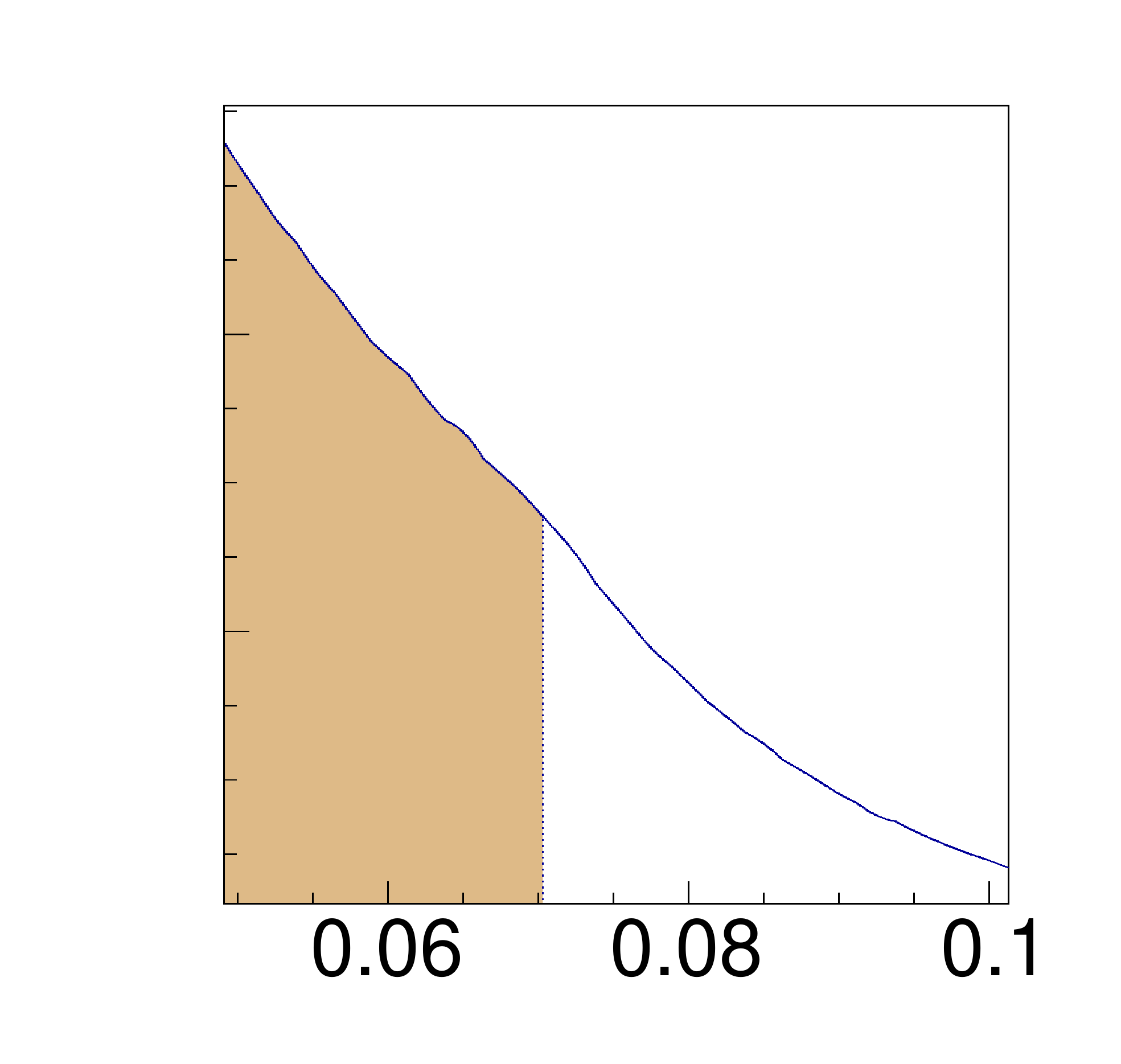}}{\includegraphics[scale=0.4]{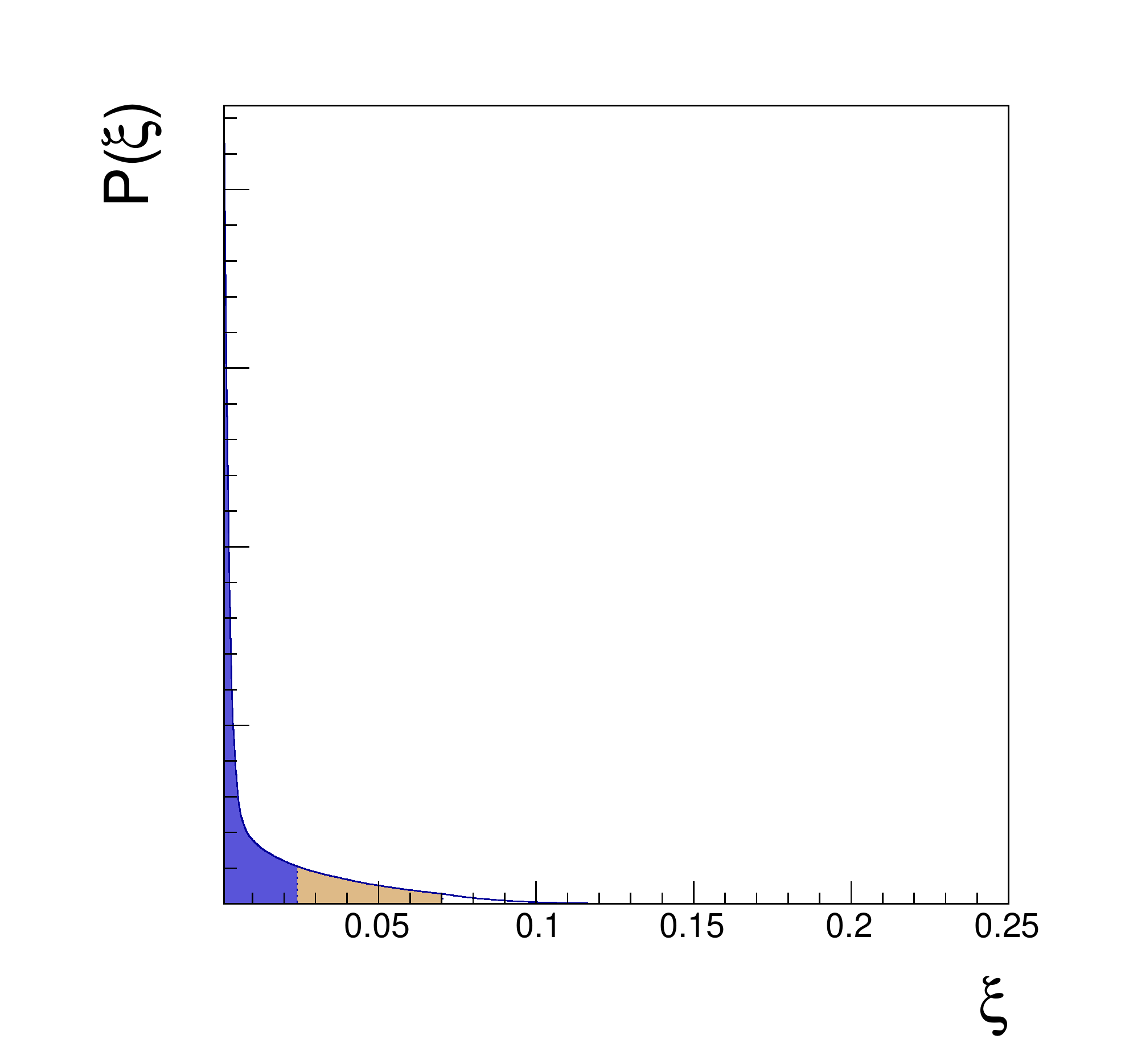}}{25pt}{35pt}&
\hspace{-10mm}\topinset{\includegraphics[scale=0.18]{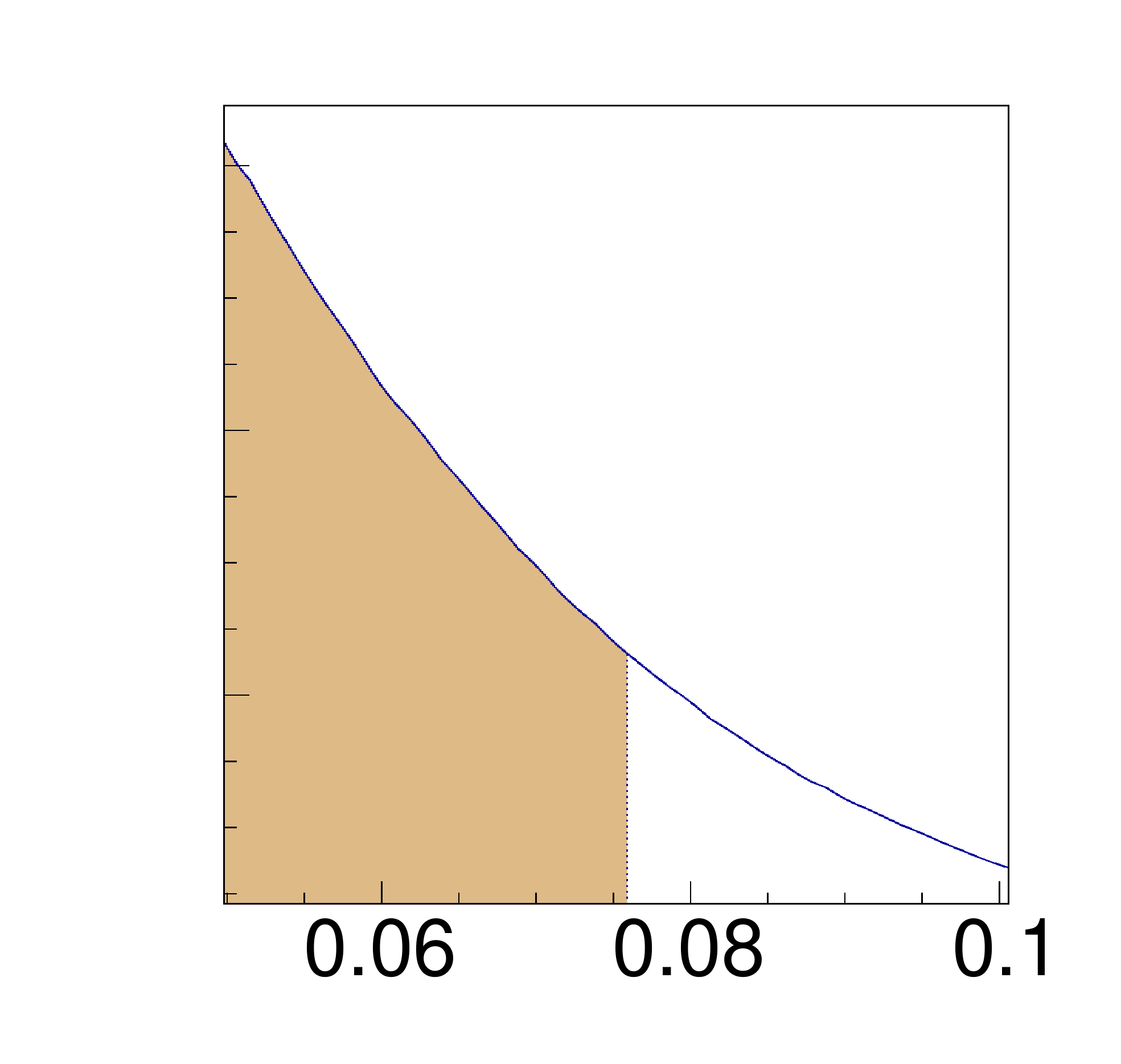}}{\includegraphics[scale=0.4]{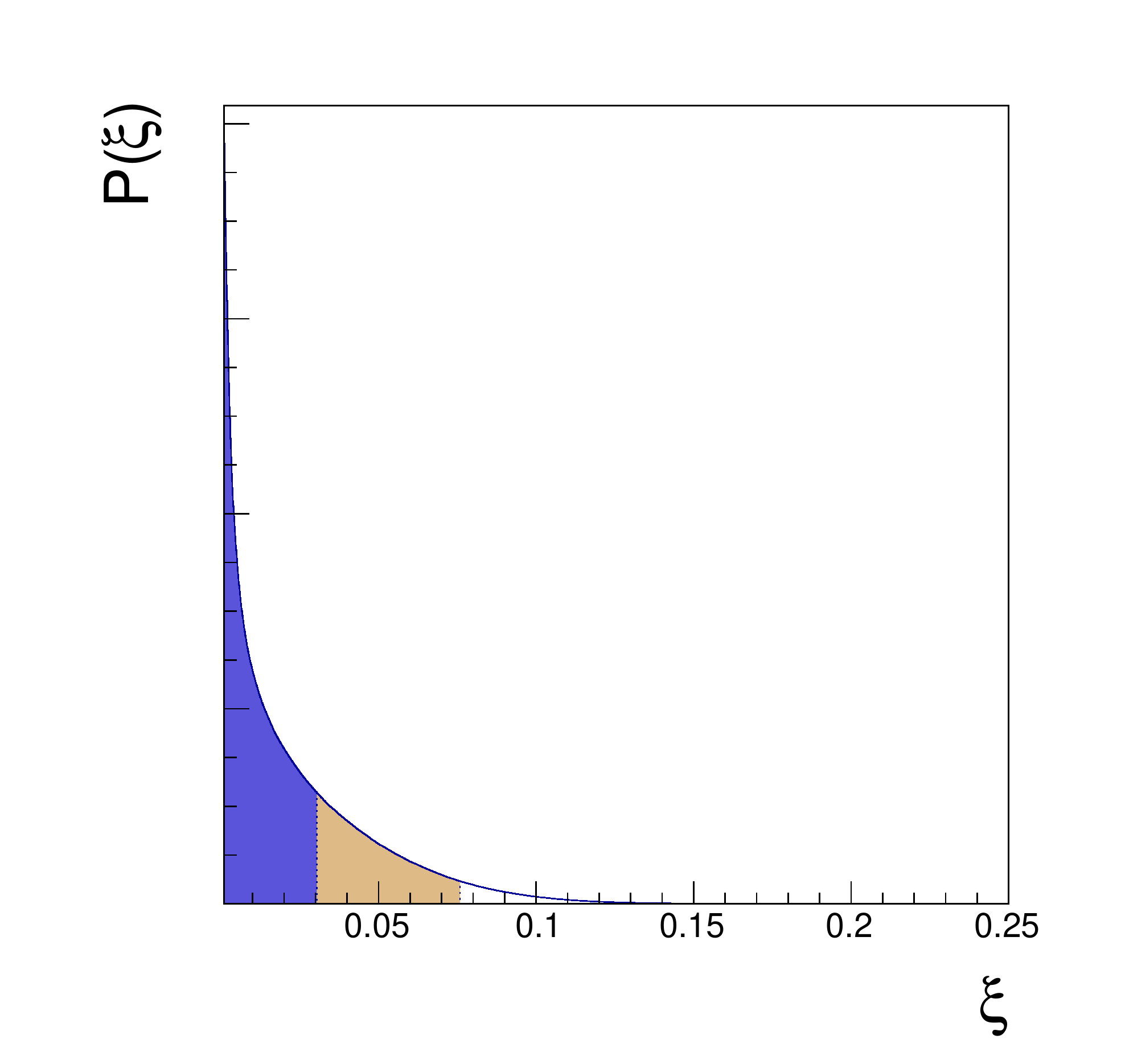}}{25pt}{35pt}\\
\end{tabular}
\caption{Fit results for the 2-site model. In the upper panel the posterior probability distributions for $\xi$ are shown when only the mass 
constraints ($m_t$, $m_h/m_t$ and 
direct searches) are considered. In the right panel an additional $\pm 30$ GeV uncertainty on the Higgs mass is included (see text for more details). 
The plots in the lower panel are the same as the corresponding plot in the upper panel once the constraints from EWPO are also considered.}
\label{fig:2sitexi}
\end{center}
\end{figure}

We now show the results from our numerical fit in Fig.~\ref{fig:2sitexi}. In the upper left panel we show the posterior of $\xi$ when only the 
mass constraints ($m_t$, $m_h/m_t$ and direct search bounds on the top partner masses) are taken into account. It is again clear from this 
figure that the mass constraints alone are enough to provide an upper 
bound on $\xi$. We would like to remark at this point that the Eq.~\eqref{eq:mh/mt} above only takes into account the fermionic contributions and 
ignores the subdominant gauge contribution which has been neglected in our analysis. Moreover, the running of the Higgs mass from the cut-off 
to the EW scale has also been neglected. In order to take into account these missing contributions in a conservative way, we also perform a fit by 
naively adding  a 30 GeV additional uncertainty to the Higgs mass. This would grant the fermionic sector the freedom to produce a Higgs mass  
sizeably bigger than the central value, thus relaxing the bound coming from the masses. The result of this fit is presented in the upper right panel of 
Fig.~\ref{fig:2sitexi}, 
which shows a clear improvement over the left plot. In the lower panel we show the same fits when the EWPO are also included. The requests from 
masses and EWPO appear to be scarcely compatible, combining in a tightly constrained final result. If no additional uncertainty on the Higgs mass 
is added (lower left panel), the 95\% probability region extends up to $\xi \sim 0.07$, with a distribution shape strongly peaked around zero. When 
the additional 30 GeV uncertainty to the Higgs mass is included the distribution shape is slightly relaxed, but still constrains $\xi$ very tightly (the lower right panel).

\section{Conclusions}

In this paper we have presented an extensive analysis of electroweak precision constraints in specific composite Higgs models. We have 
considered the $SO(5)/SO(4)$ coset including various combinations of fermionic and vectorial resonances. In particular, we have considered combinations 
of fermionic resonances living in $(\bf{2},\bf{2})$ and $(\bf{1},\bf{1})$,  and vectorial resonances in $(\bf{3},\bf{1})$ and $(\bf{1},\bf{3})$ representations of 
$SU(2)_L\otimes SU(2)_R  \sim SO(4)$. We have used a simplified effective Lagrangian describing the dynamics of such resonances at low energy and 
discussed some of its features following the approach of references \citep{PanicoGrojean} and \citep{ContinoRho}. 

We have calculated the one-loop contributions to the EWPO $\Delta \epsilon_1$, $\Delta \epsilon_3$ (thus including the light physics 
contributions from the non-standard top quark and Higgs couplings) and $\delta g_L^{(b)}$ coming from the various 
resonances.While the contributions coming from the vectorial resonances have been calculated at leading order in the elementary-composite mixings, 
the mixing effects in the fermionic sector have been resummed by numerical diagonalization. We have also presented approximate analytical expressions 
and discussed some general features of the results.

We have carried out a general NP fit to the EWPO applicable to the models under consideration and, in general, to a broader class of models producing 
negligible contributions to $\Delta \epsilon_2$ and $\delta g_R^{(b)}$. The results of this fit (shown in tables \ref{tab:obs8} and \ref{tab:corr8}) have been 
used to constrain the various NP models. We have adopted a bayesian statistical approach and, contrary to the previous literature,  performed a complete 
and systematic exploration of the parameter space using all the EWPO and the mass constraints simultaneously. 

Following our approach, at first we have studied the scenario in which only the nonlinear Higgs dynamics is considered and found a strong 95\% 
probability upper bound $\xi \lesssim 0.075$ (or equivalently 
$f \gtrsim 900\,\text{GeV}$). Although a different statistical approach has been followed, our result is generally consistent with the previous studies.  
Going further, we analysed the effect of the various resonances alone and combinations thereof.  We have shown that a scenario with a fermionic $SO(4)$ singlet 
or a single spin-1 triplet (of either  $SU(2)_L$ or $SU(2)_R$)  can considerably improve the agreement with data, relaxing the 95\% probability bound on 
$\xi$ to $\sim 0.4$ ($f \gtrsim 400\,\text{GeV}$). For the cases in which we include either both the spin-1 triplets or the combination of the fermionic 
4-plet and a single spin-1 (again either $SU(2)_L$ or $SU(2)_R$), 
we find the 95\% probability upper bound on $\xi$ to be $\sim 0.2$ ($f \gtrsim 550 \, \text{GeV}$). Finally, the two scenarios with only a fermionic 5-plet and 
only a fermionic 4-plet turn out to be very tightly constrained with the  95\% probability upper bounds on $\xi$ to be $\xi \lesssim 0.1$ and $0.02$ 
($f \gtrsim 780 \, \text{and} \, 1700\,\text{GeV}$) respectively.

We have also analysed the interesting case of the two-site model where the Higgs mass and the $\hat{S}$ parameter becomes fully calculable. 
We find that, in this case, it is rather difficult to obtain the correct Higgs mass once the experimental lower bounds on fermionic resonances are 
taken into account. In fact, a fit with only the mass constraints (i.e., $m_t$, $m_h/m_t$ and the direct search bounds) already constrains $\xi$ to be 
$\xi \lesssim 0.15$ at 95\% probability. 
The inclusion of EWPO worsens the fit further, reducing the 95\% upper bound to $\xi \lesssim 0.075$.

As far as the masses of the fermionic and spin-1 resonances are concerned, we find that the EWPO do not constrain them severely, generally allowing 
resonance masses below lower bounds from direct searches. 
Hence, fermionic top partners of mass around or below $1$ TeV and spin-1 resonances of mass around $2-3$ TeV are consistent with our fits. 
As the constraints from EWPO are not expected to improve considerably in the near future, the direct searches of resonances at the LHC 
\cite{Thamm:2015zwa} and the Higgs coupling measurements \cite{Aad:2015pla} will provide new constraints on these models.

\bigskip
\bigskip
{\bf Acknowledgments:}

We thank Roberto Contino and Luca Silvestrini for innumerable discussions and guidance throughout the course of this paper. 
The research leading to these results has received funding from the European Research Council under the European Union's
Seventh Framework Programme (FP/2007-2013) / ERC Grant Agreement n. 279972 ``NPFlavour".
\newpage
\begin{center}
\Large {\bf Appendices}
\end{center}
\appendix

\section{Mass mixings}
\label{app:mixings}

In this appendix we will present some useful formulas about the masses and the mixings of the particles in our models. We will present the most general case 
in which all the fermionic and spin-1 resonances are present. 

Spin-1 states exhibit both mass and kinetic mixings. Defining $V_0^{\mu \, T} = \rpar{W_3^\mu,B^\mu,\rho_{L\,3}^\mu,\rho_{R \,3}^\mu}$ and 
$V^{\mu \,T}_\pm = \rpar{W^\mu_\pm,\rho^\mu_{L\,\pm},\rho^\mu_{R\,\pm}}$, their mixing Lagrangian can be written as
\beq
\begin{split}
\mathcal{L}_{V}^{(mix)} =& - \frac{1}{4}V_0^{\mu \, T}\spar{-2g_{\mu\nu}\Box + 2\partial_\mu \partial_\nu}K_{V_0} V_0^\nu + \frac{1}{2} V_0^{\mu \, T} 
M_{V_0} V_{0 \mu} \\
& - \frac{1}{2}V_{+}^{\mu \, T}\spar{-2g_{\mu\nu}\Box + 2\partial_\mu \partial_\nu}K_{V_\pm} V_{-}^\nu + V_{+}^{\mu \, T} M_{V_\pm} V_{- \,\mu},
\end{split}
\eeq
where the mixing matrices are given by, 

\begin{align}
K_{V_0} =& \rpar{\begin{matrix}
1/g_{\text{el}}^2 & 0 & -2\alpha_{2_L}\cos^2\rpar{\frac{\theta}{2}} & -2\alpha_{2_R}\sin^2\rpar{\frac{\theta}{2}} \\
 & 1/g_{\text{el}}^{\prime\,2} & -2\alpha_{2_L}\sin^2\rpar{\frac{\theta}{2}} & -2\alpha_{2_R}\cos^2\rpar{\frac{\theta}{2}} \\
 & & 1/g_{\rho_L}^2 & 0\\
 & & & 1/g_{\rho_R}^2
\end{matrix}}, \\
K_{V_\pm} =& \rpar{\begin{matrix}
1/g_{\text{el}}^2 & -2\alpha_{2_L}\cos^2\rpar{\frac{\theta}{2}} & -2\alpha_{2_R}\sin^2\rpar{\frac{\theta}{2}} \\
 & 1/g_{\rho_L}^2 & 0\\
 & & 1/g_{\rho_R}^2
\end{matrix}}, \\
M_{V_0} =& \frac{1}{4} f^2 \sin^2\rpar{\theta} \rpar{\begin{matrix}
 M_{11} & M_{12} & M_{13} & M_{14} \\
        & M_{22} & M_{23} & M_{24} \\
        &        & M_{33} & M_{34} \\
        &        &        & M_{44} \\
\end{matrix}}, \\
M_{V_\pm} =& \frac{1}{4} f^2 \sin^2\rpar{\theta} \rpar{\begin{matrix}
 M_{11} & M_{13} & M_{14} \\
        & M_{33} & M_{34} \\
        &        & M_{44} \\
\end{matrix}},
\end{align}
and 
\begin{align}
M_{11} &=  \rpar{1 + a_{\rho_L}^2\cot^2\rpar{\frac{\theta}{2}} + a_{\rho_R}^2\tan^2\rpar{\frac{\theta}{2}}}, &
M_{12} &= - \rpar{1 - a_{\rho_L}^2 - M_{\rho_R}^2}, \nn \\
M_{13} &= - \csc^2\rpar{\frac{\theta}{2}}a_{\rho_L}^2, &
M_{14} &= - \sec^2\rpar{\frac{\theta}{2}}a_{\rho_R}^2, \nn \\
M_{22} &= \rpar{1 + M_{\rho_L}^2\tan^2\rpar{\frac{\theta}{2}} + M_{\rho_R}^2\cot^2\rpar{\frac{\theta}{2}}}, &
M_{23} &= -\sec^2\rpar{\frac{\theta}{2}}a_{\rho_L}^2, \\
M_{24} &= -\csc^2\rpar{\frac{\theta}{2}}\rpar{a_{\rho_L}^2}, &
M_{33} &= 4 \csc^2\rpar{\theta}a_{\rho_L}^2 , \nn \\
M_{34} &= 0, &
M_{44} &= 4 \csc^2\rpar{\theta}a_{\rho_R}^2. \nn
\end{align}

On the fermionic side only mass mixings appear. Defining $U=\rpar{t,T,X^{2/3},\tilde{T}}$ and $D=\rpar{b,B}$ we have
\beq
\mathcal{L}_\Psi^{(mix)} = -\overline{U}_L M_U U_R -\overline{D}_L M_D D_R + \rm h.c. \, ,
\eeq
where the mass matrices $M_U$ and $M_D$ are given by, 
\begin{align}
M_U =& \rpar{\begin{matrix}
0 & -y_{L4} f \sin^2 \rpar{\frac{\theta}{2}} & -y_{L4} f \cos^2 \rpar{\frac{\theta}{2}} & \frac{1}{\sqrt{2}} y_{L1} f \sin \rpar{\theta} \\
 \frac{1}{\sqrt{2}} y_{R4} f \sin \rpar{\theta} & M_4 & 0 & 0 \\
-\frac{1}{\sqrt{2}} y_{R4} f \sin \rpar{\theta} & 0 & M_4 & 0 \\
-y_{R1} f \cos \rpar{\theta} & 0 & 0 & M_1 
\end{matrix}},\\
M_D =& \rpar{\begin{matrix}
0 & -y_{L4} f \\
0 & M_4
\end{matrix}}.
\end{align}

Below, we also report the expressions for the physical masses at the leading order in $\xi$ and for $\alpha_{2_L} = \alpha_{2_R} =0 $, 
\begin{align}
M_Z^2 &= \frac{1}{4} f^2 \xi \rpar{\frac{g_{\text{el}}^2 g_{\rho_L}^2}{g_{\text{el}}^2 + g_{\rho_L}^2}+\frac{g_{\text{el}}^{\prime\,2} 
g_{\rho_R}^2}{g_{\text{el}}^{\prime\,2} + g_{\rho_R}^2}} + \mathcal{O}\rpar{\xi^2}, & 
M_{\rho_L}^0 &= M_{\rho_L}^\pm = M_{\rho_L}^2 \rpar{1 + \frac{g_{\text{el}}^2}{g_{\rho_L}^2}} + \mathcal{O}\rpar{\xi}, \nn \\
M_W^2 &= \frac{1}{4} f^2 \xi \frac{g_{\text{el}}^2 g_{\rho_L}^2}{g_{\text{el}}^2 + g_{\rho_L}^2} + \mathcal{O}\rpar{\xi^2}, &
M_{\rho_R}^0 &= M_{\rho_R}^2 \rpar{1 + \frac{g_{\text{el}}^{\prime\,2}}{g_{\rho_R}^2}} + \mathcal{O}\rpar{\xi}, \nn \\  
m_t^2 &= \frac{1}{2} f^2 \xi \frac{\rpar{g_1 \, y_{L4} \, y_{R4} - g_4 \, y_{L1} \, y_{R1} }^2}{\rpar{g_1^2 + y_{R1}^2} \rpar{g_4^2 + y_{L4}^2}} + 
\mathcal{O}\rpar{\xi^2}, &
M_{\rho_R}^\pm &= M_{\rho_R}^2 + \mathcal{O}\rpar{\xi}, \\
m_T^2 &= m_B^2 = M_4^2\rpar{1 + \frac{y_{L4}^2}{g_4^2}} + \mathcal{O}\rpar{\xi}, &
m_{X^{2/3}}^2 &= m_{X^{5/3}}^2 = M_4^2 + \mathcal{O}\rpar{\xi}, \nn \\
& & \hspace{-4.2cm}m_{\widetilde{T}}^2 = M_1^2\rpar{1 + \frac{y_{R1}^2}{g_1^2}} + \mathcal{O}\rpar{\xi}. \nn
\end{align}

%
\section{Light physics contributions to the $\epsilon$ parameters}
\label{app:epslightpart}

In this appendix we show the expressions for the top quark and Higgs boson contributions to $\epsilon_1$ and $\epsilon_3$.

The top quark contribution is model dependent because the coupling of the physical top quark to the physical electroweak gauge bosons 
depends on the particular choice of the representations of the composite fermions. 
In order to be model independent, we will present the formulas assuming some general coupling between the top quarks and the gauge bosons. 
Keeping the contributions to $\hat{T}$ and $\hat{S}$ separate, they can be written as,
\beq
\begin{split}
\epsilon_1 \big|_{\rm top} =& \; \hat{T}\big|_{\rm top} + \frac{3}{8\pi^2}\frac{t}{\sqrt{4t-1}}\arctan\rpar{\frac{\sqrt{4t-1}}{1-2t}} \rpar{\rpar{1-2t}
\rpar{g_{LZ}^2 + g_{RZ}^2} - 2 \rpar{1-6t}g_{LZ} \, g_{RZ}} \\
& \frac{3}{16\pi^2}\spar{\rpar{\frac{2}{3}-2t}\rpar{g_{LZ}^2 + g_{RZ}^2} + 12 \, t \, g_{LZ} \, g_{RZ}} \, ,
\end{split}
\label{eq:eps1 top}
\eeq
\beq
\begin{split}
\epsilon_3 \big|_{\rm top} =& \; \hat{S}\big|_{\rm top} + \frac{1}{8\pi^2}\frac{c}{s}\sqrt{4t-1}\arctan\rpar{\frac{\sqrt{4t-1}}{1-2t}}\rpar{\rpar{1-t}g_\alpha + 3t g_\beta} \\
&-\frac{3}{8\pi^2}c^2\frac{t}{\sqrt{4t-1}}\arctan\rpar{\frac{\sqrt{4t-1}}{1-2t}}\spar{ \rpar{1-2t}\rpar{g_{LZ}^2+g_{RZ}^2} -2\rpar{1-6t} g_{LZ} \, g_{RZ} } \\
& + \frac{1}{8\pi^2}c^2\rpar{\rpar{1-3t}\rpar{g_{LZ}^2 + g_{RZ}^2} + 18 \, t \, g_{LZ} \, g_{RZ}} \\
& +\frac{1}{48\pi^2}\frac{c}{s}\rpar{\rpar{13-12t}g_\alpha + 3(12t-1) g_\beta} + c^2 e_4 \big|_{top} \, ,
\end{split}
\label{eq:eps3 top}
\eeq
where $\hat{T}\big|_{\rm top}$ and $\hat{S}\big|_{\rm top}$ are given by,
\begin{align}
\hat{T}\big|_{\rm top} =& \frac{3t}{16\pi^2c^2} \spar{ 4 \log\rpar{\frac{\Lambda}{m_t}} \rpar{g_{L1}^2 + g_{R1}^2 -\rpar{g_{L3} - g_{R3}}^2 } + g_{L1}^2 + g_{R1}^2 } \, ,\\
\hat{S}\big|_{\rm top} =& \frac{1}{16\pi^2}\frac{c}{s} \rpar{ 4 \log\rpar{\frac{\Lambda}{m_t}} g_\alpha - g_\alpha + g_\beta  } \, .
\end{align}

In the above,  $t=m_t^2/M_Z^2$ and $g_{L(R)i}$ indicates the coupling of the left (right) chirality top pair with the $i$-th EW gauge boson (for example, 
in the SM, $g_{LZ} = -g/c\rpar{1/2 - 2/3s^2}$). The quantities $g_\alpha$ and $g_\beta$ correspond to the combinations,
\begin{align}
g_\alpha &= g_{L3} \, g_{LB} + g_{R3} \, g_{RB} \, , \\
g_\beta &= g_{L3} \, g_{RB} + g_{R3} \, g_{LB} \, .
\end{align}

While $\hat{S}\big|_{\rm top}$ and $\hat{T}\big|_{\rm top}$ were largely used in the literature, the remaining contributions in 
Eq.\eqref{eq:eps1 top}-\eqref{eq:eps3 top}  have been so far neglected.
Note that the above expressions include the full top quark contribution, hence in order to compute  $\Delta \epsilon_{1 \, , 3}$, one has to 
subtract the SM contribution obtained from the above expressions by setting $g_{L(R)i}$ to their SM values.
For this reason, we also kept the dependence on $e_4$ explicit in $\epsilon_3$  as it will always cancel when subtracting the SM result. 
This is because $e_4$ is computed from the photon self energies and in order to respect the exact $U(1)_Q$ symmetry photon couplings remain unchanged.

On the other hand, the effects of the nonlinear Higgs dynamics are only dictated by the choice of the coset, i.e. $SO(5)/SO(4)$ in our case. 
These were first computed in \cite{RychkovSO(5)} and we report them below for convenience, 
\begin{align}
\Delta \epsilon_1 \big|_H &= -\frac{3g^{\prime \, 2}_{\text{el}}}{32 \pi^2} \xi \spar{ \log{\rpar{\frac{\Lambda}{M_Z}}} + f_1 (h) }, 
\label{eq:Deps1 Higgs}\\
\Delta \epsilon_3 \big|_H &= \frac{g^2}{96 \pi^2} \xi \spar{ \log{\rpar{\frac{\Lambda}{M_Z}}} + f_3(h) },
\label{eq:Deps3 Higgs}
\end{align}
where $h=M_H^2/M_Z^2$ and the loop functions are given by \cite{ContinoSalvarezza}
\begin{align}
\begin{split}
f_{1}(h) & =\frac{1}{s^{2}}\left(-\frac{5c^{2}}{12}+\frac{h^{2}}{6}-\frac{7h}{12}+\frac{31}{18}\right) \\
           & - \frac{\log(h)}{12s^{2}\left(c^{2}-h\right)} \left[\left(c^{2}+5\right)h^{3}-\left(5c^{2}+12\right)h^{2}+2\left(9c^{2}+2\right)h-4c^{2}-h^{4}\right]\\
           & -\frac{c^{4}}{s^{2}\left(h-c^{2}\right)}\log(c)+\frac{h\left(h^{3}-7h^{2}+20h-28\right)}{6s^{2}\sqrt{(4-h)h}}\, \arctan\left(\sqrt{\frac{4}{h}-1}\right) \, , \\
\end{split} 
\\
\begin{split}
f_{3}\left(h\right) & =\left(-h^{2}+3h-\frac{31}{6}\right)+\frac{1}{4}\left(2h^{3}-9h^{2}+18h-12\right)\log(h) \\
                           & -\frac{\left(2h^{3}-13h^{2}+32h-36\right)h}{2\sqrt{(4-h)h}}\, \arctan\left(\sqrt{\frac{4}{h}-1}\right)\, .
\end{split}
\end{align}

The above functions evaluate to approximately $f_1(h) \simeq -0.48$ and $f_3(h) \simeq 0.72$.

%
\providecommand{\href}[2]{#2}\begingroup\raggedright\endgroup

%
\end{document}